\documentclass[preprint2,tighten,margin=0.5in]{aastex62_reep}

\usepackage{epstopdf}
\usepackage{subfigure}  
\usepackage{amsmath}
\usepackage{hyperref}

\usepackage{longtable}

\usepackage{fancyhdr}   
\fancypagestyle{specialfooter}{%
 \fancyhf{}
 
 \fancyfoot[C]{Distribution Statement A.  Approved for public release.  Distribution unlimited.}  }

\shorttitle{EVE Irradiance Scalings}
\shortauthors{Reep et al.}

\begin{document}

\title{Solar Flare Irradiance: Observations and Physical Modeling}

\author[0000-0003-4739-1152]{Jeffrey W. Reep}
\author{David E. Siskind}
\author[0000-0001-6102-6851]{Harry P. Warren}

\affiliation{Space Science Division, Naval Research Laboratory, Washington, DC 20375, USA}

\email{jeffrey.reep@nrl.navy.mil}

\begin{abstract}
We examine SDO/EVE data to better understand solar flare irradiance, and how that irradiance may vary for large events.  We measure scaling laws relating GOES flare classes to irradiance in 21 lines measured with SDO/EVE, formed across a wide range of temperatures, and find that this scaling depends on the line formation temperature.  We extrapolate these irradiance values to large events, exceeding X10.  In order to create full spectra, however, we need a physical model of the irradiance.  We present the first results of a new physical model of solar flare irradiance, NRLFLARE, that sums together a series of flare loops to calculate the spectral irradiance ranging from the X-rays through the far ultraviolet ($\approx 0$ to $1250$ \AA), constrained only by GOES/XRS observations.  We test this model against SDO/EVE data.  The model spectra and time evolution compares well in high temperature emission, but cooler lines show large discrepancies.  We speculate that the discrepancies are likely due to both a non-uniform cross section of the flaring loops as well as opacity effects.  We then show that allowing the cross-sectional area to vary with height significantly improves agreement with observations, and is therefore a crucial parameter needed to accurately model the intensity of spectral lines, particularly in the transition region from $4.7 \lesssim \log{T} \lesssim 6$.    
\end{abstract}


\nopagebreak

\section{Introduction}
\label{sec:intro}
\thispagestyle{specialfooter}   

Measuring solar irradiance variation due to solar flares is critical to understanding the connection between the Sun and the Earth's ionosphere-thermosphere-mesosphere (ITM) system.  The irradiance of many spectral lines and continuum emission increases sharply during flares, which can directly impact the ITM \citep{hayes2017,qian2019}.  Further, since the optical depth of the ITM is strongly wavelength dependent, it is crucial to understand how the irradiance increase scales with wavelength for different size flares in order to understand how variations in solar irradiance affect the ITM.  

Empirical modeling has been used to predict irradiance scalings for flares.  For example, the Flare Irradiance Spectral Model (FISM) uses measurements of the irradiance to predict a spectrum from 1\,\AA\ to 1050\,\AA, including both a daily background component \citep{chamberlin2007} as well as a flare component \citep{chamberlin2008,chamberlin2020}.  While these models compare well in general when compared against observed spectra, they are limited by their inability to extrapolate beyond the historical database.  

These limitations suggest that physical models may offer an important advance in flare modeling.  There are many advantages offered by a physical model of irradiance.  Irradiance observations cover a broad wavelength and temperature range, and observe the whole flare rather than a small field-of-view, as in spectral observations with the Hinode EUV Imaging Spectrometer (EIS) or IRIS.  Through direct comparison to observed spectra, we can test the basic assumptions of the model at many temperatures simultaneously to better understand the energy release and transport processes in flares.  For example, in this paper, we will describe the need for a non-constant loop cross-sectional area to reproduce line intensities.  Second, the synthetic spectra can be calculated at any cadence, spectral resolution, or spatial resolution down to the scales of the simulation itself, which are often significantly finer than can be resolved by observations.  In this work, we synthesize spatially-unresolved spectra at a 1 s cadence and spectral resolution as small as 0.01\,\AA\ in the X-rays.  Third, the model can be extrapolated to unobserved flare sizes, for example, to Carrington-like events or flares intense enough to saturate GOES/XRS (\textit{e.g.} 28 October 2003 saturated XRS-A and 4 November 2003 saturated both channels).  Synthetic spectra for X50 or X100 flares can then be used to predict their impacts on the ITM.  Fourth, the model can be used to predict unobserved or poorly observed wavelength ranges, such as the range near 10--50\,\AA, for which only sparse spectrally-resolved observations exist.

Physical models such as the NRLEUV model, based on measurements of emission measure distributions with extreme ultraviolet (EUV) spectra, provide a physically-motivated spectrum \citep{warren1998,warren2006}, but only during quiescent (non-flaring) time periods.  In this work, we introduce a new model, named NRLFLARE, which calculates the full flare spectrum at high cadence and spectral resolution.  This model is driven by hydrodynamic simulations of flaring loops composing the event, constrained by the soft X-ray emission measured by NOAA's Geostationary Orbital Environmental Satellites (GOES) X-ray Sensors (XRS).  In a previous paper, \citet{reep2020} used the \texttt{ebtel++} zero-dimensional model \citep{barnes2016a} to simulate the loops, showing that it can reproduce accurately both X-ray lightcurves and spectra.  While the 0D model is convenient for running a vast number of simulations, it does not account for the spatial variation of the loop properties, simulating only a coronal average, and therefore lacks modeling of more detailed physical processes known to occur in flares.  In this work, we therefore employ the more detailed field-aligned \texttt{HYDRAD} model \citep{bradshaw2003, bradshaw2013}, as explained in Section \ref{sec:modeling}.  

Unfortunately, building a physical model of the irradiance is fraught with difficulties: all of the important physics must be accounted for and treated correctly in the modeling.  Although the modeling of flares has progressed significantly in recent years, there are still many outstanding issues (loop geometry, wave motions, optical depth, late phase heating, etc.), and each of these contribute to errors in the synthetic irradiance.  In this first attempt at creating a physical model of irradiance, we show that the model can reproduce much of the behavior of the flare, but that the ``standard'' methodology of simulating flaring loops is deficient in reproducing the time series of many spectral lines, particularly at cooler temperatures.  In particular, a non-uniform loop cross-sectional area \citep{emslie1992,mikic2013} is required to reproduce emission simultaneously in hot and cool lines.  

In this work, we synthesize spectra for a few flares of various GOES class and duration, and compare synthetic light curves to observations in various spectral lines.  In \ref{sec:eve}, we first examine how the observed irradiance in these lines vary with flare class, and show that there is a relation between an ion's formation temperature and the irradiance.  In \ref{sec:modeling}, we explain the model, how we synthesize spectra, and then examine three flares in detail.  We show that in all of the flares examined, there is a discrepancy between the model and observations which requires a non-uniform loop cross-sectional area to resolve.

\section{Scaling of Spectral Line Irradiance with Flare Class}
\label{sec:eve}

To better understand how spectral irradiance varies across flare classes, we first examine observations across a broad sample of flares.  The EUV Variability Experiment (EVE; \citealt{woods2012}) onboard the Solar Dynamics Observatory (SDO; \citealt{pesnell2012}) provides direct measurements of the total solar irradiance across a wide spectral range.  EVE has two spectrographs, Multiple Extreme ultraviolet Grating Spectrographs (MEGS), labeled A and B, with wavelength coverage between 50--370\,\AA\ and 350--1050\,\AA, respectively.  EVE has measured the solar irradiance continuously since its launch in early 2010, although MEGS-A failed in May 2014, reducing the wavelength coverage since then.  

We have measured the irradiance variation of 21 spectral lines with SDO/EVE, version 6 data, for flares that occurred between 1 May 2010 and 13 May 2014 (before the failure of MEGS A).  We search all events in the GOES catalogue for that period, and then prune the events based on a few criteria.  We throw out an event if any of the following apply: (1) there are any data gaps in the GOES/XRS data, (2) the irradiance in any of the EVE lines gives a bad value (negative, for example), (3) any basic values such as the full-width-at-half-maximum (FWHM) duration of the event are undefined.  

The flares range in class from B1 to X8.  The spectral lines range in formation temperature from the cool chromosphere ($\approx$ 10 kK) through hot flaring temperatures ($\approx$ 20 MK).  We background subtract a pre-flare EVE spectrum averaged over one minute, and then measure the irradiance (photons\,s$^{-1}$\,cm$^{-2}$) over the width of each line.   

In Figure \ref{fig:eve}, we plot the maximum irradiance of 21 spectral lines in each event (blue) as a function of GOES class.  In each case, we fit a line in log-log space (equivalently, a power law fit) to the data (blue) with a Theil-Sen non-parametric regressor \citep{theil1992,sen1968}, which is more robust than least squares and insensitive to outliers, with the slope indicated in each case with 1-$\sigma$ uncertainty.  That is, we fit a function of the form $\log{I} = a + b \log{F_{\text{1--8\,\AA}}}$, or equivalently, $I = 10^{a} F_{\text{1--8\,\AA}}^{b}$, for $I$ the irradiance of a spectral line and $F_{\text{1--8\,\AA}}$ the peak GOES/XRS-B irradiance.  Between \ion{Fe}{18} and \ion{Fe}{23}, the distributions may be bimodal, and therefore we show the fits for all flares (blue) and for flares only above C-class have been fit (red).  In the case of Fe XXIV 192\,\AA, we only fit the slope for flares above M-class since the line is blended with weaker Fe XI and Fe XII lines, which likely dominate the signal in smaller and cooler flares (perhaps even in the gradual phase of larger flares).  
\begin{figure*}
    \centering
    \includegraphics[width=0.32\textwidth]{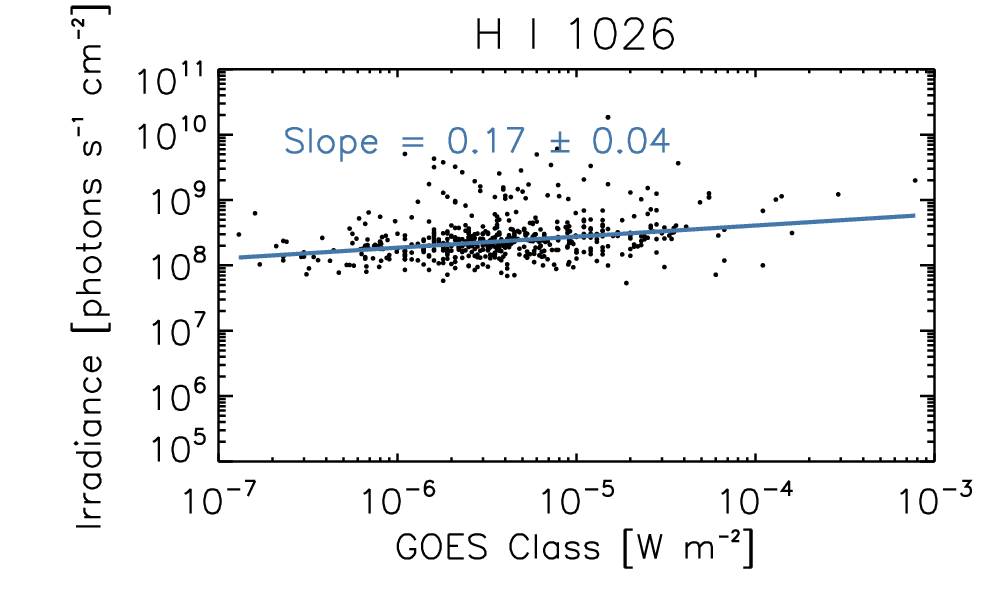}
    \includegraphics[width=0.32\textwidth]{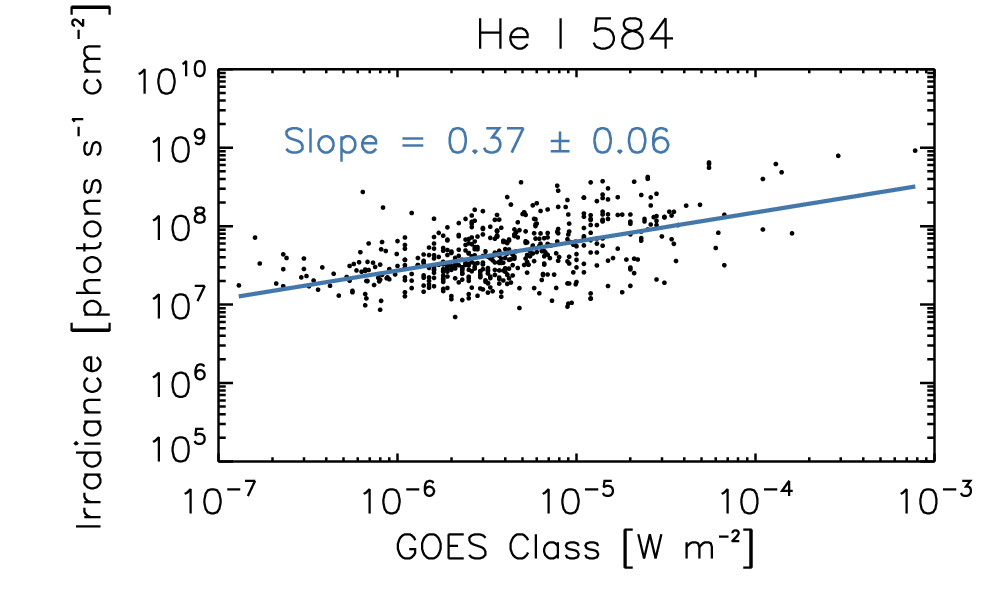}
    \includegraphics[width=0.32\textwidth]{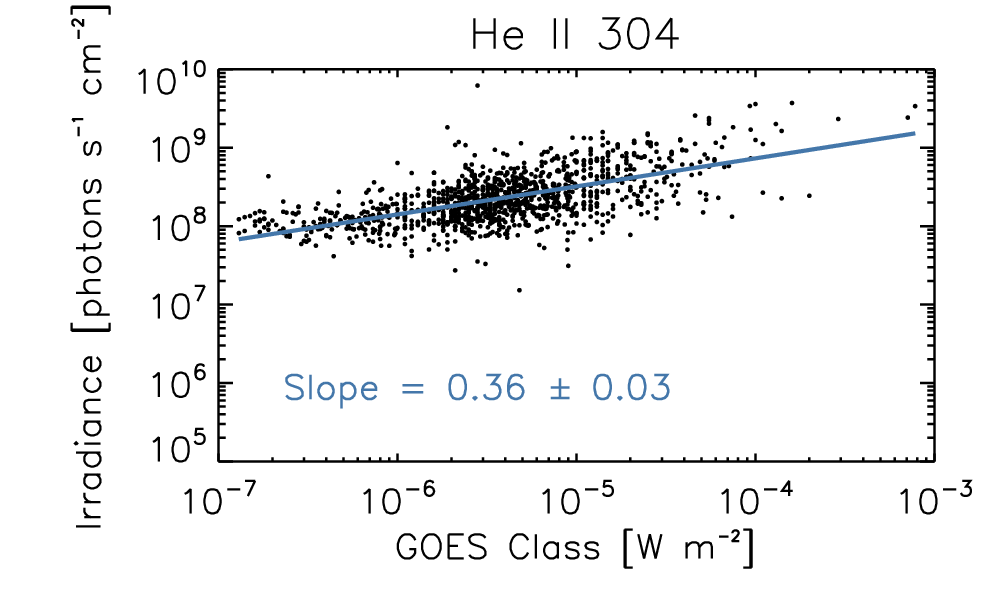}
    \includegraphics[width=0.32\textwidth]{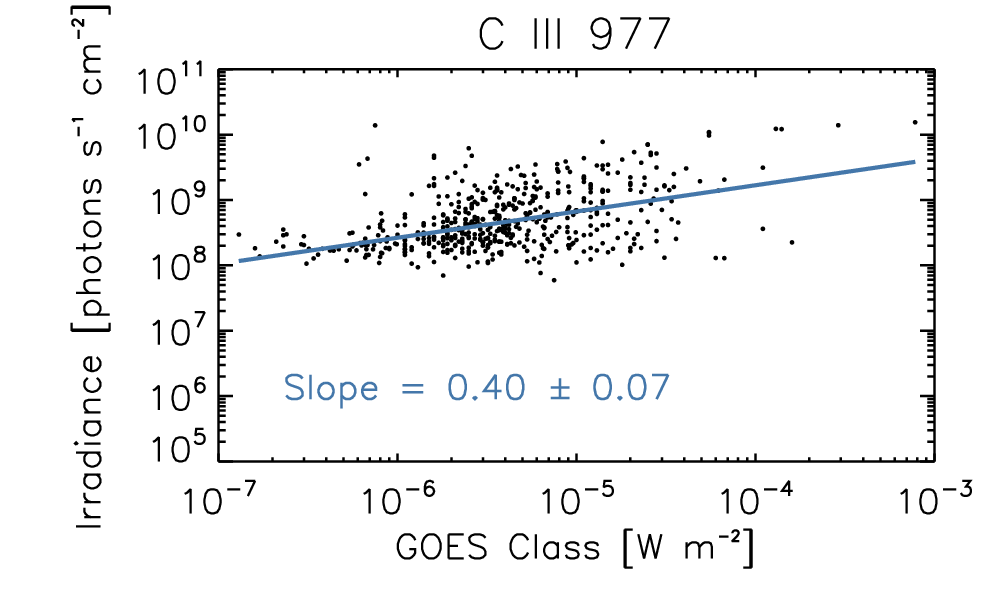}
    \includegraphics[width=0.32\textwidth]{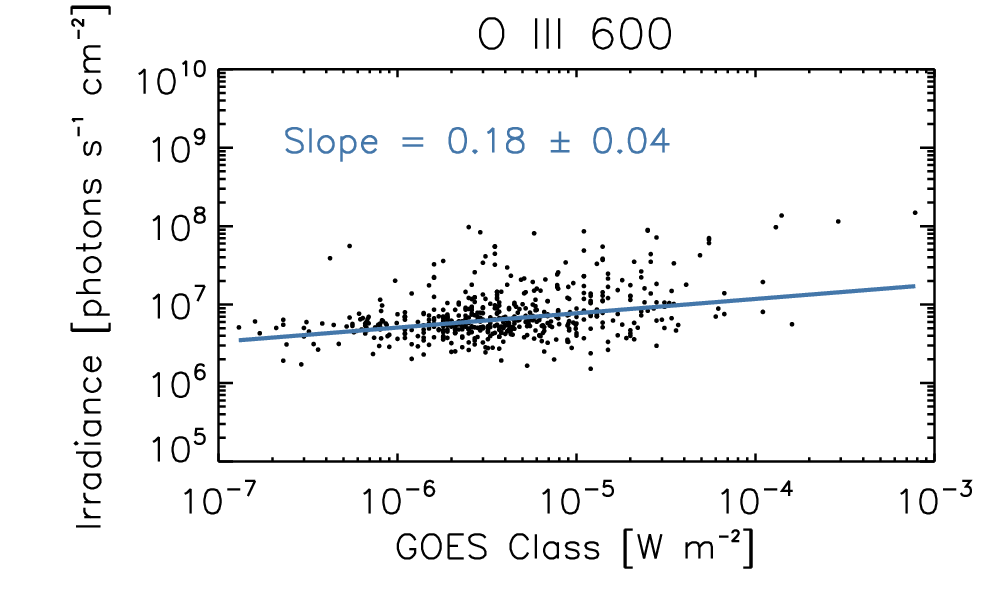}
    \includegraphics[width=0.32\textwidth]{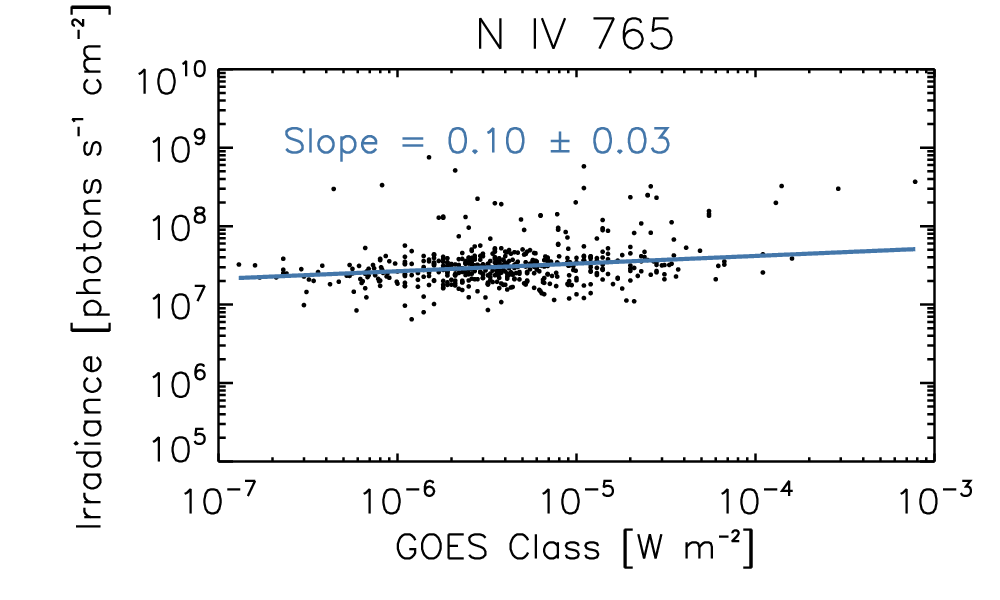}
    \includegraphics[width=0.32\textwidth]{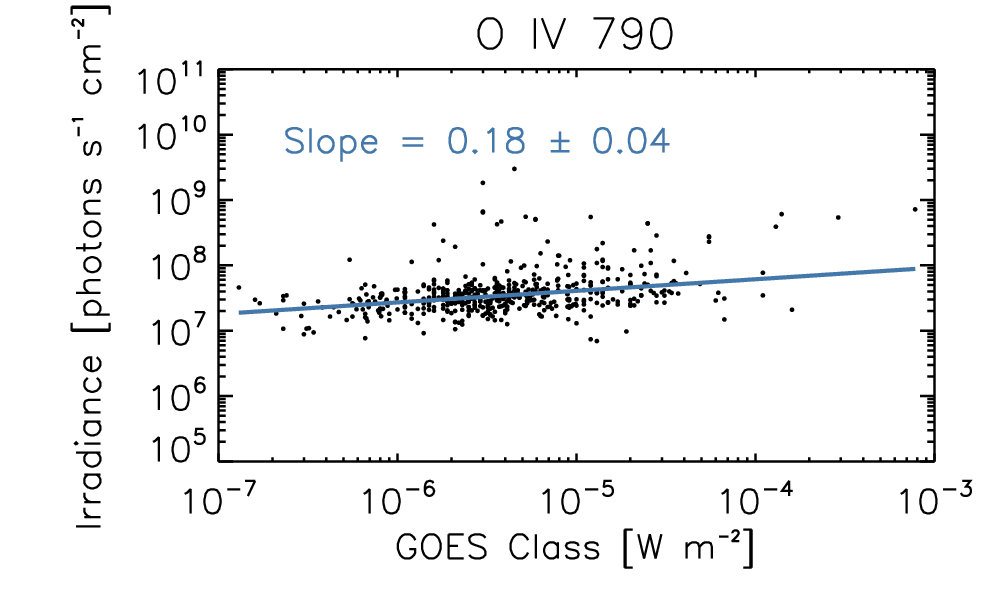}
    \includegraphics[width=0.32\textwidth]{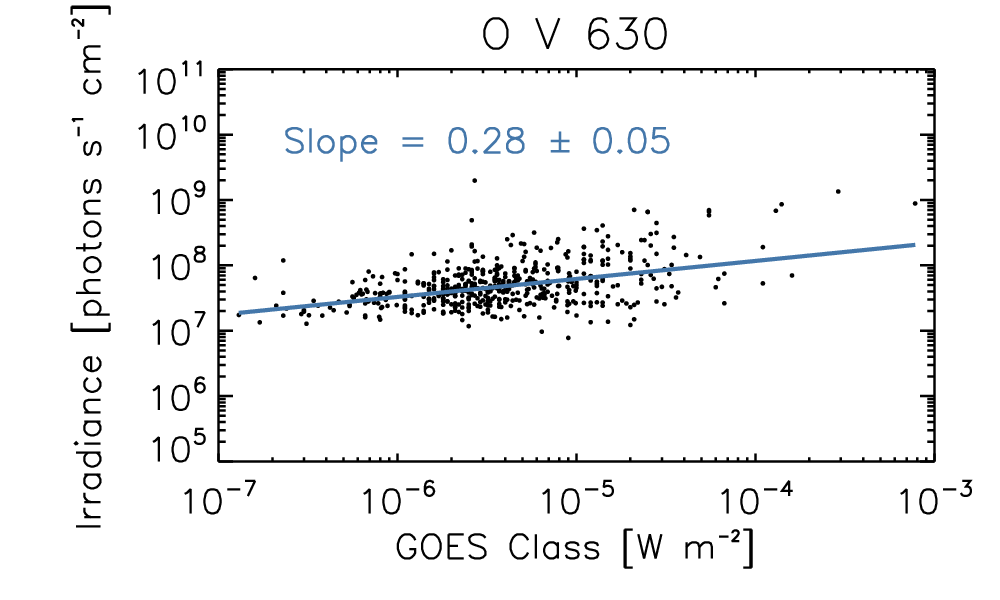}
    \includegraphics[width=0.32\textwidth]{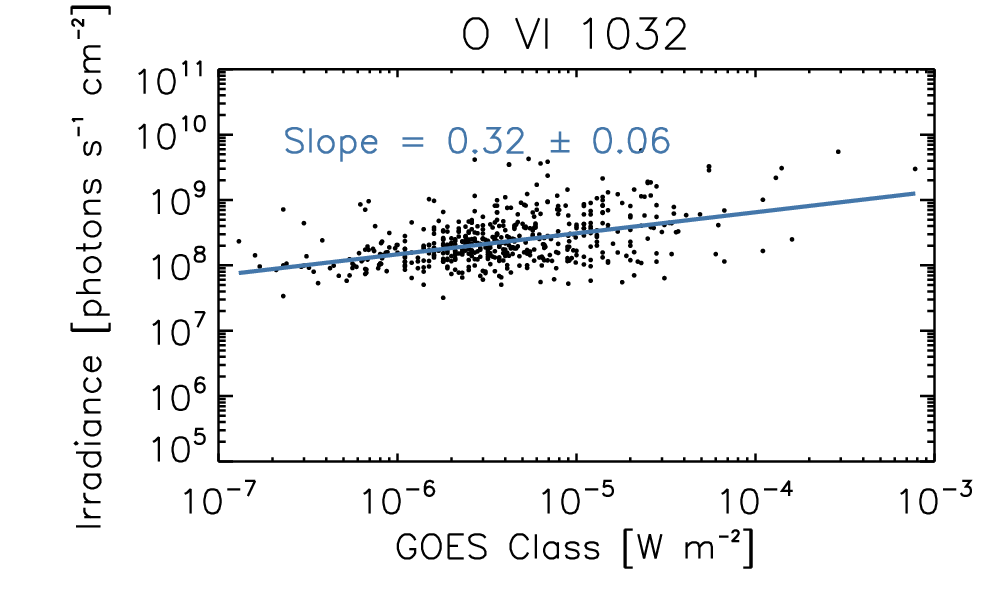}
    \includegraphics[width=0.32\textwidth]{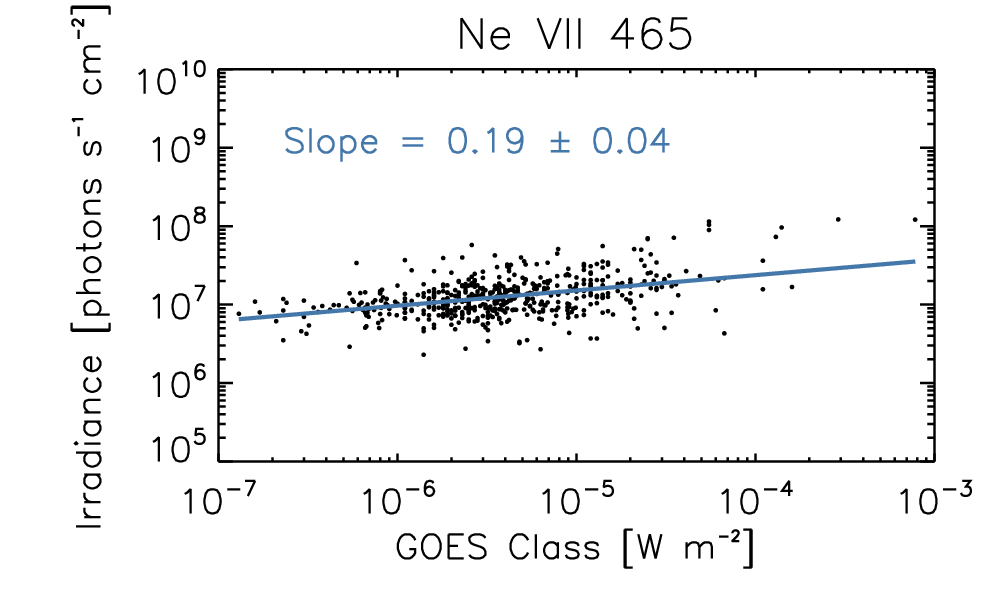}
    \includegraphics[width=0.32\textwidth]{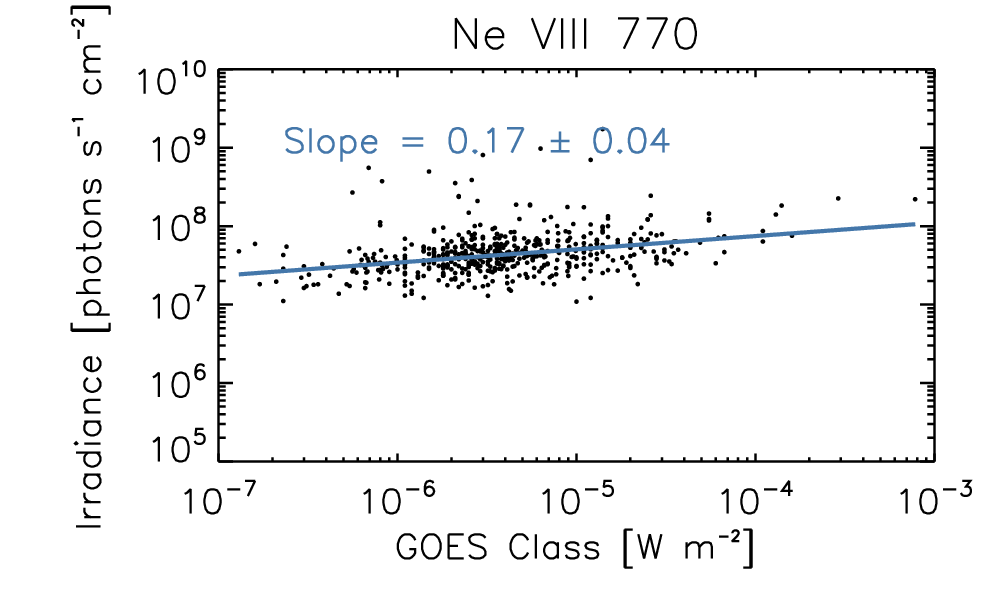}
    \includegraphics[width=0.32\textwidth]{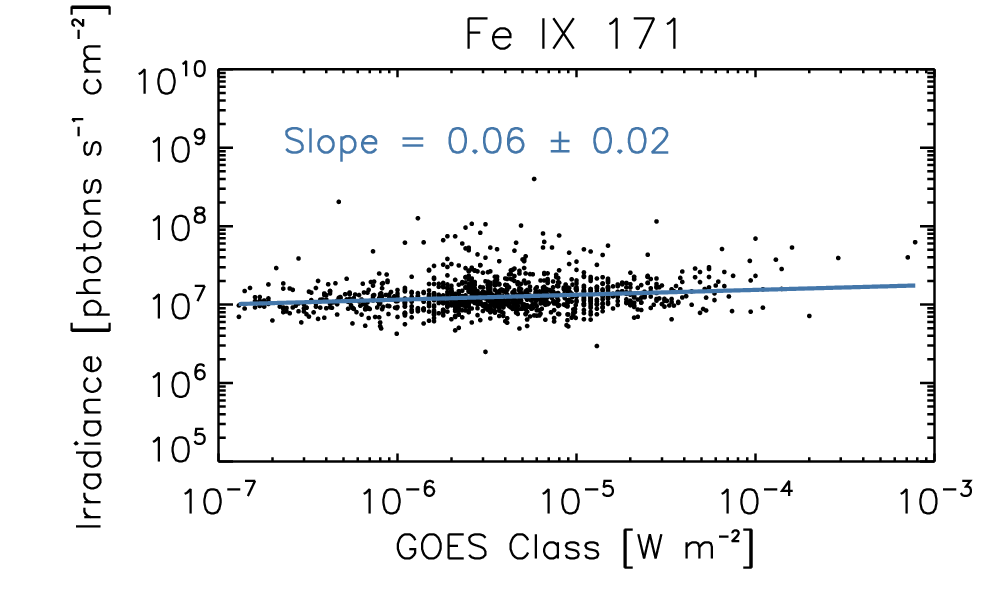}
    \caption{Background-subtracted irradiance values for chromospheric and transition region lines measured with SDO/EVE as a function of GOES class.  The blue line marks a Theil-Sen linear regression fit to the data in log-log space, with $\pm 1\sigma$ uncertainty indicated.}
\end{figure*}
\addtocounter{figure}{-1}
\begin{figure*} 
    \centering
    \includegraphics[width=0.32\textwidth]{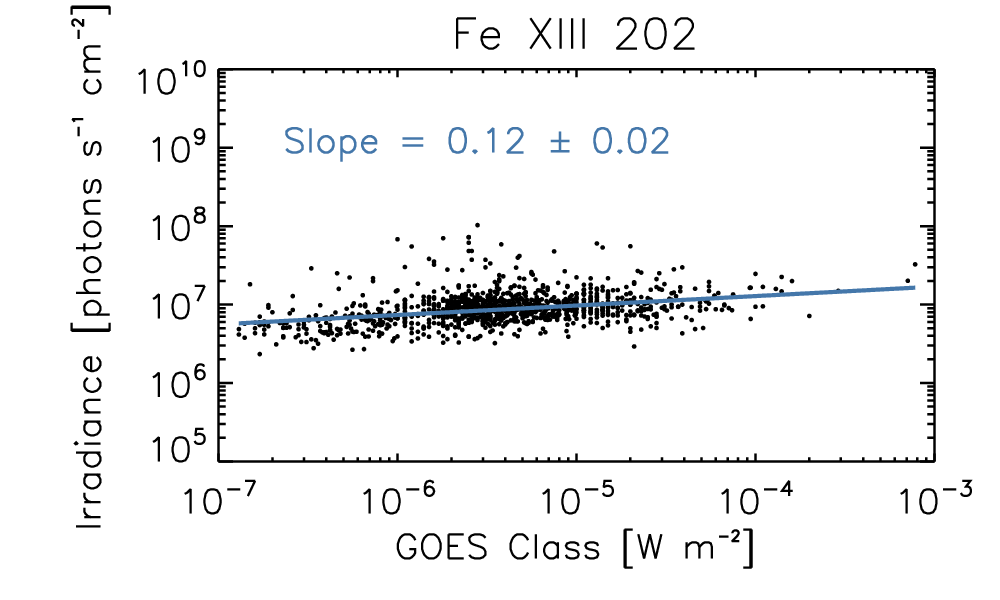}
    \includegraphics[width=0.32\textwidth]{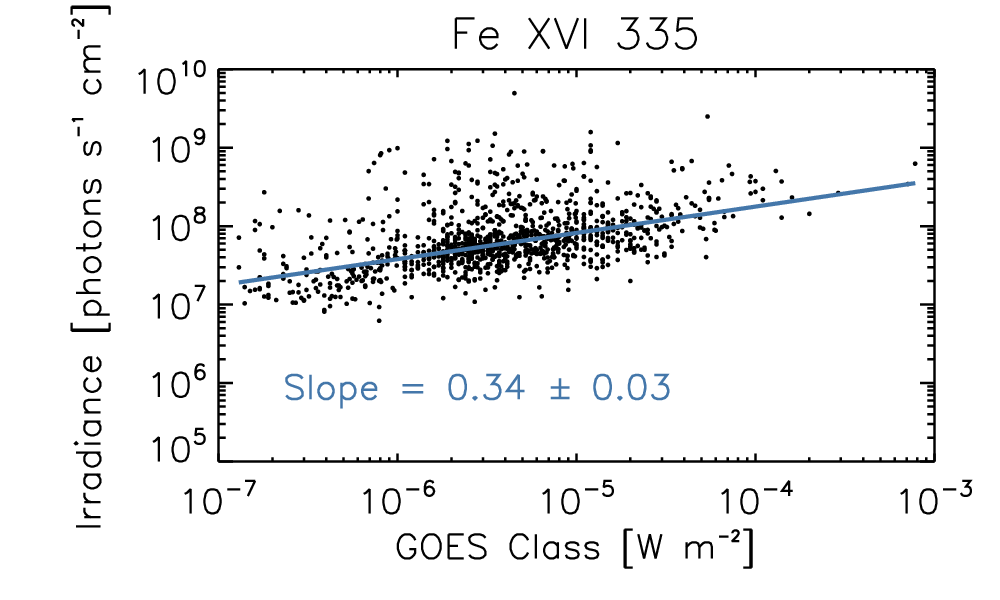}
    \includegraphics[width=0.32\textwidth]{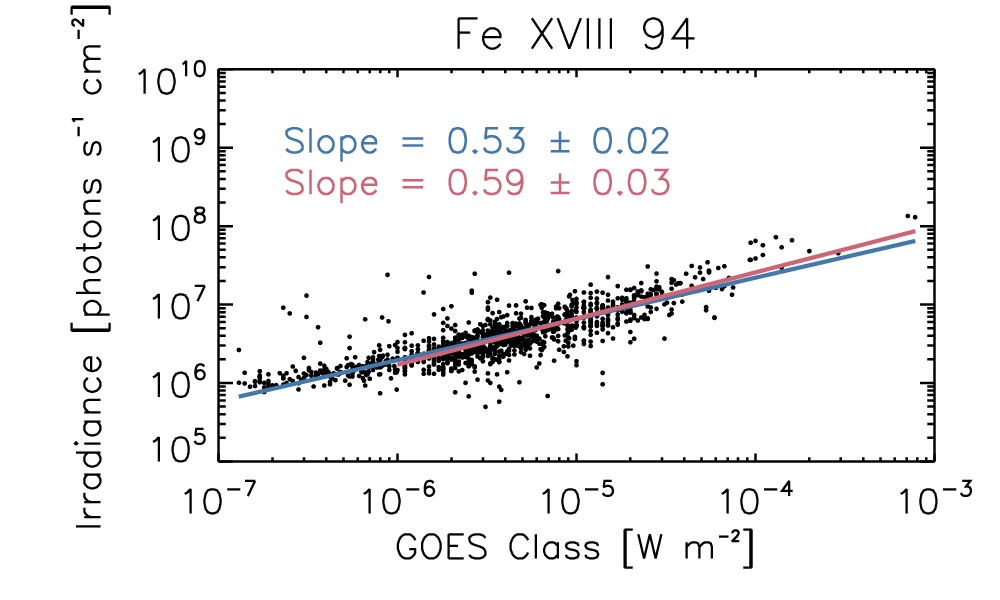}
    \includegraphics[width=0.32\textwidth]{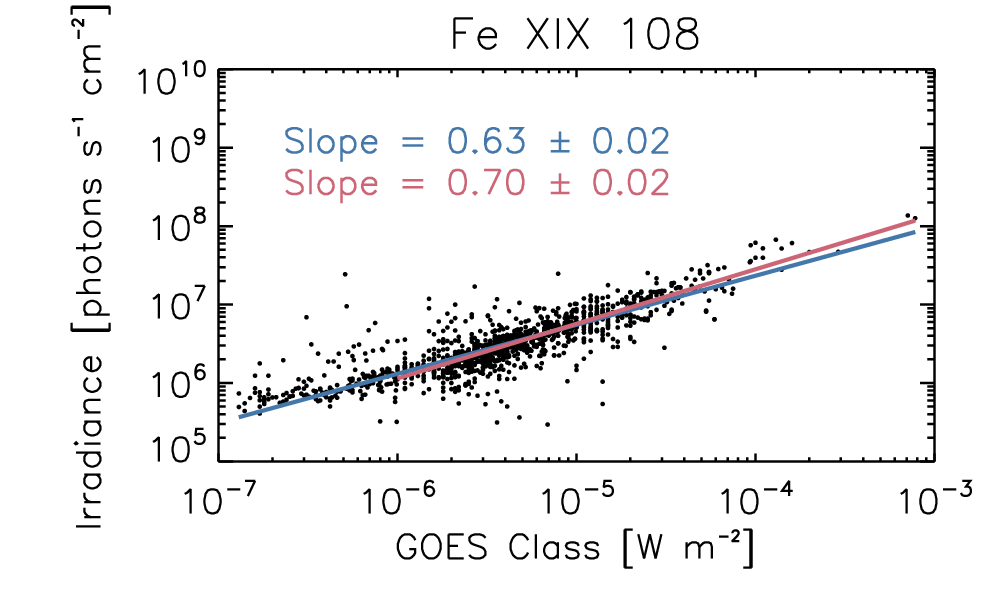}
    \includegraphics[width=0.32\textwidth]{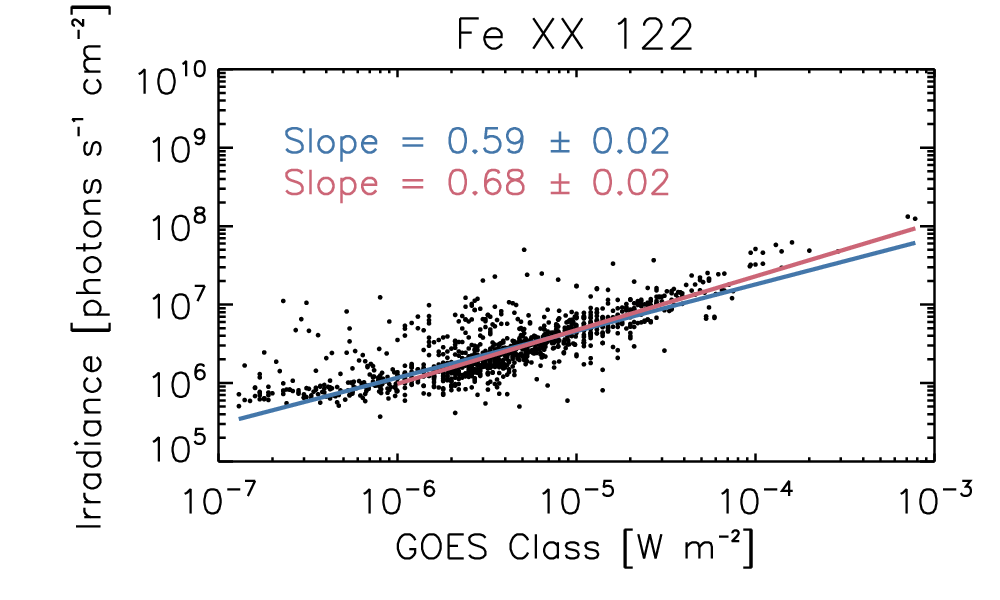}
    \includegraphics[width=0.32\textwidth]{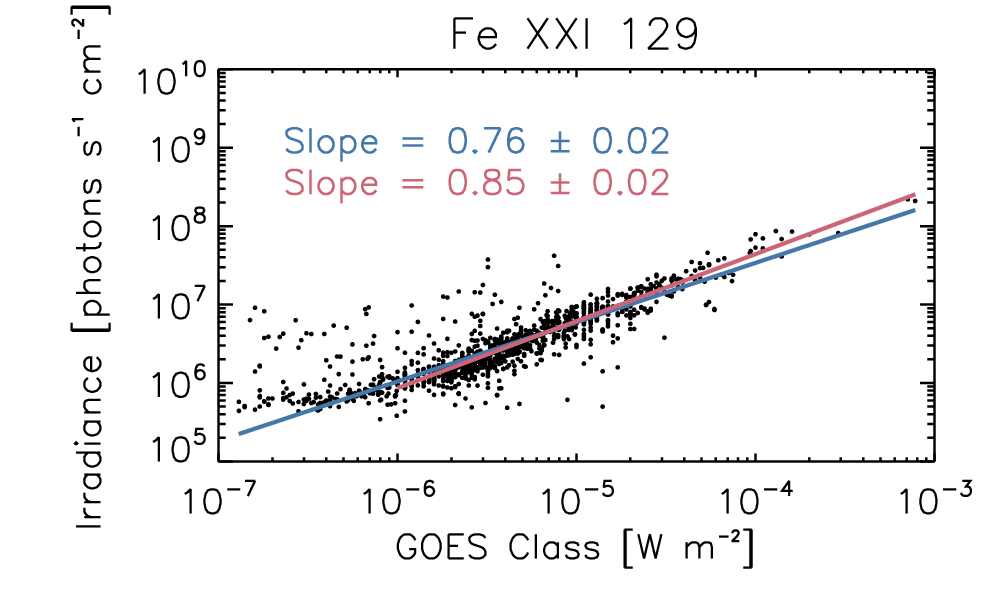}
    \includegraphics[width=0.32\textwidth]{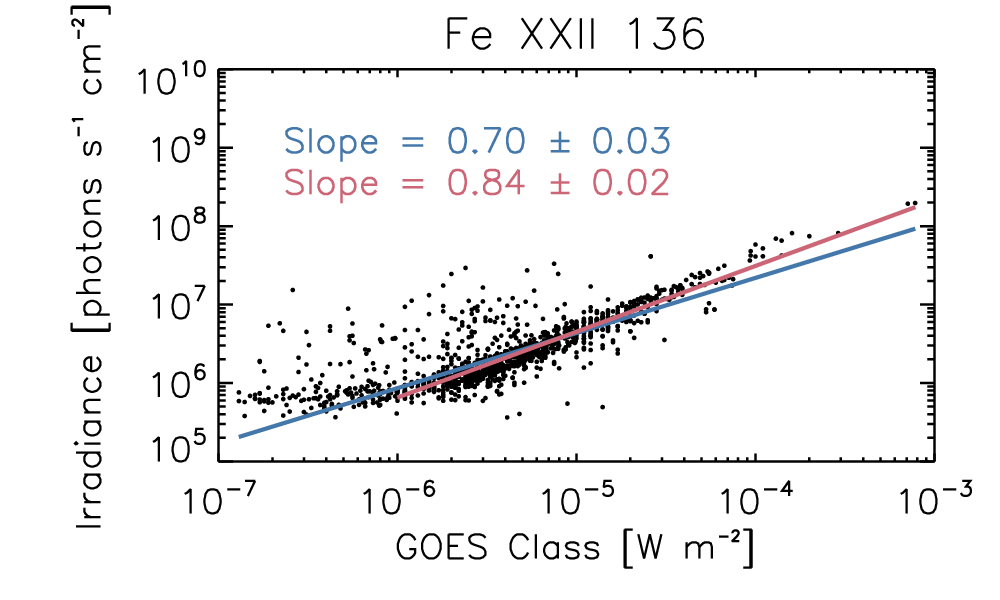}
    \includegraphics[width=0.32\textwidth]{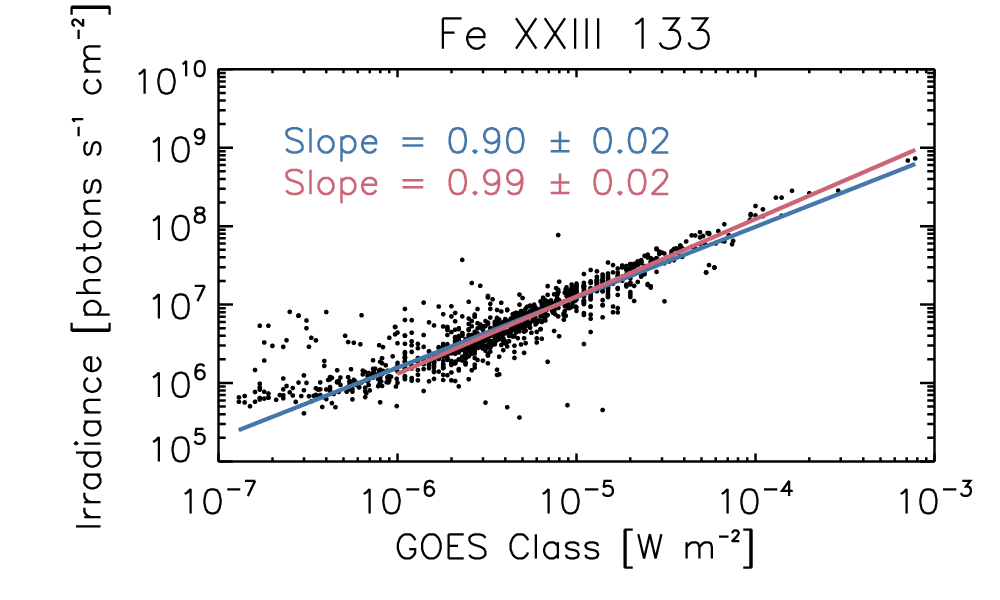}
    \includegraphics[width=0.32\textwidth]{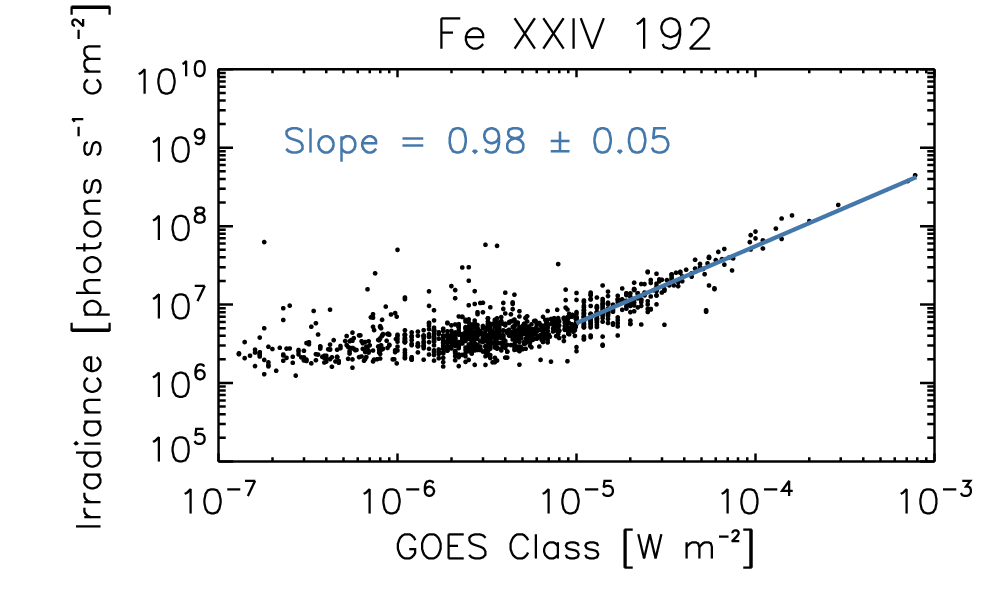}
    \caption{Continued.  Background-subtracted irradiance values for coronal and flaring lines measured with SDO/EVE as a function of GOES class.  The red lines mark the fits only for flares above C-class.  The trend for Fe XXIV 192\,\AA\ has only been fit for M and X-class flares because of the blends with Fe XI and Fe XIV.  \label{fig:eve}}
\end{figure*}

\begin{figure}
    \centering
    \includegraphics[width=0.5\textwidth]{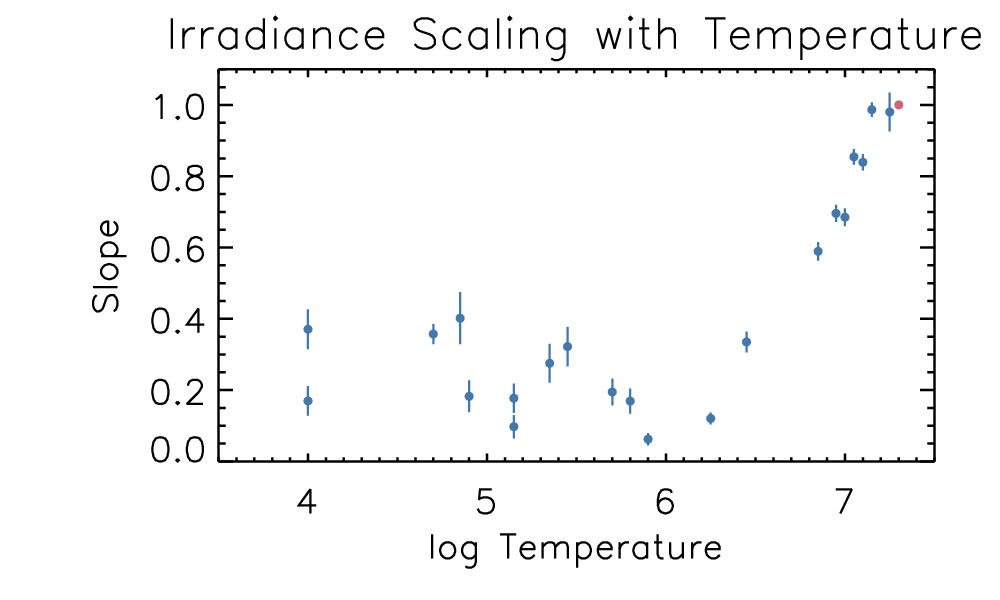}
    \caption{The slope of the irradiance scaling as a function of line formation temperature.  The errors are the $\pm 1 \sigma$ confidence interval.  Each blue dot is one spectral line measured with SDO/EVE, while the red dot indicates GOES XRS-B (which by definition has a slope of 1.0).  \label{fig:scaling}}
\end{figure}

It is immediately apparent that the slope found from the linear regression varies as a function of formation temperature.  At higher temperatures, the slope approaches 1, while at cooler temperatures it ranges between about 0.1 and 0.4.  In Figure \ref{fig:scaling}, we show the measured slopes as a function of formation temperature of the lines (blue dots).  The peak formation temperatures were taken from the CHIANTI atomic database \citep{dere2019}.  The errors are the $\pm 1$-$\sigma$ uncertainties measured from the fits.  The red dot shows the GOES/XRS value at approximately 20 MK, which by definition must scale with a slope of 1.  It is certain that the coolest lines in the sample, particularly Lyman-$\beta$, He I 584, and He II 304 are not optically thin, and thus would be subject to center-to-limb variation which we have not accounted for here (\textit{e.g.} \citealt{qian2010}).  See, also, the study of Lyman-$\alpha$ irradiance variability by \citet{milligan2020}.  

Finally, using the fitted scalings, we can linearly extrapolate the irradiance values of each spectral line.  In Table \ref{table:extrap}, we show the extrapolated values, in photons\,s$^{-1}$cm$^{-2}$, for each line at five different flare sizes: X1, X10, X25, X50, and X100.  For the hottest lines, we use the fits only above C-class to extrapolate.  In the table, we also indicate the approximate range of the extrapolation, calculated from the uncertainties in the slopes.  As noted above, one reason for deriving a physical model of irradiance is to be able to extrapolate full spectra to unobserved flare scales, so this table gives a baseline comparison for a set of well-observed spectral lines.  Extrapolations may be unreliable, however, and therefore need to be confirmed with observations.
\begin{table*}[]
    \centering
    \footnotesize
    \begin{tabular}{l c c c | c c c c c}
         Ion & $\log{T}$ & $\lambda$ [\AA] & Slope & X1 & X10 & X25 & X50 & X100 \\ \hline
H I & 4.0 & 1026 & $0.17 \pm 0.04$ & 4.07E+08 & 6.01E+08 & 7.02E+08 & 7.90E+08 & 8.89E+08 \\
 & & & & & $^{-24.8\%}_{+33.1\%}$ & $^{-21.9\%}_{+28.2\%}$ & $^{-19.7\%}_{+24.6\%}$ & $^{-17.3\%}_{+21.0\%}$ \\
He I & 4.0 & 584 & $0.37 \pm 0.06$ & 1.49E+08 & 3.51E+08 & 4.93E+08 & 6.37E+08 & 8.24E+08 \\
 & & & & & $^{-31.9\%}_{+47.6\%}$ & $^{-28.3\%}_{+40.2\%}$ & $^{-25.5\%}_{+34.8\%}$ & $^{-22.6\%}_{+29.6\%}$ \\
He II & 4.7 & 304 & $0.36 \pm 0.03$ & 7.31E+08 & 1.66E+09 & 2.31E+09 & 2.96E+09 & 3.79E+09 \\
 & & & & & $^{-17.6\%}_{+21.4\%}$ & $^{-15.5\%}_{+18.3\%}$ & $^{-13.8\%}_{+16.0\%}$ & $^{-12.1\%}_{+13.8\%}$ \\
C III & 4.85 & 977 & $0.40 \pm 0.07$ & 1.68E+09 & 4.24E+09 & 6.12E+09 & 8.09E+09 & 1.07E+10 \\
 & & & & & $^{-39.6\%}_{+67.4\%}$ & $^{-35.4\%}_{+56.3\%}$ & $^{-32.1\%}_{+48.5\%}$ & $^{-28.5\%}_{+41.0\%}$ \\
O III & 4.90 & 600 & $0.18 \pm 0.04$ & 1.18E+07 & 1.79E+07 & 2.12E+07 & 2.41E+07 & 2.73E+07 \\
 & & & & & $^{-26.5\%}_{+36.2\%}$ & $^{-23.4\%}_{+30.7\%}$ & $^{-21.0\%}_{+26.7\%}$ & $^{-18.6\%}_{+22.9\%}$ \\
N IV & 5.15 & 765 & $0.10 \pm 0.03$ & 4.17E+07 & 5.22E+07 & 5.70E+07 & 6.10E+07 & 6.53E+07 \\
 & & & & & $^{-20.3\%}_{+24.4\%}$ & $^{-17.9\%}_{+20.8\%}$ & $^{-16.0\%}_{+18.2\%}$ & $^{-14.0\%}_{+15.6\%}$ \\
O IV & 5.15 & 790 & $0.18 \pm 0.04$ & 6.13E+07 & 9.22E+07 & 1.08E+08 & 1.23E+08 & 1.39E+08 \\
 & & & & & $^{-24.7\%}_{+32.8\%}$ & $^{-21.8\%}_{+27.9\%}$ & $^{-19.5\%}_{+24.3\%}$ & $^{-17.2\%}_{+20.8\%}$ \\
O V & 5.35 & 630 & $0.28 \pm 0.05$ & 1.17E+08 & 2.20E+08 & 2.83E+08 & 3.43E+08 & 4.15E+08 \\
 & & & & & $^{-31.2\%}_{+43.1\%}$ & $^{-27.7\%}_{+36.4\%}$ & $^{-24.9\%}_{+31.6\%}$ & $^{-22.1\%}_{+27.0\%}$ \\
O VI & 5.45 & 1032 & $0.32 \pm 0.06$ & 6.49E+08 & 1.36E+09 & 1.83E+09 & 2.29E+09 & 2.86E+09 \\
 & & & & & $^{-31.8\%}_{+47.9\%}$ & $^{-28.3\%}_{+40.4\%}$ & $^{-25.5\%}_{+35.0\%}$ & $^{-22.5\%}_{+29.8\%}$ \\
Ne VII & 5.70 & 465 & $0.19 \pm 0.04$ & 2.37E+07 & 3.71E+07 & 4.44E+07 & 5.08E+07 & 5.81E+07 \\
 & & & & & $^{-22.8\%}_{+28.1\%}$ & $^{-20.1\%}_{+24.0\%}$ & $^{-18.0\%}_{+20.9\%}$ & $^{-15.8\%}_{+17.9\%}$ \\
Ne VIII & 5.80 & 770 & $0.17 \pm 0.04$ & 7.47E+07 & 1.10E+08 & 1.29E+08 & 1.45E+08 & 1.63E+08 \\
 & & & & & $^{-21.5\%}_{+28.4\%}$ & $^{-19.0\%}_{+24.2\%}$ & $^{-17.0\%}_{+21.1\%}$ & $^{-14.9\%}_{+18.1\%}$ \\
Fe IX & 5.90 & 171 & $0.06 \pm 0.02$ & 1.54E+07 & 1.78E+07 & 1.88E+07 & 1.96E+07 & 2.05E+07 \\
 & & & & & $^{-11.2\%}_{+12.4\%}$ & $^{-9.8\%}_{+10.7\%}$ & $^{-8.7\%}_{+9.4\%}$ & $^{-7.6\%}_{+8.1\%}$ \\
Fe XIII & 6.25 & 202 & $0.12 \pm 0.02$ & 1.28E+07 & 1.69E+07 & 1.89E+07 & 2.05E+07 & 2.23E+07 \\
 & & & & & $^{-10.9\%}_{+12.4\%}$ & $^{-9.6\%}_{+10.7\%}$ & $^{-8.5\%}_{+9.4\%}$ & $^{-7.4\%}_{+8.1\%}$ \\
Fe XVI & 6.45 & 335 & $0.34 \pm 0.03$ & 1.77E+08 & 3.84E+08 & 5.22E+08 & 6.58E+08 & 8.30E+08 \\
 & & & & & $^{-18.3\%}_{+23.0\%}$ & $^{-16.0\%}_{+19.6\%}$ & $^{-14.3\%}_{+17.2\%}$ & $^{-12.6\%}_{+14.8\%}$ \\
Fe XVIII & 6.85 & 94 & $0.59 \pm 0.03$ & 2.57E+07 & 9.99E+07 & 1.71E+08 & 2.58E+08 & 3.88E+08 \\
 & & & & & $^{-16.3\%}_{+20.0\%}$ & $^{-14.3\%}_{+17.1\%}$ & $^{-12.8\%}_{+15.0\%}$ & $^{-11.2\%}_{+12.9\%}$ \\
Fe XIX & 6.95 & 108 & $0.70 \pm 0.02$ & 2.82E+07 & 1.40E+08 & 2.64E+08 & 4.28E+08 & 6.93E+08 \\
 & & & & & $^{-15.1\%}_{+18.6\%}$ & $^{-13.3\%}_{+15.9\%}$ & $^{-11.8\%}_{+13.9\%}$ & $^{-10.4\%}_{+12.0\%}$ \\
Fe XX & 7.0 & 122 & $0.68 \pm 0.02$ & 2.29E+07 & 1.11E+08 & 2.08E+08 & 3.34E+08 & 5.37E+08 \\
 & & & & & $^{-15.8\%}_{+19.4\%}$ & $^{-13.9\%}_{+16.6\%}$ & $^{-12.4\%}_{+14.6\%}$ & $^{-10.8\%}_{+12.5\%}$ \\
Fe XXI & 7.05 & 129 & $0.85 \pm 0.02$ & 4.40E+07 & 3.14E+08 & 6.88E+08 & 1.24E+09 & 2.25E+09 \\
 & & & & & $^{-14.0\%}_{+17.0\%}$ & $^{-12.3\%}_{+14.6\%}$ & $^{-10.9\%}_{+12.8\%}$ & $^{-9.6\%}_{+11.1\%}$ \\
Fe XXII & 7.10 & 136 & $0.84 \pm 0.02$ & 3.11E+07 & 2.15E+08 & 4.63E+08 & 8.28E+08 & 1.48E+09 \\
 & & & & & $^{-14.6\%}_{+18.1\%}$ & $^{-12.8\%}_{+15.5\%}$ & $^{-11.4\%}_{+13.6\%}$ & $^{-10.0\%}_{+11.7\%}$ \\
Fe XXIII & 7.15 & 133 & $0.99 \pm 0.02$ & 1.23E+08 & 1.20E+09 & 2.95E+09 & 5.85E+09 & 1.16E+10 \\
 & & & & & $^{-13.2\%}_{+16.0\%}$ & $^{-11.5\%}_{+13.7\%}$ & $^{-10.3\%}_{+12.1\%}$ & $^{-9.0\%}_{+10.4\%}$ \\
Fe XXIV & 7.25 & 192 & $0.98 \pm 0.05$ & 5.55E+07 & 5.30E+08 & 1.30E+09 & 2.57E+09 & 5.07E+09 \\
 & & & & & $^{-31.4\%}_{+49.8\%}$ & $^{-27.8\%}_{+42.0\%}$ & $^{-25.1\%}_{+36.3\%}$ & $^{-22.2\%}_{+30.9\%}$ \\

    \end{tabular}
    \caption{Empirically extrapolated irradiance, in photons\,s$^{-1}\,$cm$^{-2}$, for each spectral line used in this work derived from the observed scaling.  The lines are sorted by their peak formation temperature.  The extrapolations are shown for flares of GOES classes X10, X25, X50, and X100, along with the interpolated X1 values.  The $\pm$ 1-$\sigma$ ranges (percent difference) of the extrapolations are also indicated.\label{table:extrap}}
\end{table*}
\section{Full spectrum hydrodynamic modeling}
\label{sec:modeling}

\subsection{Methodology} 

In this Section, we describe a method for modeling the flare irradiance, where many of the basic features follow \citet{reep2020}.  In that work, the 0D \texttt{ebtel++} code \citep{barnes2016a} was used to model the flaring loops, primarily because the code can run many thousands of simulations within seconds.  However, the code does not treat the full temperature distribution of a coronal loop, nor does it model the chromosphere, an important source of radiation in flares.  In this work, therefore, we now switch to the magnetic field-aligned HYDrodynamics and RADiation (\texttt{HYDRAD}) code, which treats the hydrodynamics of a two-fluid plasma constrained to flow along a magnetic flux tube \citep{bradshaw2013}.  The code includes a full treatment of thermal conduction, including flux limiting terms, a full optically thin radiative loss calculation, optically thick chromospheric radiation using the treatment of \citet{carlsson2012}, and an approximate treatment to the non-local-thermodynamic-equilibrium (NLTE) of hydrogen level populations and electron density \citep{reep2019}.  

We have run a series of \texttt{HYDRAD} simulations to create a database of loop simulations with various heating profiles and loop lengths.  We heat each loop with an electron beam, with maximum energy fluxes ranging between $10^{9}$ and $10^{11}$ erg s$^{-1}$ cm$^{-2}$, with no coronal background heating term.  The loop lengths range between $10$ and $100$ Mm.  Each simulation is run for 10,000 seconds of simulation time.  From each simulation, we synthesize the irradiance spectra as a function of time, and from those spectra, we synthesize the GOES/XRS and SDO/EVE light curves.  We are not focused on the dynamics of each individual loop in this work, though many such papers have been written in the literature, both with \texttt{HYDRAD} simulations (\textit{e.g.} \citealt{delzanna2011,bradshaw2019,mandage2020}) or with similar codes (\textit{e.g.} \citealt{emslie1992,kowalski2017,tei2018,kerr2021})  For these simulations, we assume that the plasma is only heated by an electron beam, that the heating is impulsive, and we have not accounted for late-phase heating \citep{qiu2016}.  We do not attempt to reproduce non-thermal bremsstrahlung emission, so the electron beam parameters are not constrained directly in this work.

As in \citet{warren2006} and \citet{reep2020}, we construct the flare model from a series of loops being heated in succession, where the heating rate and volume for each are constrained using the GOES/XRS light curves.  The ratio of the two XRS channels is a proxy for temperature \citep{garcia1994}, and can therefore be used to estimate the heating rates, and similarly the magnitudes are a proxy for emission measure (EM).  \citet{warren2004} derived scaling laws to approximate the heating rates and volume, which we use:
\begin{align}
F_{\text{1--8\,\AA}} &\approx 3.7 \times 10^{-35} \Bigg( \frac{E L}{V}\Bigg)^{7/2} \frac{V}{L^{2}} \nonumber \\ 
F_{\text{0.5--4\,\AA}}   &\approx 4.4 \times 10^{-42} \Bigg( \frac{E L}{V}\Bigg)^{9/2} \frac{V}{L^{2}}
\end{align}
\noindent for $E$ the energy release, $L$ the loop length, and $V$ the volume.

We use the same chromospheric footpoint separation model in \citet{reep2020}, which was based on observations by \citet{asai2004} and \citet{hinterreiter2018}.  In short, the loop lengths are determined by the duration of the flare, which is based on observations relating the flare duration to the ribbon separation \citep{toriumi2017}.  The geometry is crucial because the duration of a flare is unrelated to the flare's class \citep{reep2019b}, but the duration of a flare is directly proportional to the footpoint separation \citep{toriumi2017}, which was shown to be consistent with ongoing reconnection well beyond the impulsive phase in flares \citep{reep2017}.  Additionally, the period of quasi-periodic pulsations (QPPs) in the X-rays is not related to flare class, but is related to the ribbon separation \citep{hayes2020}.  These suggest that the formation of QPPs and the duration of a flare itself are both directly related to the magnetic reconnection event driving the energy release, a topic which requires significantly more work.  In \citet{reep2020}, the QPP periods were measured directly and used to estimate the time between energization of successive loops.  In this work, we use the scaling laws relating flare duration to QPP period from \citet{hayes2020} to approximate the time between loop energization.

These geometric assumptions are all empirical, and not constrained based on images or extrapolations of the actual flare geometry.  Furthermore, they are not unique, and other permutations with better constraints might be able to reproduce the XRS light curves better than found here.  While imaging observations do exist for features like loop lengths, they have other limitations like line-of-sight and saturation issues.  It may be important to therefore use 3D magnetohydrodynamic (MHD) simulations to better understand the evolution of the magnetic topology to constrain these models in the future, such as in \citet{longcope2009,guidoni2016}.

For each simulation in our database, we forward model the emission at a one-second cadence.  We use the CHIANTI atomic database \citep{dere1997}, version 10 \citep{delzanna2020}, which calculates spectra at a specified spectral resolution given the emission measure as an input.  At each time step of a simulation, we calculate the emission measure in a \texttt{HYDRAD} grid cell:
\begin{equation}
    EM_{i}(s) = n_{e}(s)\ n_{H}(s)\ A(s)\ ds 
    \label{eqn:em}
\end{equation}
\noindent where the electron density $n_{e}$, hydrogen density $n_{H}$, and loop cross-sectional area $A$ are tabulated along the field-aligned coordinate $s$, and multiplied with the width of the grid cell $ds$.  Additionally, we use the electron temperature $T_{e}(s)$ in the grid cell as input to CHIANTI to calculate the ionization balance in that grid cell, \textit{i.e.}, we assume equilibrium ionization here.  We thus use this EM and the temperature to calculate the emission from each grid cell, and then sum over all grid cells at each time step to synthesize the total irradiance spectrum at each time step.  We perform the entire calculation twice for each \texttt{HYDRAD} simulation: once where $A(s)$ is assumed uniform along the loop, and once where $A(s)$ varies along the loop with empirical corrections.  We then synthesize the full spectra, from 0.01 \AA\ to 1250 \AA\ for each \texttt{HYDRAD} simulation.  Using these spectra, we directly calculate the synthetic GOES/XRS light curves.  Finally, for a given NRLFLARE model, we add together the irradiance spectra of a series of \texttt{HYDRAD} loop simulations that approximately reproduce the observed GOES/XRS light curves \citep{warren2006, reep2020}, and use the total spectrum to calculate SDO/EVE light curves.

In this first iteration of the NRLFLARE model, we have attempted to include much of the physics relevant to simulating solar flares, but by necessity have omitted some features that could impact the spectra.  We use ionization equilibrium, where the temperature is used to determine the ionization fractions.  Non-equilibrium ionization is likely important at low densities, or when the temperature changes rapidly such as during the impulsive phase, and therefore ought to be examined.  We assume that each flare occurs at disk center, that is, we do not include center-to-limb variation effects, which are known to affect line intensities, even in coronal lines \citep{thiemann2018}.  We also do not constrain the heating parameters directly through measurements from X-ray spectrometers like RHESSI \citep{lin2002} or MinXSS \citep{moore2018}, which can only be done when observations are available.  Finally, the model is constrained entirely by soft X-ray time series, which may not be sufficient to adequately describe the lower atmosphere or cooler plasma.  One advantage, however, is that GOES has provided a near continuous data stream since the 1970s, and therefore we can synthesize any event over that time period, regardless of other observations.  Though these assumptions will need to be examined and improved in the future, as we will show in the next section, the model still is capable of reproducing many observed features of flares.

\subsection{Modeled Events}

We have modeled nine flares using this methodology, listed in Table \ref{table:events}.  We have chosen events covering a range of GOES classes, full-width-at-half-maximum (FWHM) durations, and peak GOES temperatures (measured from the ratio of XRS-A to XRS-B, \citealt{white2005}).  All nine events had continuous MEGS-A coverage, but only four had continuous MEGS-B coverage.  We present the light curves, observed and modeled, for those four events in this section.
\begin{table*}[]
    \centering
    \begin{tabular}{l c c c c}
         Date & GOES class & GOES Temperature & XRS-B FWHM & XRS-A FWHM  \\ 
               &   & [MK] & [min] & [min] \\ \hline 
         12 Jun 2010 UT 00:30 & M2.9 & 20.7 & 5.5 & 3.8 \\
         15 Dec 2010 UT 06:27 & C7.8 & 13.4 & 11.8 & 7.1 \\
         03 Nov 2011 UT 20:16 & X2.9 & 20.0 & 10.3 & 5.9 \\
         07 Mar 2012 UT 00:03 & X7.8 & 26.2 & 22.9 & 15.8 \\
         14 Mar 2012 UT 15:08 & M4.1 & 16.0 & 19.6 & 14.5 \\
         17 May 2012 UT 01:24 & M7.3 & 15.8 & 38.8 & 27.0 \\
         29 Oct 2013 UT 21:43 & X3.4 & 32.5 & 10.8 & 8.0 \\
         25 Feb 2014 UT 00:40 & X7.1 & 30.1 & 16.7 & 12.1 \\
         15 May 2014 UT 05:14 & C4.7 & 11.4 & 19.0 & 8.5
    \end{tabular}
    \caption{The nine flares modeled in this work.  Four of these events had both MEGS-A and B coverage, and the light curves for these are presented in this section. \label{table:events}}
\end{table*}

We first discuss the X7.8\footnote{GOES/XRS values were recalibrated by NOAA in May 2020.  Previously, flare classes were listed in terms of a scaled value, where the XRS-B 1--8 \AA\ channel was reduced by a factor of 0.7 (0.85 in XRS-A) to give agreement with the oldest GOES satellites.  In May 2020, NOAA dropped this scaling factor and now reports XRS data in true irradiance units.  This flare has a true irradiance of $7.8 \times 10^{-4}$ W m$^{-2}$, corresponding to X7.8, but has previously been reported as X5.4 due to the 0.7 scaling factor.  See the XRS User's Guide, Section 2.2: \url{https://data.ngdc.noaa.gov/platforms/solar-space-observing-satellites/goes/goes16/l2/docs/GOES-R_XRS_L2_Data_Users_Guide.pdf}} flare that occurred on 7 March 2012 UT00:03, with a FWHM of 23 min in XRS-B and 16 min in XRS-A.  Observations of this flare with SDO/EVE were previously analyzed by \citet{delzanna2013}.  It is the largest flare in our data set, and crucially has continuous coverage with both MEGS-A and B.  The GOES temperature peaked at about 26 MK.

\begin{figure*}
\includegraphics[width=0.96\textwidth]{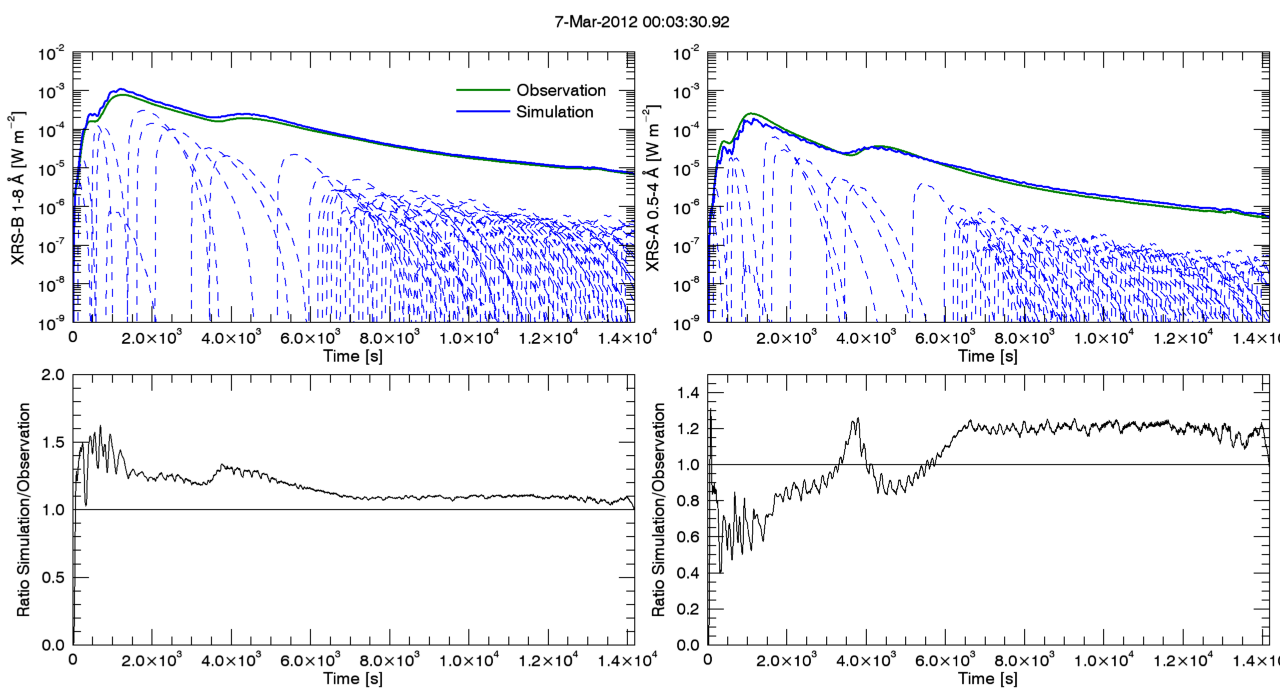}
\caption{The modeled GOES/XRS light curves for the 7 March 2012 X7.8 flare, showing both channels, as well as the ratio of modeled/observed intensity.  Because the number of \texttt{HYDRAD} simulations is more limited, the fits are slightly worse than those in \citet{reep2020}.  XRS-B is slightly overestimated during the impulsive phase and XRS-A is slightly underestimated, meaning that the temperature at this time is slightly underestimated.\label{fig:goesX7}}
\end{figure*}

In Figure \ref{fig:goesX7}, we show the modeled GOES/XRS light curves.  The left column shows XRS-B at top, where the green curve is the background-subtracted observation, the solid blue is the total modeled curve, and the dashed blue curves are the light curves of individual loops.  At bottom, we show the ratio of model/observation, showing that the emission is over-estimated by a factor of about 1.5 during the impulsive phase, and 1.1 during the gradual phase.  Similarly, the right column shows XRS-A, where we now slightly underestimate the emission during the impulsive phase.  The ratio of XRS-A to XRS-B is a proxy for the flare temperature since XRS-A is sensitive to somewhat hotter plasma than XRS-B, so this means that the temperature of the flare is thus slightly underestimated during the impulsive phase in the model.  The number of \texttt{HYDRAD} simulations that match both the heating rate and loop length necessary to fit this is limited and computationally expensive, so it is difficult to get perfect fits to the temperature (compared to the \texttt{ebtel++} case in \citealt{reep2020}).    

\begin{figure*}
    \centering
    \includegraphics[width=0.32\textwidth]{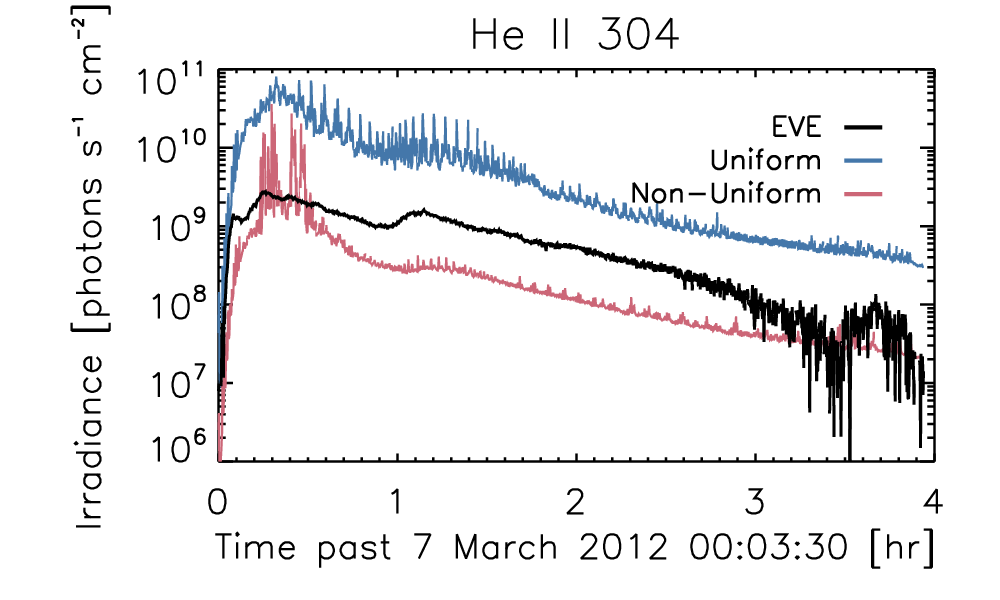}
    \includegraphics[width=0.32\textwidth]{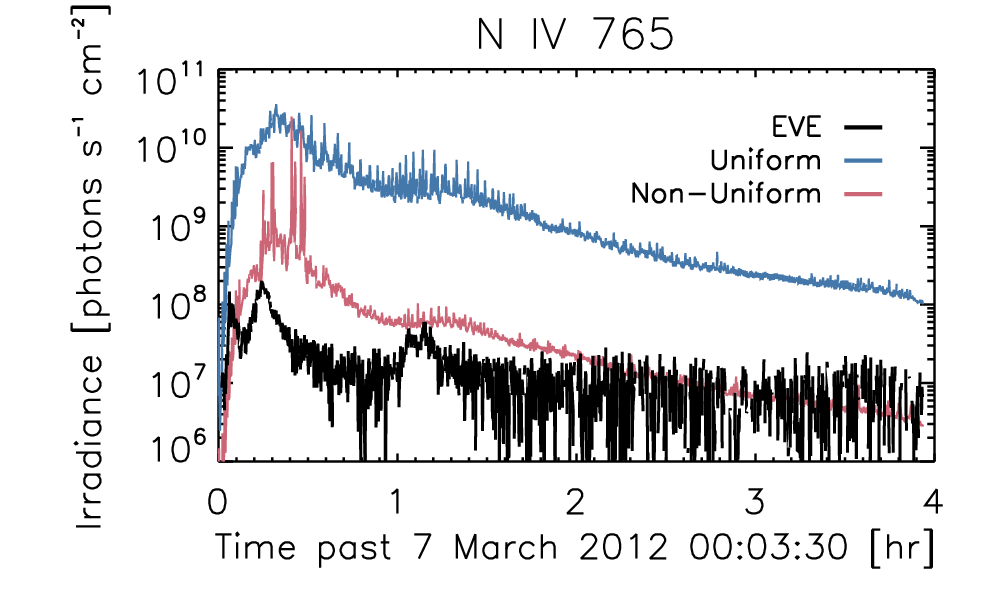}
    \includegraphics[width=0.32\textwidth]{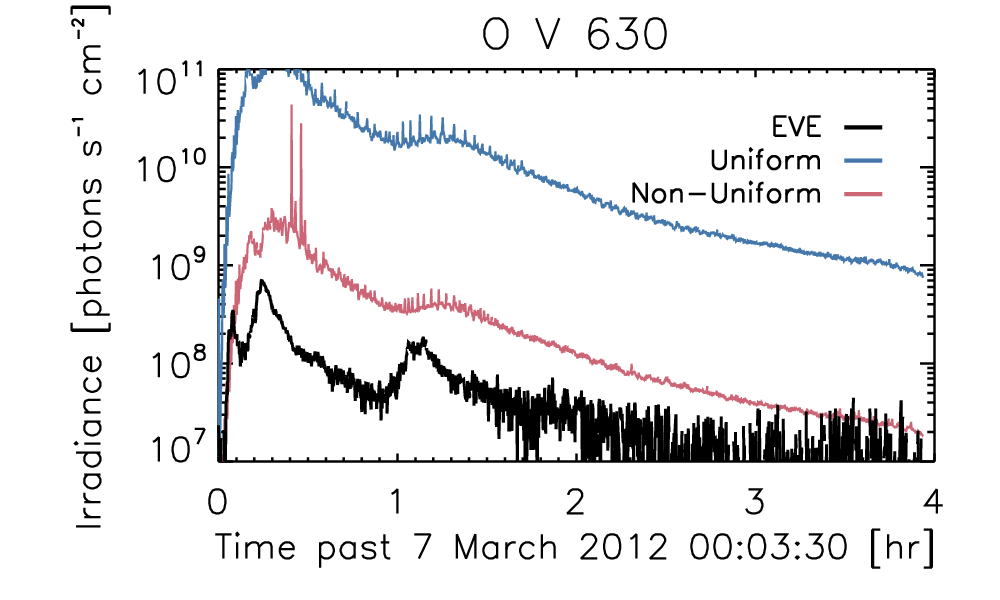}
    \includegraphics[width=0.32\textwidth]{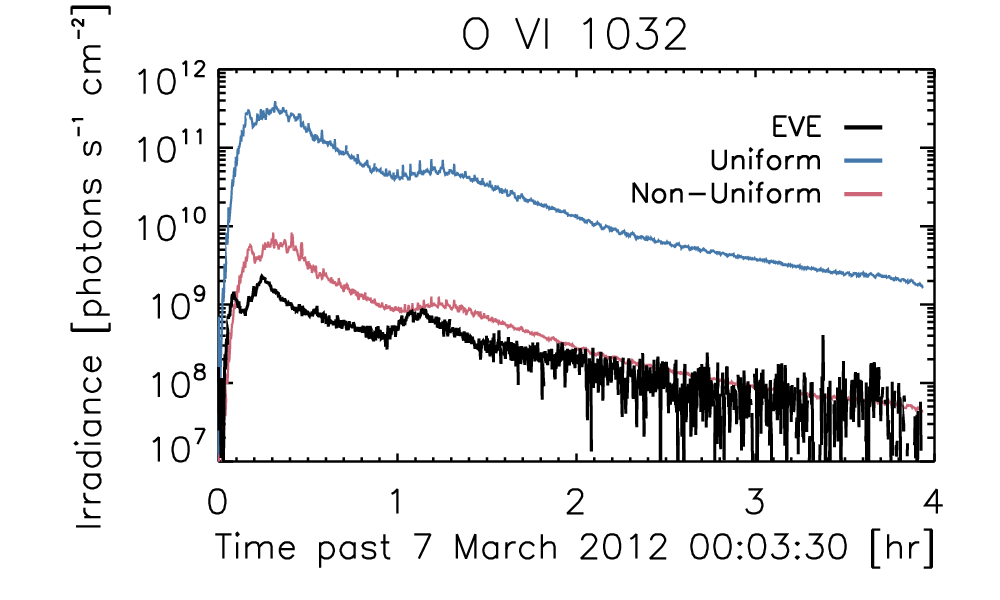}
    \includegraphics[width=0.32\textwidth]{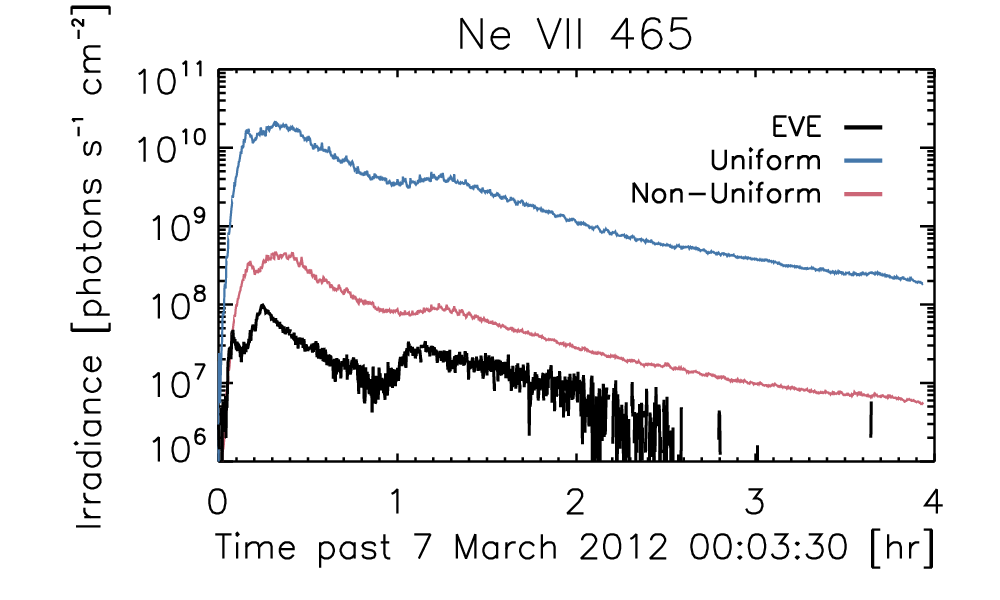}
    \includegraphics[width=0.32\textwidth]{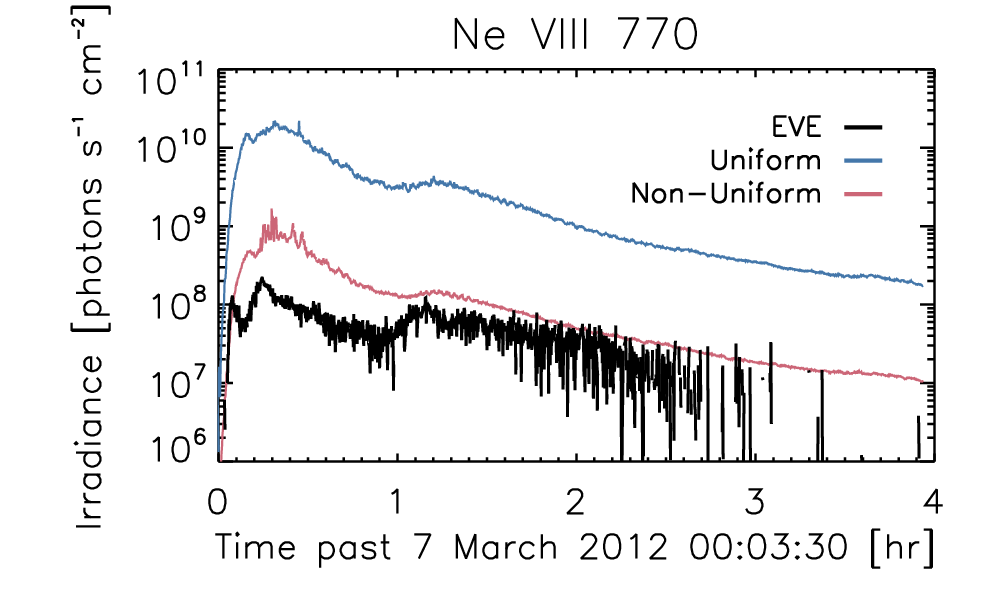}
    \includegraphics[width=0.32\textwidth]{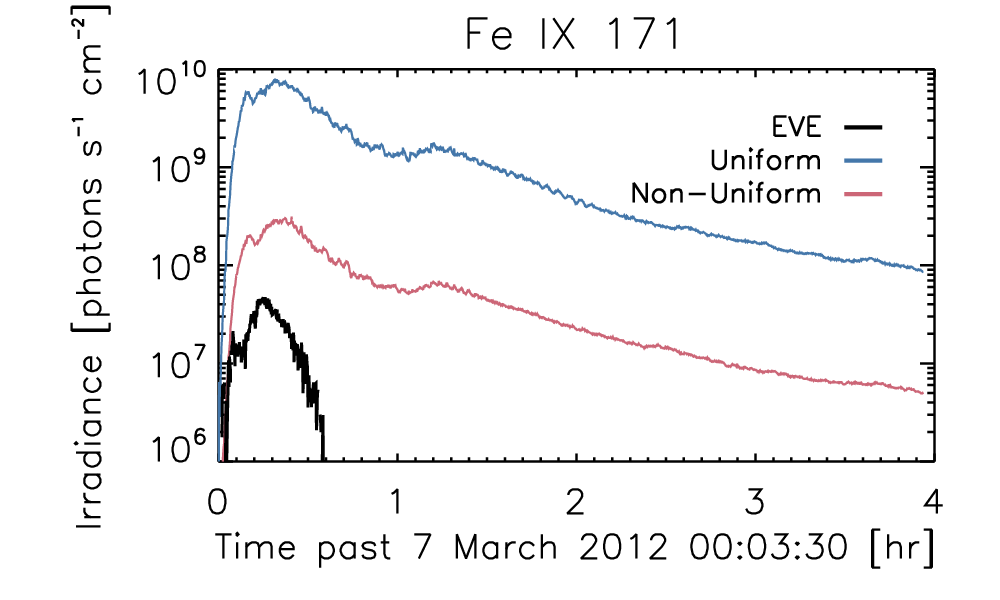}
    \includegraphics[width=0.32\textwidth]{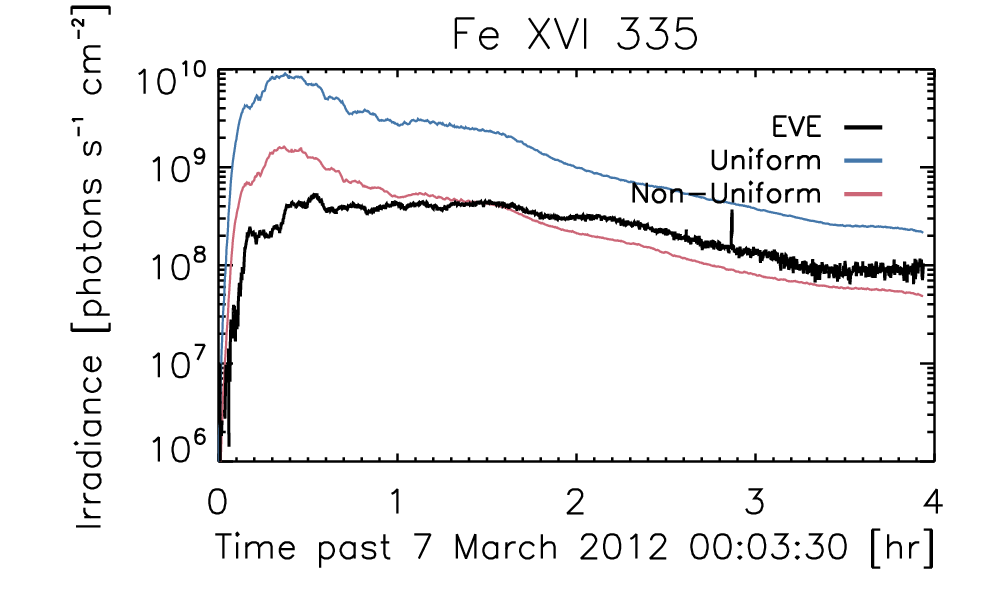}
    \includegraphics[width=0.32\textwidth]{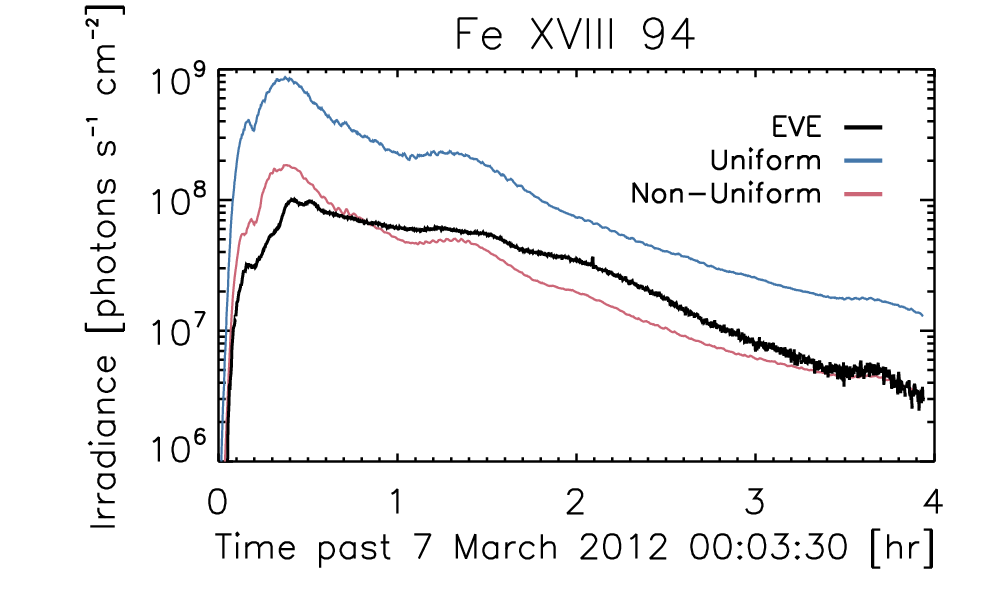}
    \includegraphics[width=0.32\textwidth]{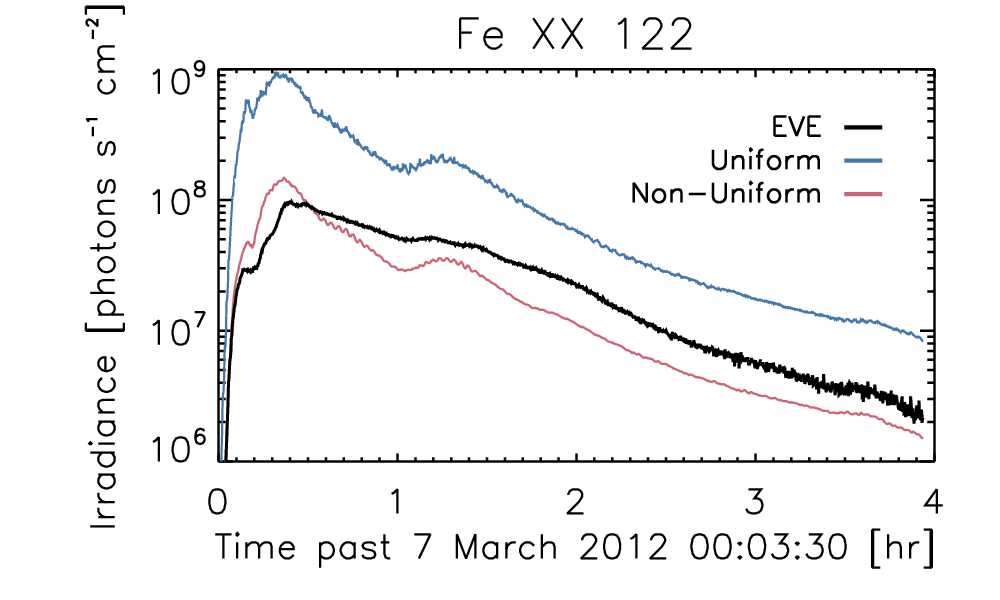}
    \includegraphics[width=0.32\textwidth]{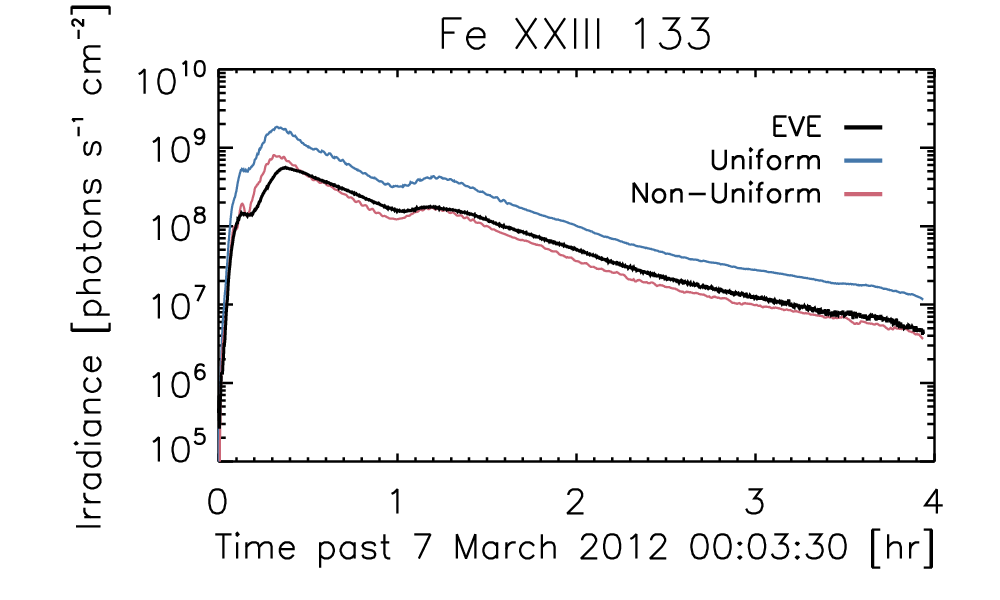}
    \includegraphics[width=0.32\textwidth]{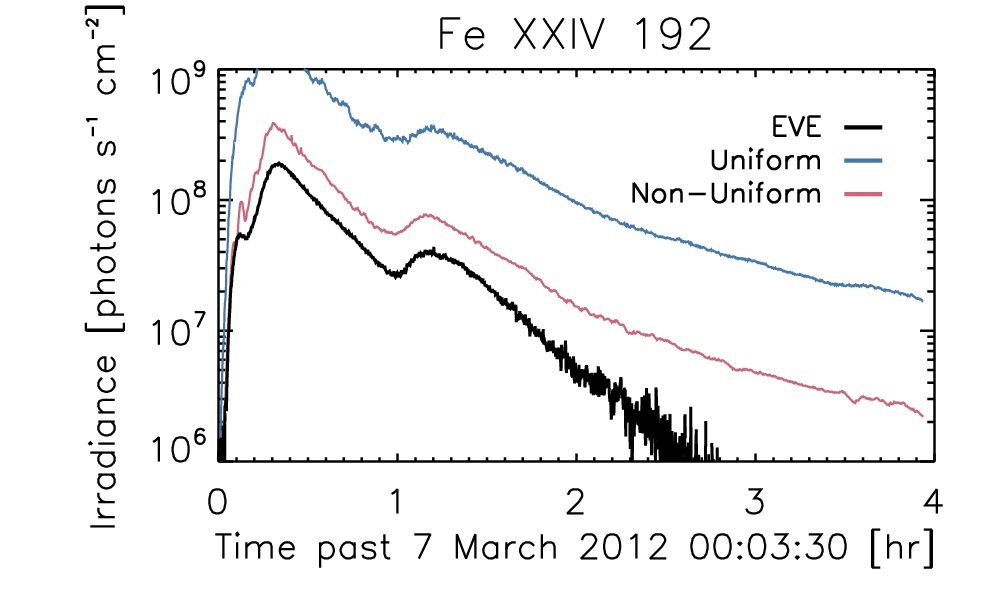}
    \caption{Modeled and observed SDO/EVE light curves for the 07 March 2012 X7.8 flare at 1 \AA\ spectral resolution and 10 s cadence.  The black curves show the EVE data, blue the NRLFLARE model with a uniform cross-sectional area, and red a non-uniform area scaling based on error analysis (see text).  \label{fig:X7}}
\end{figure*}

In Figure \ref{fig:X7}, we show a comparison of the SDO/EVE observed and modeled light curves for 12 of the spectral lines analyzed in \ref{sec:eve}.  The lines are arranged by formation temperature, from coolest to hottest.  The black curves show the observed SDO/EVE light curves, at 1 \AA\ spectral resolution and 10 s cadence.  We have run NRLFLARE with two different cross-sectional area functions for comparison.  The blue curves show the calculation assuming that the loops composing the flare have uniform cross-section, as done commonly in flare modeling studies.  This assumption over-estimates the irradiance at all wavelengths, and in some of the cooler lines by more than an order of magnitude.  The second case, the red curves, show a relative area scaling based on error analysis for the set of nine modeled flares, which we explain in Section \ref{subsec:errors}.  Note that there was a coronal mass ejection (CME) associated with this flare that caused dimming in the \ion{Fe}{9} 171 \AA\ emission, as seen in the figure, which the model is incapable of reproducing directly.  The intensities are similarly overestimated in the three other events presented in this Section, though they were not associated with a CME.

\begin{figure*}
    \centering
    \includegraphics[width=0.98\textwidth]{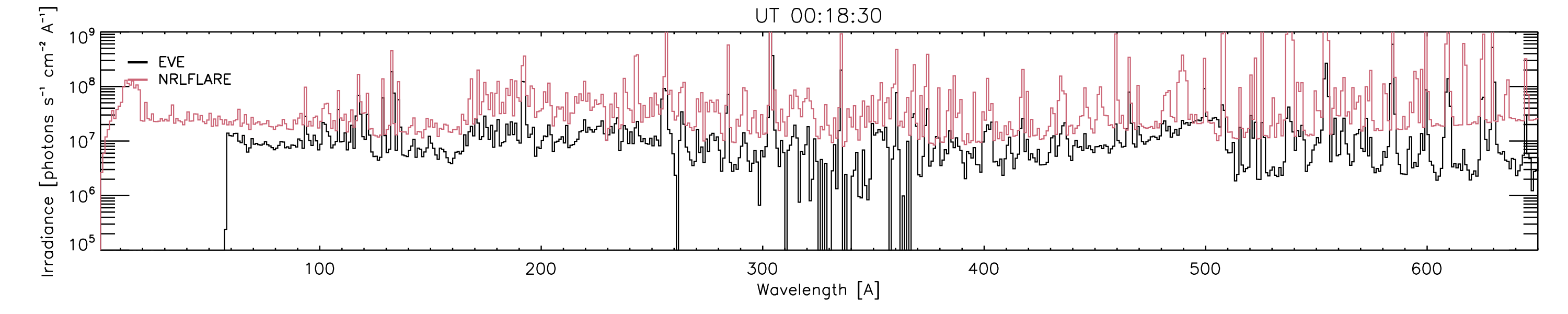}
    \includegraphics[width=0.98\textwidth]{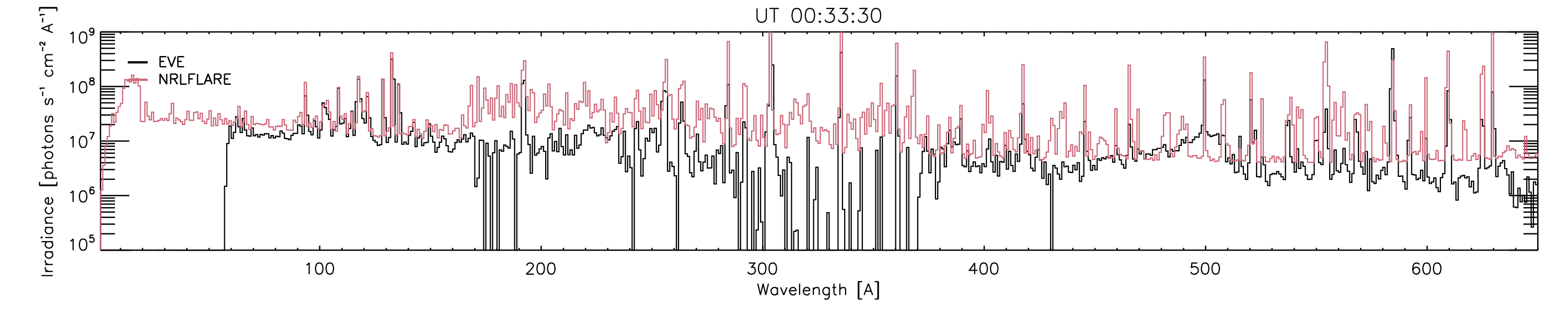}
    \includegraphics[width=0.98\textwidth]{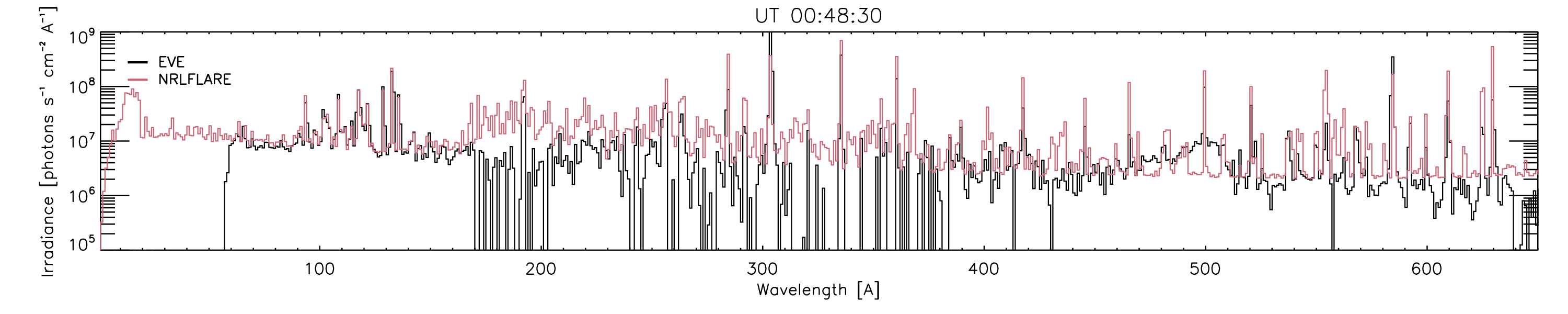}
    \caption{Modeled and observed SDO/EVE spectra for the 07 March 2012 X7.8 flare at 1 \AA\ spectral resolution and 10 s cadence.  The black curves show the EVE data and red the NRLFLARE model with a non-uniform cross-sectional area.  We show the spectra 15, 30, and 45 minutes into the flare, for wavelengths from 1 to 650 \AA. The agreement is best in hot temperatures and short wavelengths, but NRLFLARE often overestimates emission.  \label{fig:spectra}}
\end{figure*}

In Figure \ref{fig:spectra}, we additionally show example spectra comparing the background-subtracted EVE spectra with the non-uniform NRLFLARE spectra.  As in the lightcurves, we have binned to 1 \AA\ spectral resolution and 10 second cadence.  We show this at 15, 30, and 45 minutes into the flare, for wavelengths from 1 to 650 \AA.  We can see that the best agreement between model and observations is at the shortest wavelengths, very roughly corresponding to the hottest plasma.  At longer wavelengths, NRLFLARE generally over-estimates the observed emission.  We can also see that the helium continuum near 500 \AA\ is significantly underestimated, since this emission is optically thick (the Lyman continuum near 900 \AA, not shown, is also significantly underestimated).  Proper calculation of optically thick radiation will be crucial to implement in future versions of NRLFLARE in order to treat the continuum emission and lines like Lyman-$\alpha$ properly.

There is a large discrepancy between the SDO/EVE light curves and the uniform area curves in blue, particularly in the transition region lines such as \ion{O}{5} 630 \AA.  The general trends of the light curves are similar, but the cooler lines are too intense by more than an order of magnitude.  In these simulations, we use assumptions common in the literature of flares: that the plasma is heated by a strong electron beam, that the loop is semi-circular with uniform cross-section, and that the transition region and coronal emission is optically thin.  It is apparent that at least one of these assumptions is incorrect.

We hypothesize that the discrepancy as well as the relation found in Figure \ref{fig:scaling} imply that the cross-sectional areas of the loops composing the flares are not constant.  First, we know that there is no relation between the duration of a flare and its GOES class \citep{reep2019b}, which is primarily because longer duration flares are composed of longer loops, and the average loop length does not depend on the class \citep{toriumi2017}.  Second, while the temperature does scale with the GOES class, it is a rather weak scaling \citep{feldman1995}, so that the temperature difference between \textit{e.g.} X- and M-class flares is of order 10--20\%, with considerable scatter.  Finally, the analysis in Section 5 of \citet{reep2020} implies that the major difference between flares of different classes is that the physical volume increases substantially with class (between 70 and 100\% of the total energy increase is due to a larger volume).  From these three points, we conclude that the scaling in intensity is primarily due to a physical volume, but not the loop lengths, so the differences are due to cross-sectional area.  Furthermore, if the area were uniform along the length of the loops, then this would imply that the slopes in Figure \ref{fig:scaling} would be approximately 1 at all temperatures.  Therefore, this appears to be an indication of a non-uniform cross-section, though we discuss other possibilities in Section \ref{sec:discussion}.   \citet{warren2010} reached a similar conclusion from discrepancies between observations and modeling of active region footpoint emission.  The effects of a non-uniform cross-section would both directly change the volume at different heights in the atmosphere, as well as strongly impacting the hydrodynamics, leading to a different distribution of temperatures and densities along the loop structure itself and extending the cooling timescale of the loop \citep{emslie1992,mikic2013}, although this is relatively poorly studied.

\begin{figure*}
    \centering
    \includegraphics[width=0.32\textwidth]{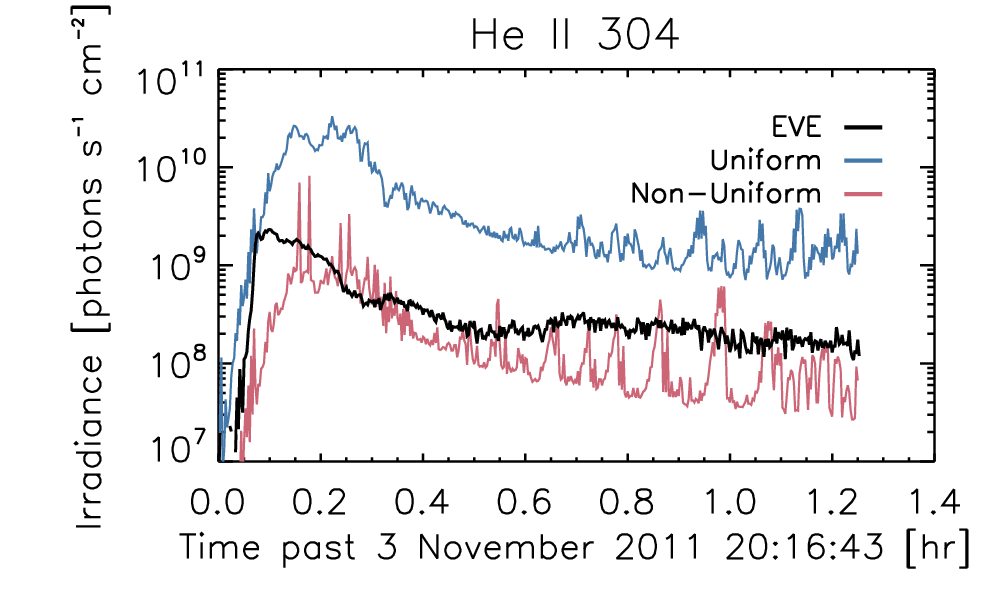}
    \includegraphics[width=0.32\textwidth]{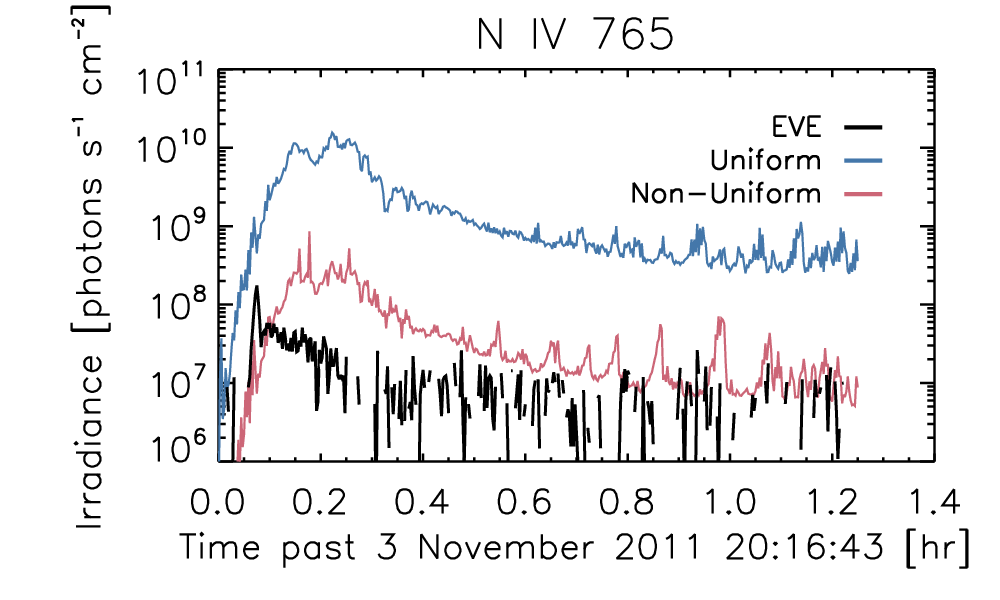}
    \includegraphics[width=0.32\textwidth]{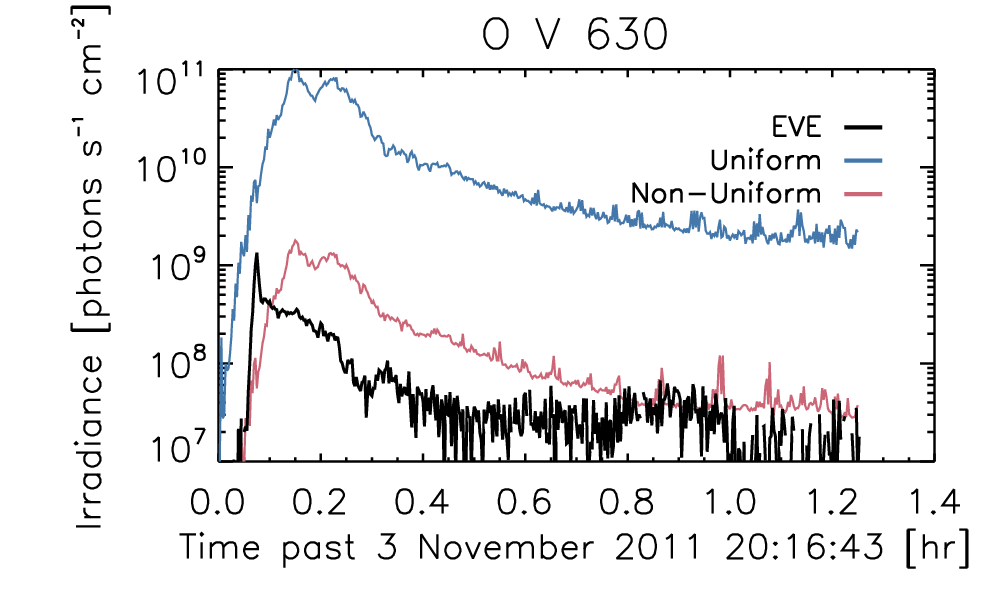}
    \includegraphics[width=0.32\textwidth]{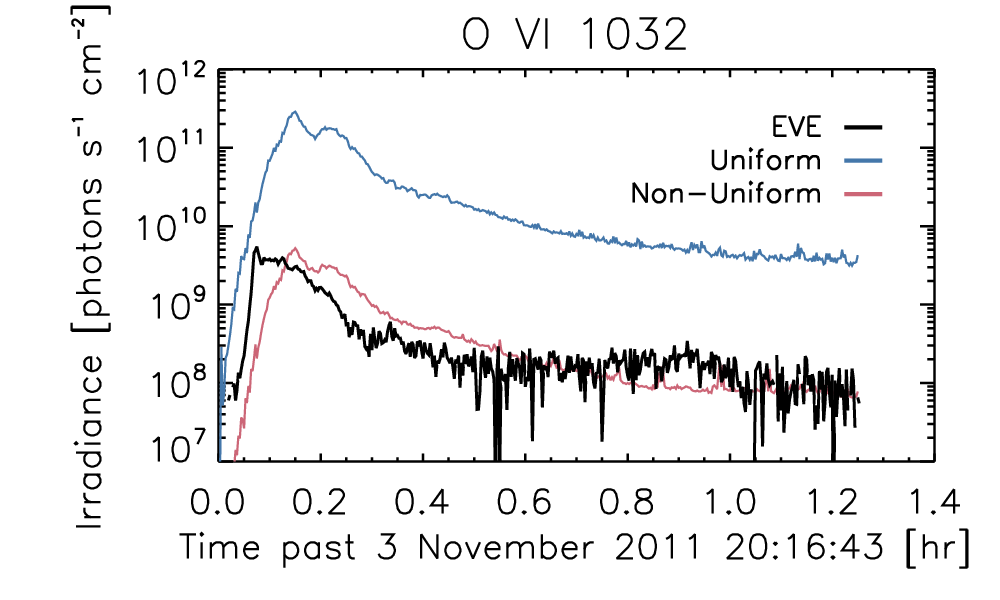}
    \includegraphics[width=0.32\textwidth]{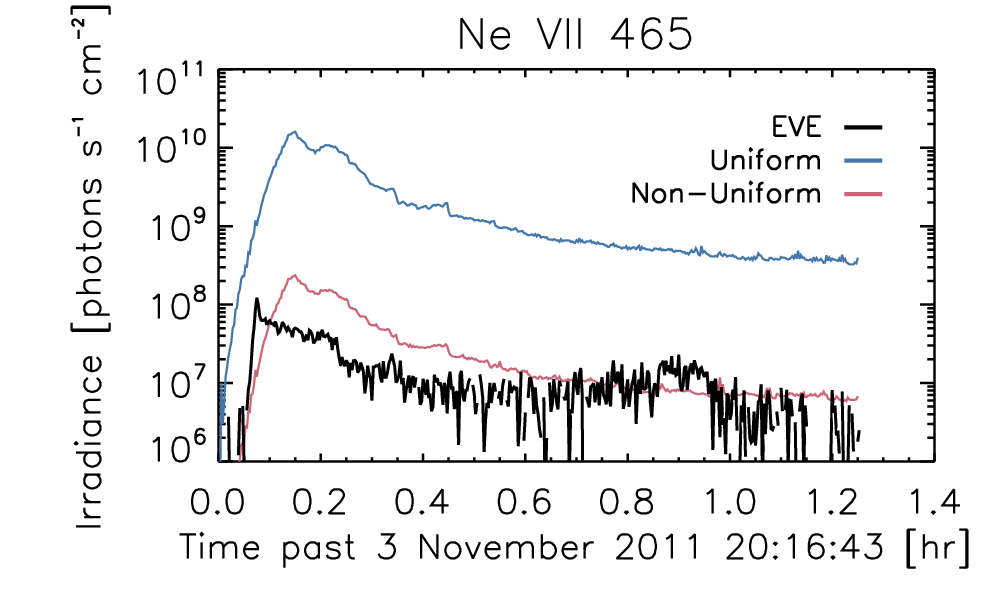}
    \includegraphics[width=0.32\textwidth]{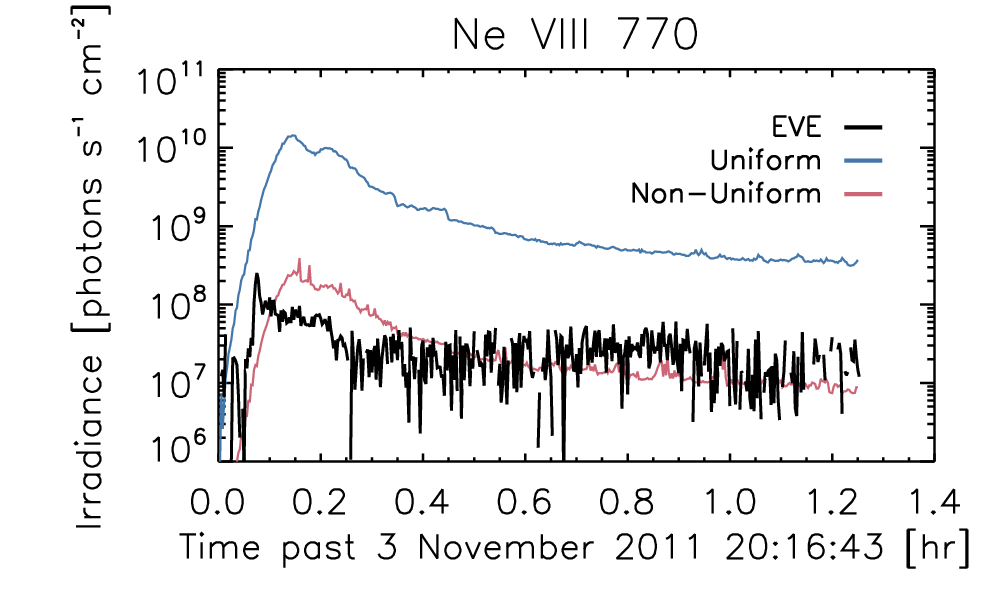}
    \includegraphics[width=0.32\textwidth]{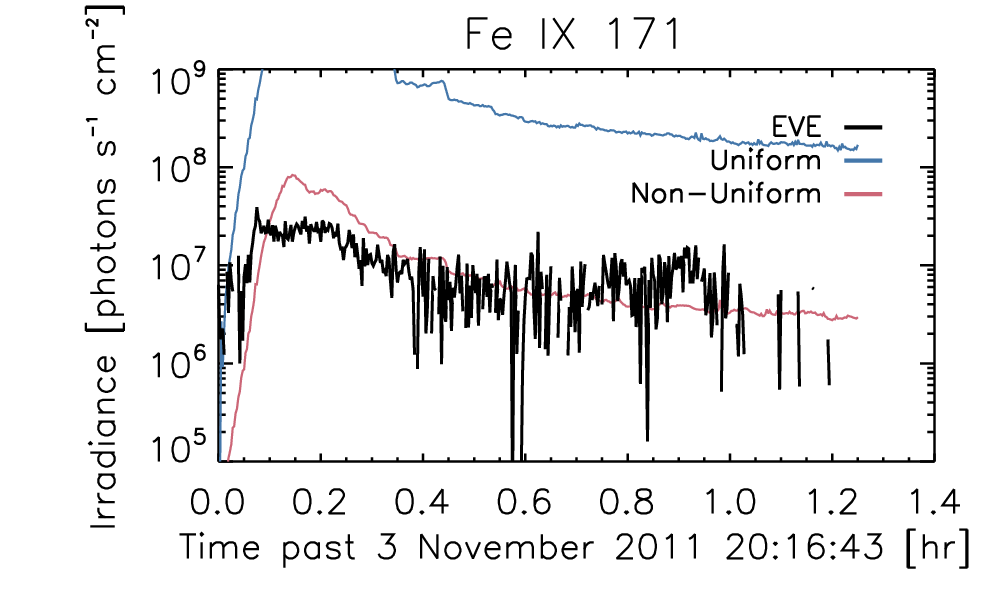}
    \includegraphics[width=0.32\textwidth]{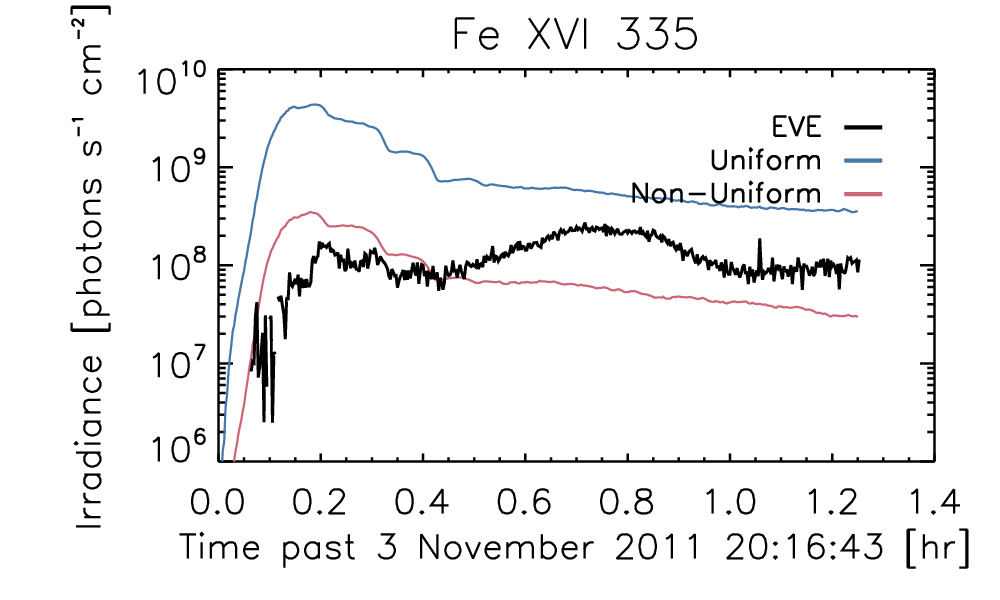}
    \includegraphics[width=0.32\textwidth]{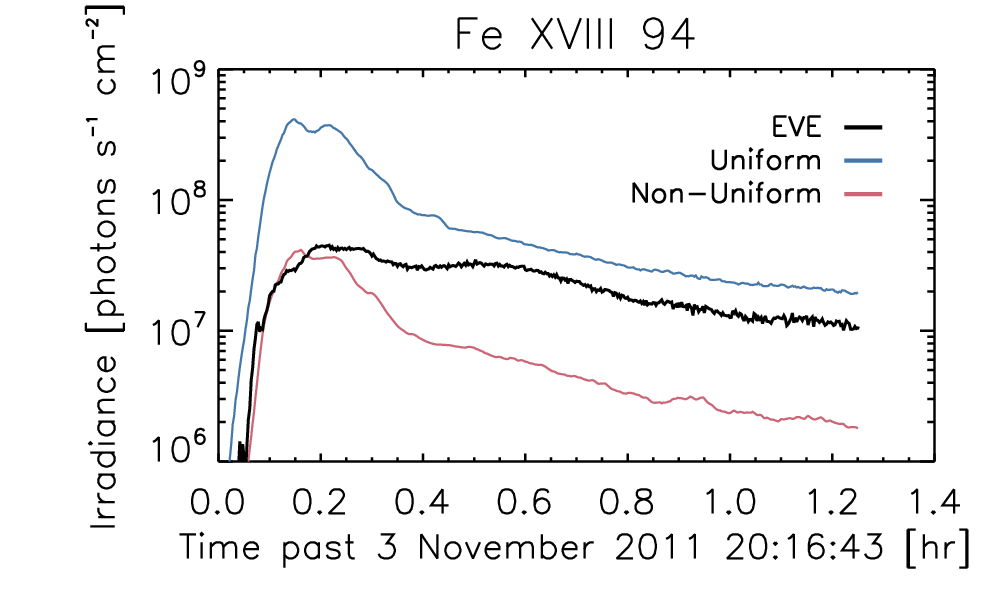}
    \includegraphics[width=0.32\textwidth]{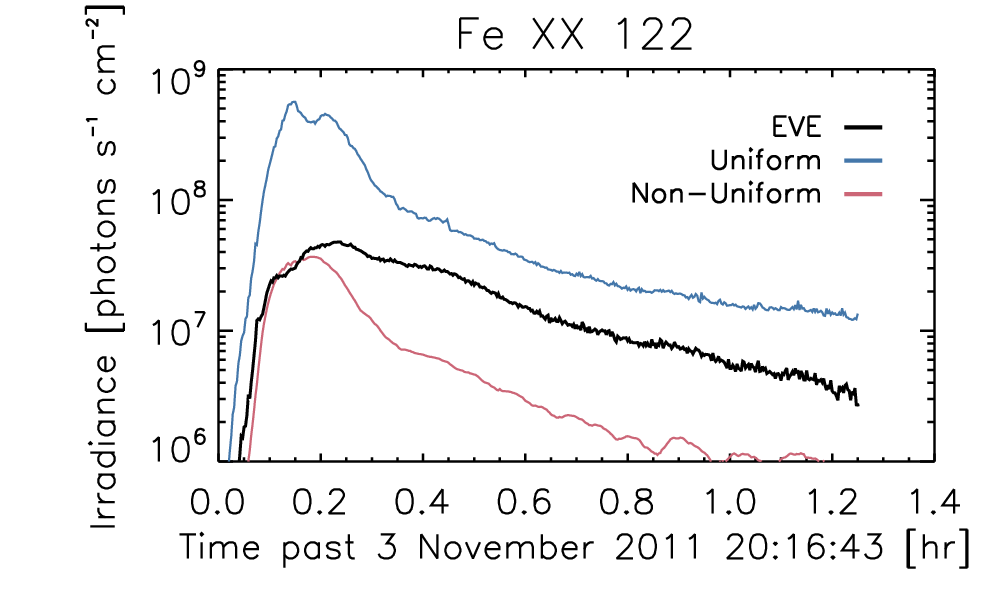}
    \includegraphics[width=0.32\textwidth]{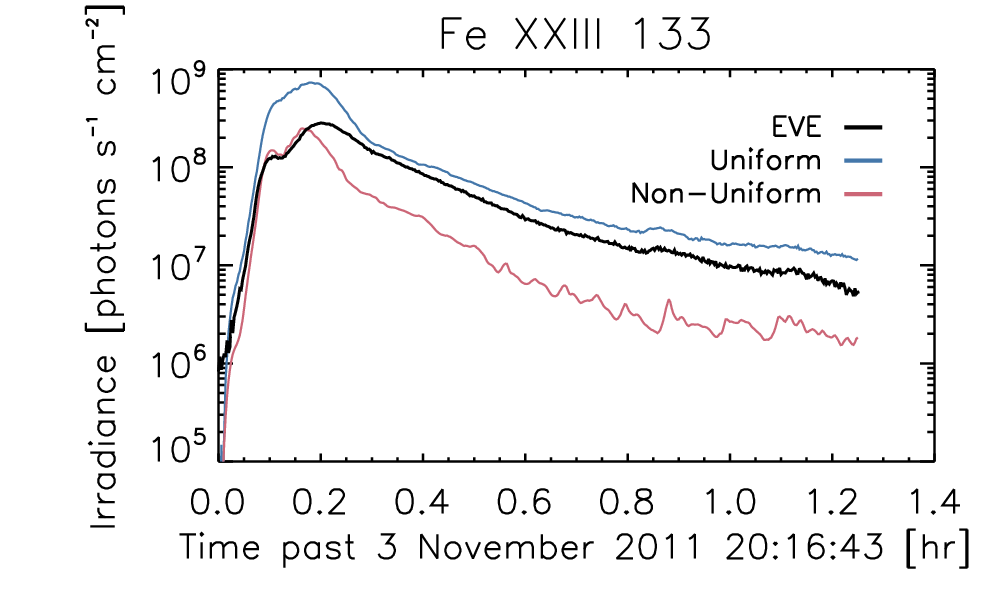}
    \includegraphics[width=0.32\textwidth]{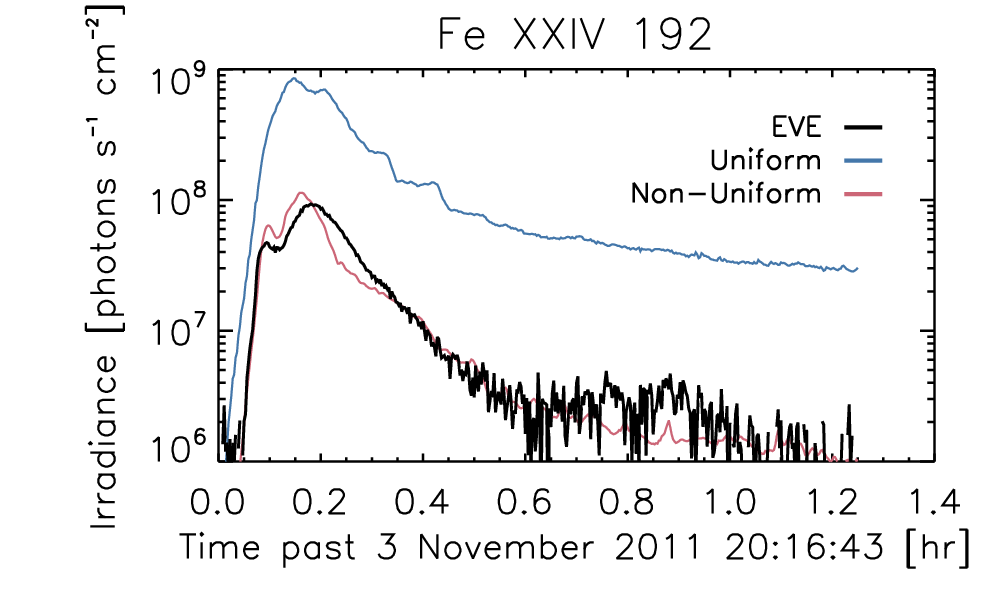}
    \caption{Modeled and observed SDO/EVE light curves for the 3 November 2011 X2.9 flare.  Similar to Figure \ref{fig:X7}.  \label{fig:X29}}
\end{figure*}

We present the results for three more events.  In Figure \ref{fig:X29}, we show the same lines for the 3 November 2011 X2.9 flare, with a peak GOES temperature of about 20 MK.  In this case, we similarly find that the uniform area model (blue) overestimates the observed irradiance during the impulsive phase, though the hotter lines (\ion{Fe}{23} 133 \AA) have good agreement in the gradual phase.  The error-based scaling of the area reproduces the peak intensities of the impulsive phase better in all lines, underestimates the intensity during the gradual phase in \ion{Fe}{18} 94 \AA and \ion{Fe}{23} 133 \AA, but reproduces the intensities at all times in \textit{e.g.} \ion{Fe}{24} 192 \AA\ and \ion{O}{6} 1032 \AA.  In this case, there was no CME-related dimming in \ion{Fe}{9} 171 \AA.  

\begin{figure*}
    \centering
    \includegraphics[width=0.32\textwidth]{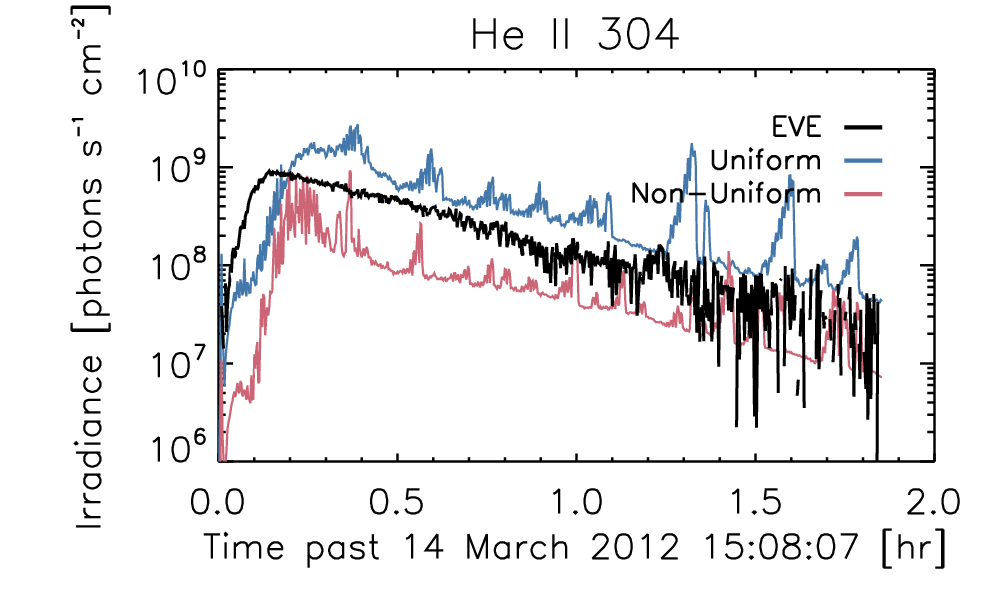}
    \includegraphics[width=0.32\textwidth]{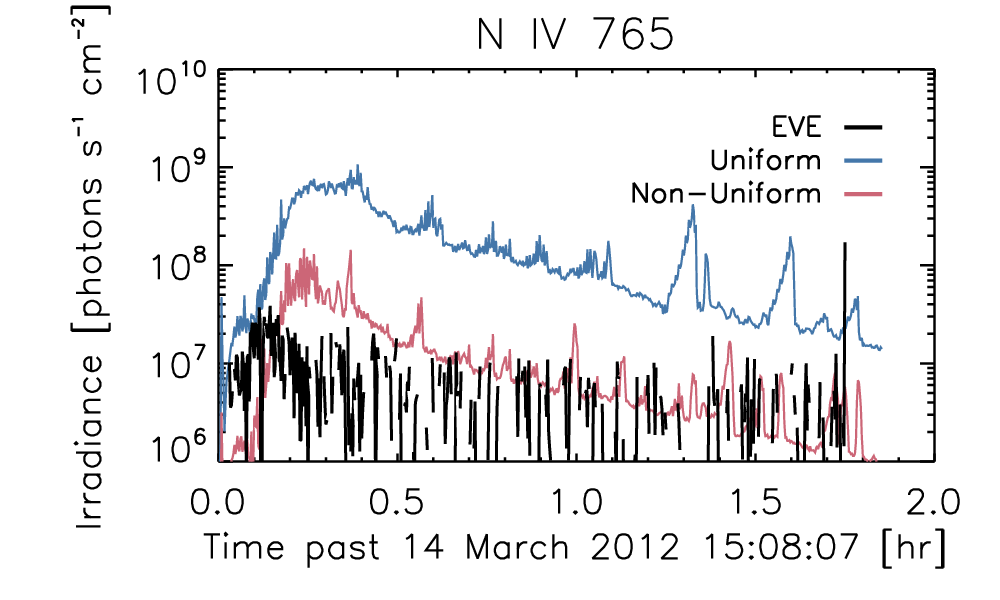}
    \includegraphics[width=0.32\textwidth]{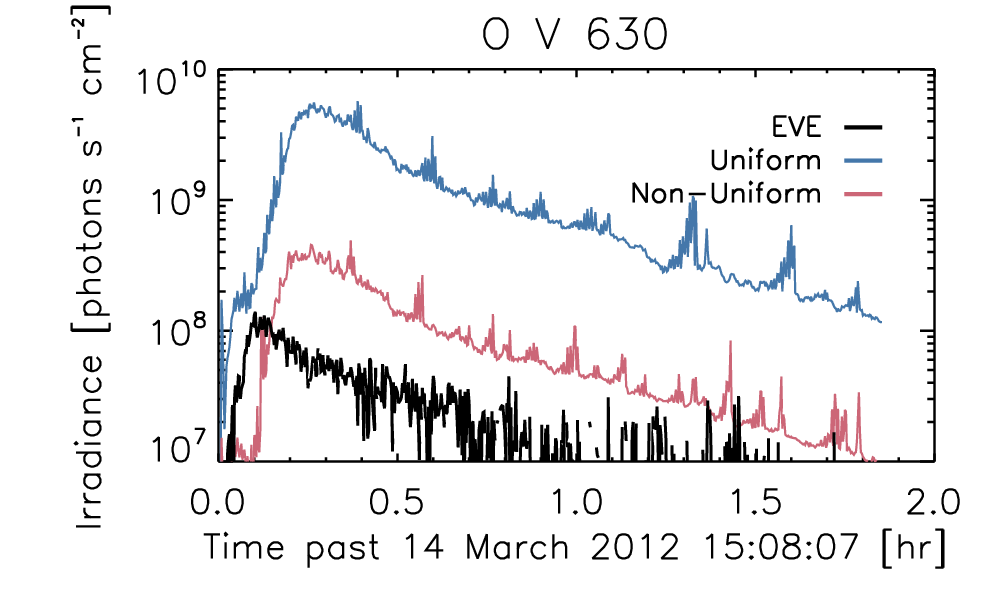}
    \includegraphics[width=0.32\textwidth]{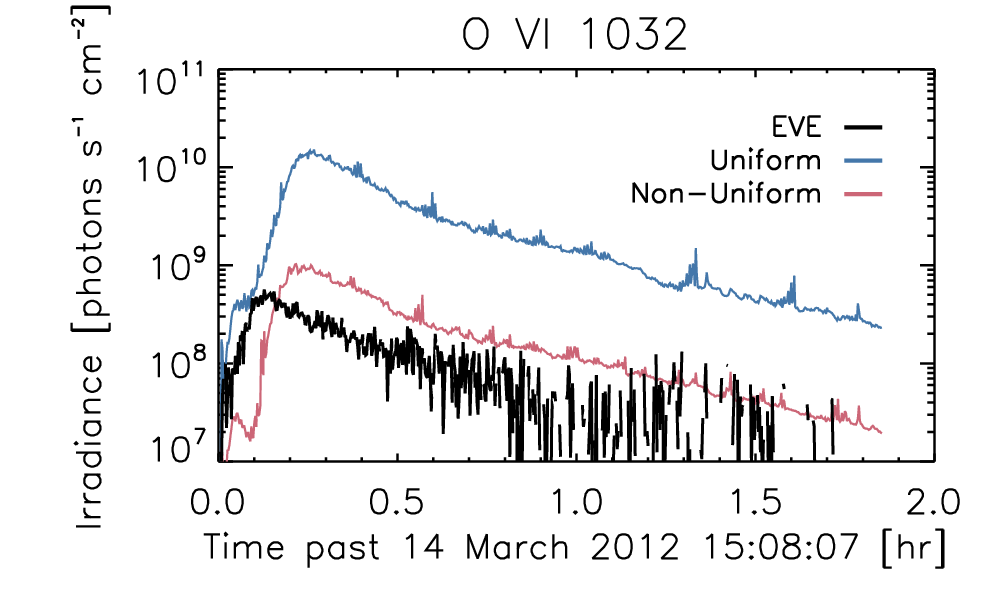}
    \includegraphics[width=0.32\textwidth]{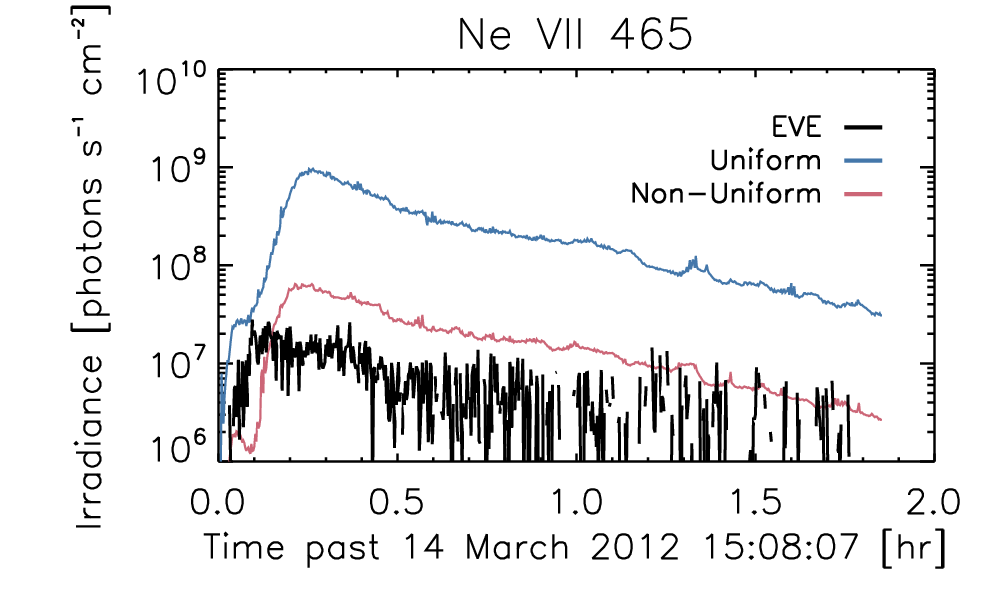}
    \includegraphics[width=0.32\textwidth]{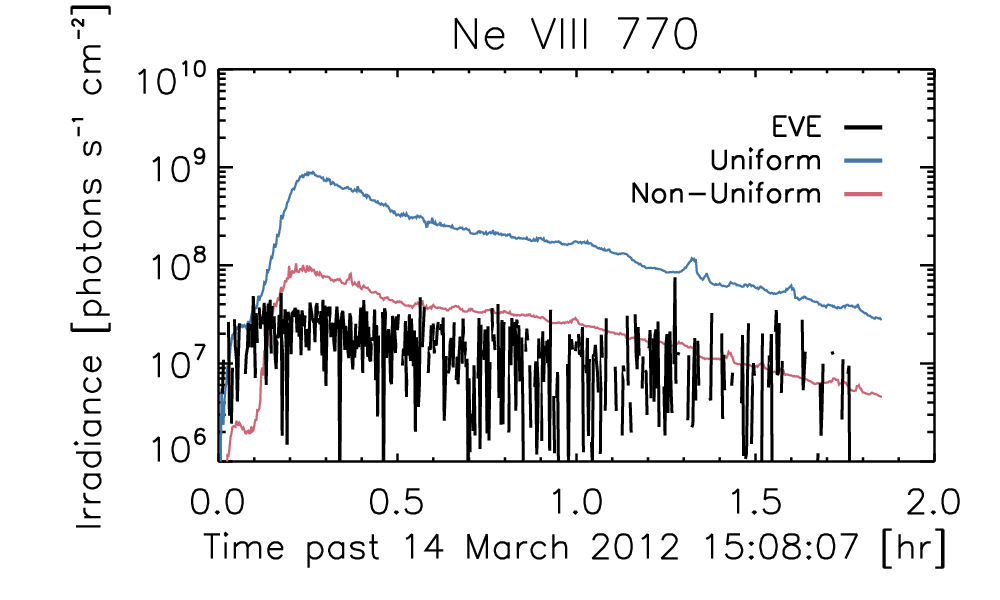}
    \includegraphics[width=0.32\textwidth]{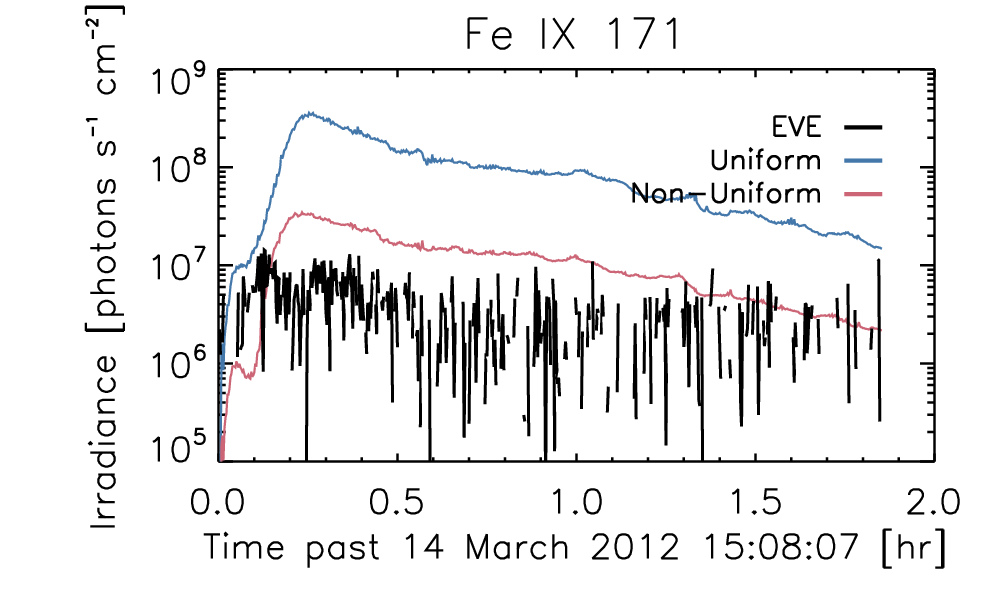}
    \includegraphics[width=0.32\textwidth]{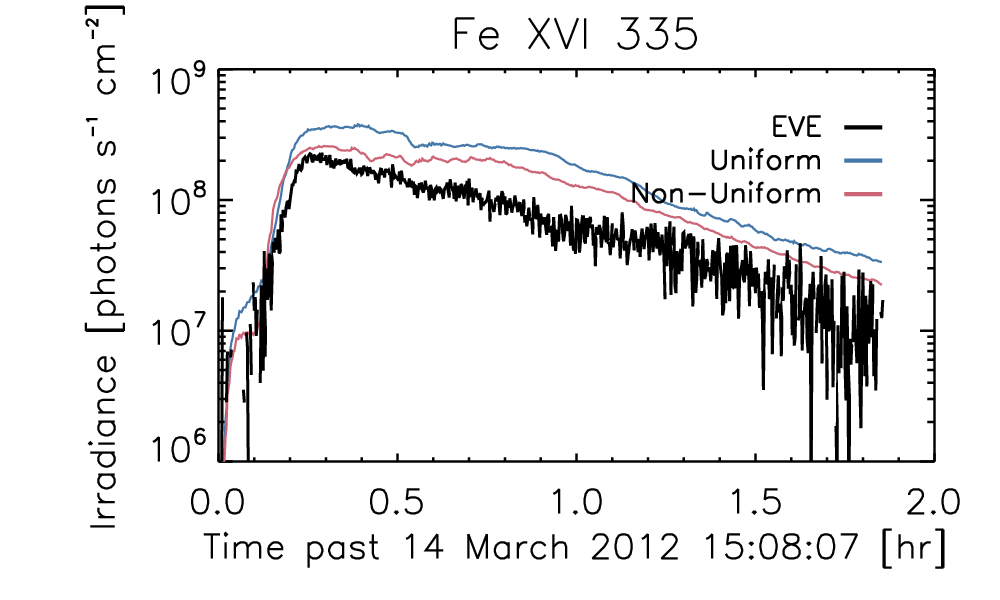}
    \includegraphics[width=0.32\textwidth]{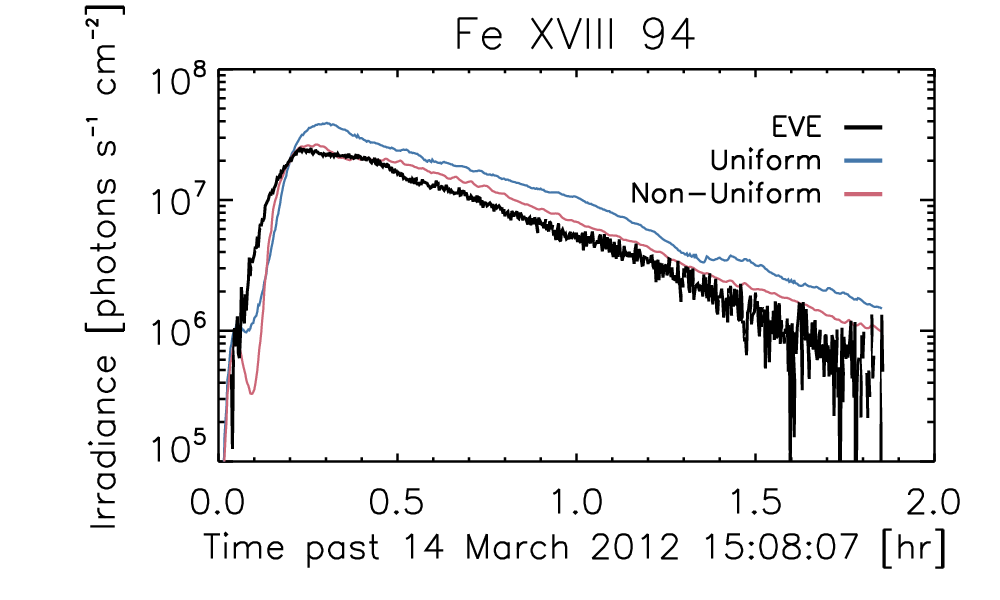}
    \includegraphics[width=0.32\textwidth]{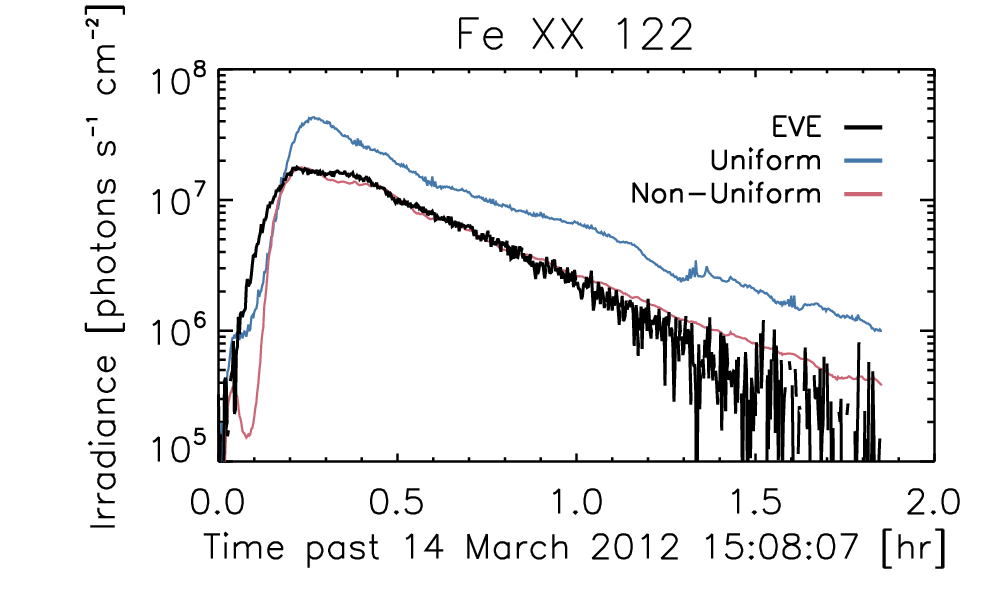}
    \includegraphics[width=0.32\textwidth]{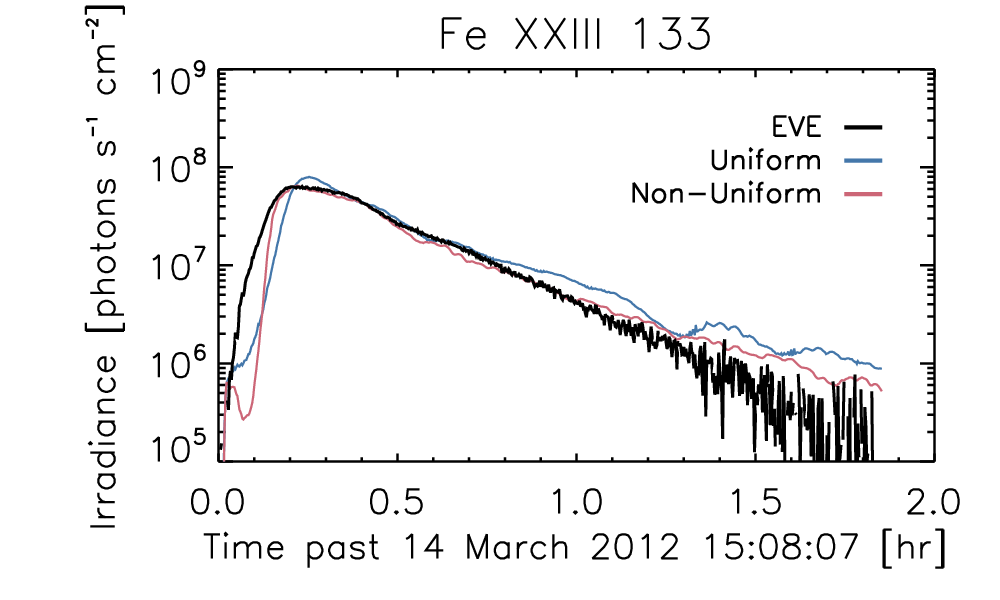}
    \includegraphics[width=0.32\textwidth]{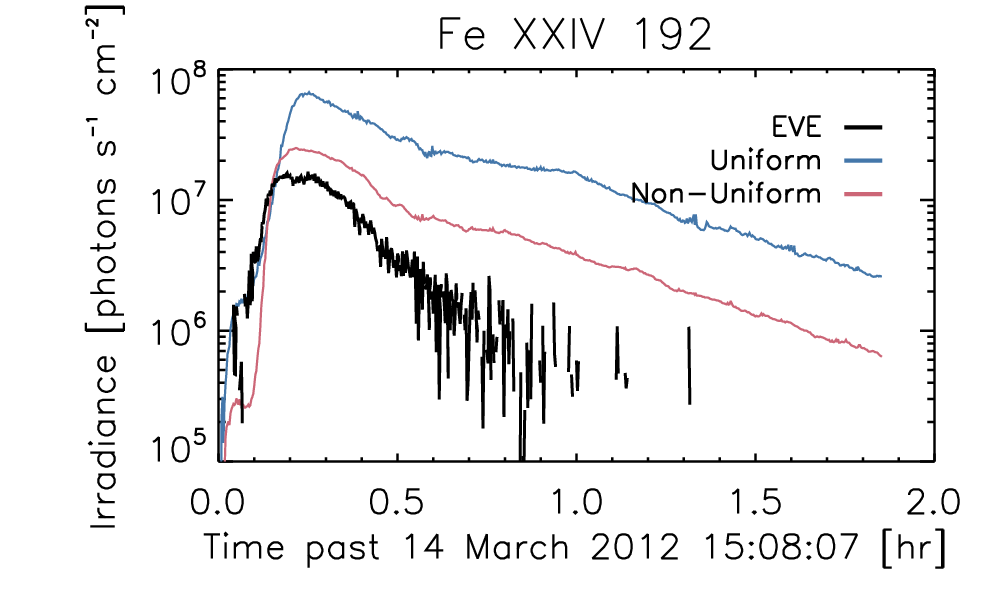}
    \caption{Modeled and observed SDO/EVE light curves for the 14 March 2012 M4.1 flare.  Similar to Figure \ref{fig:X7}.  \label{fig:M41}}
\end{figure*}

In Figure \ref{fig:M41}, we present the light curves for the 14 March 2012 M4.1 flare, with a peak GOES temperature of about 16 MK.  In this case, the lines from \ion{Fe}{16} through \ion{Fe}{23} are well-reproduced with the error-based area scaling.  The cooler lines are still somewhat overestimated even with this empirical correction, but the general evolution of the light curves is reasonable in all lines.

\begin{figure*}
    \centering
    \includegraphics[width=0.32\textwidth]{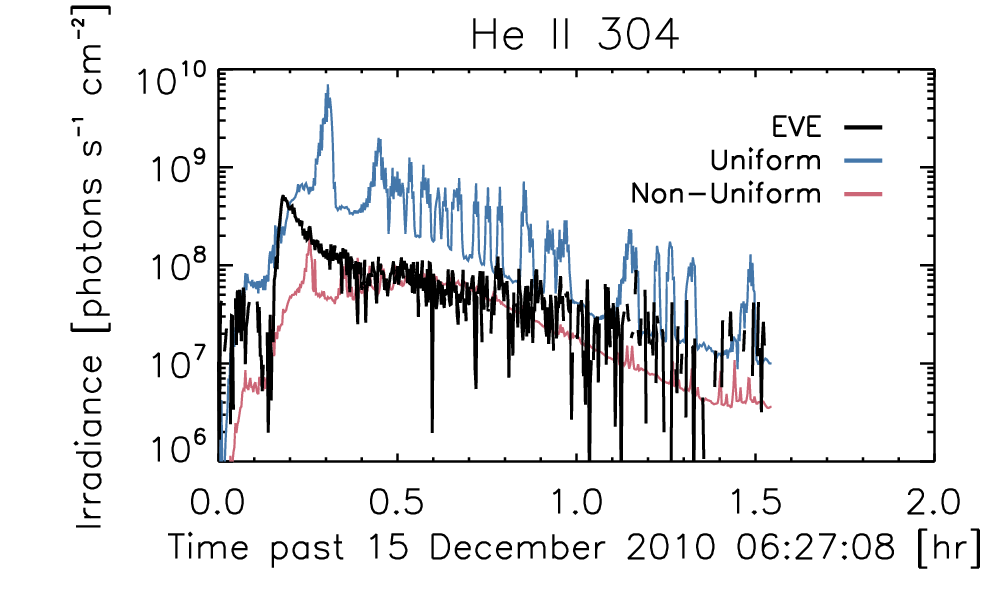}
    \includegraphics[width=0.32\textwidth]{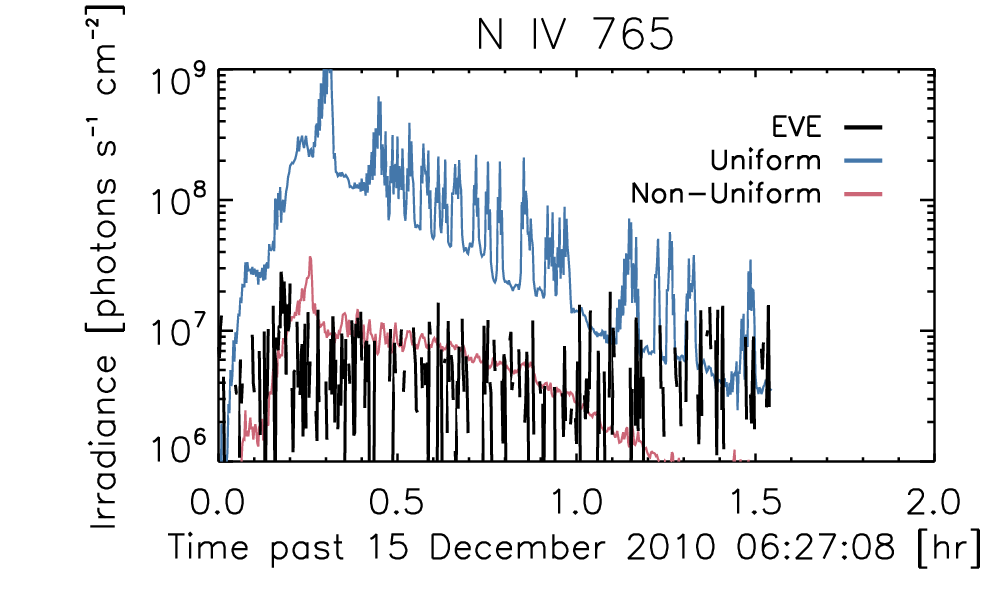}
    \includegraphics[width=0.32\textwidth]{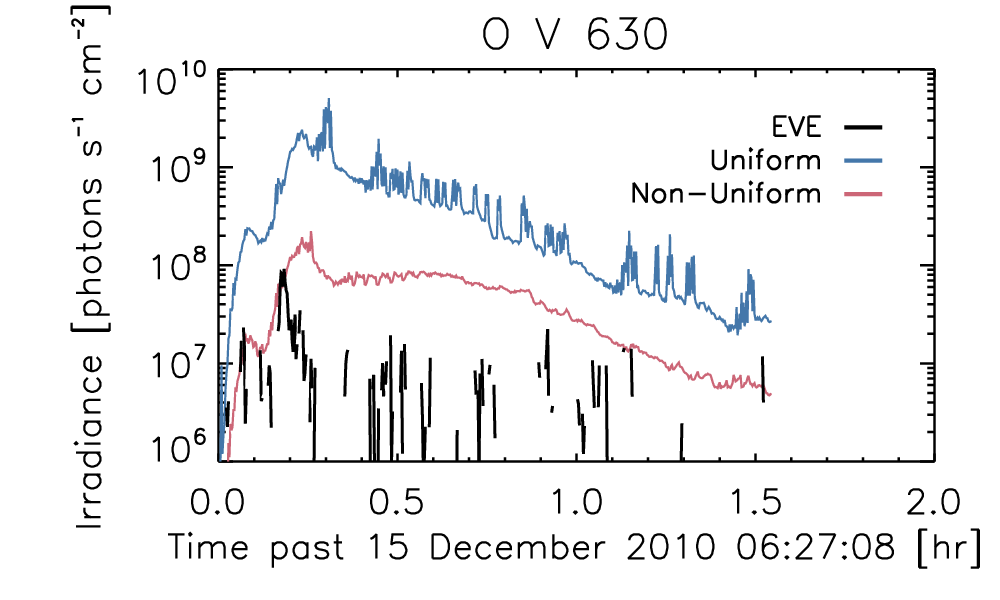}
    \includegraphics[width=0.32\textwidth]{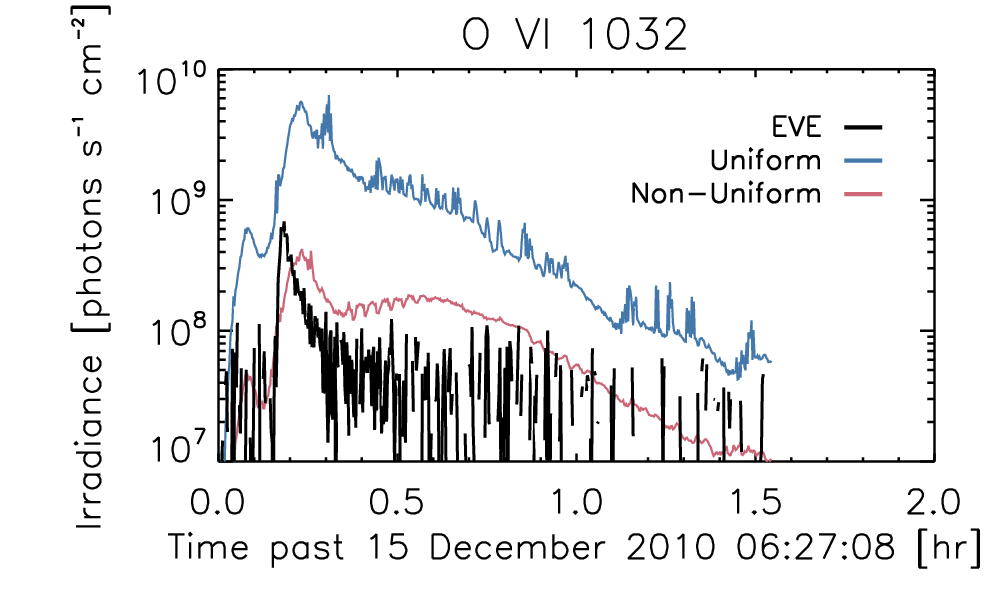}
    \includegraphics[width=0.32\textwidth]{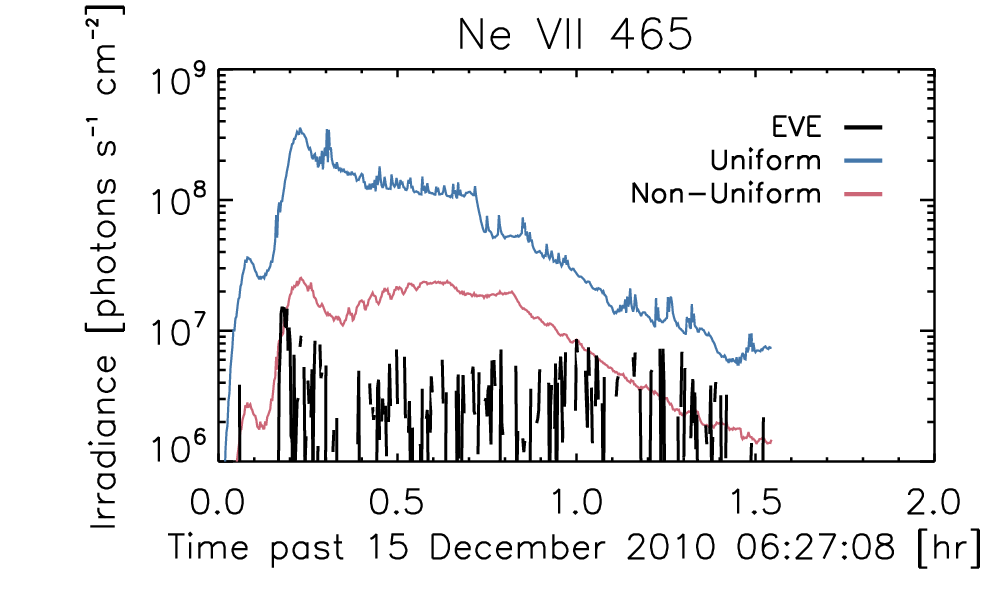}
    \includegraphics[width=0.32\textwidth]{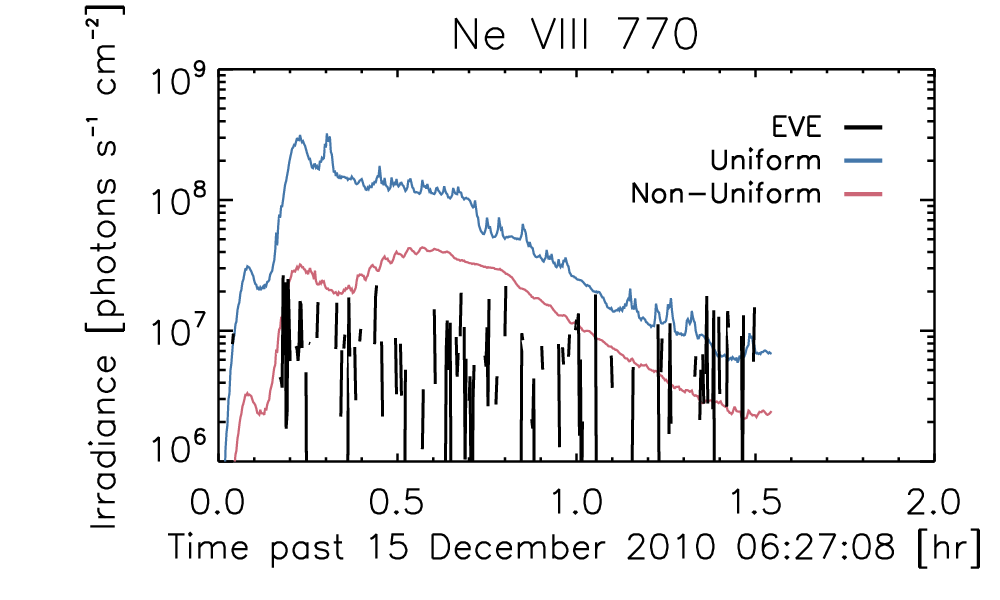}
    \includegraphics[width=0.32\textwidth]{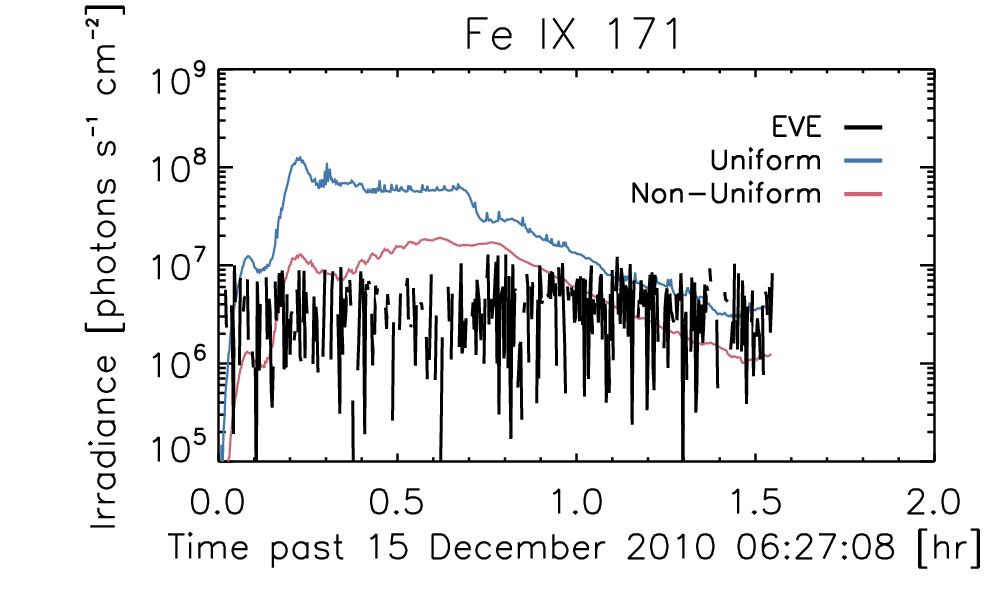}
    \includegraphics[width=0.32\textwidth]{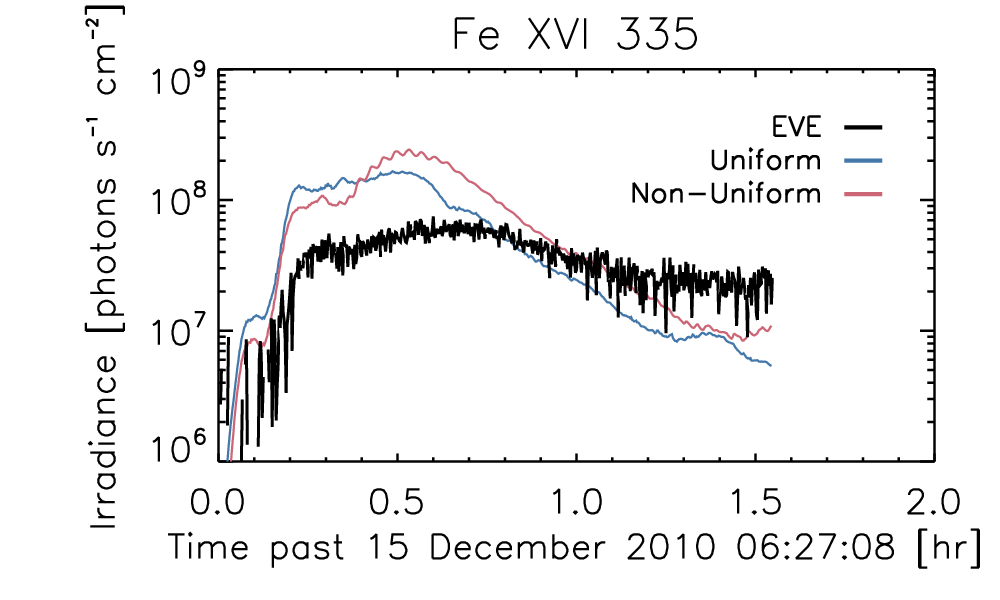}
    \includegraphics[width=0.32\textwidth]{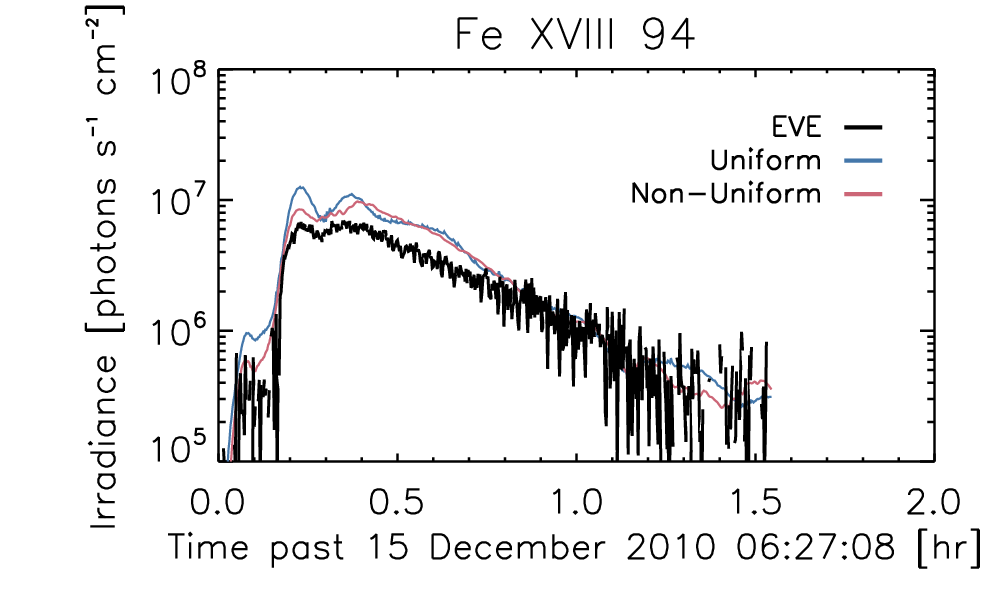}
    \includegraphics[width=0.32\textwidth]{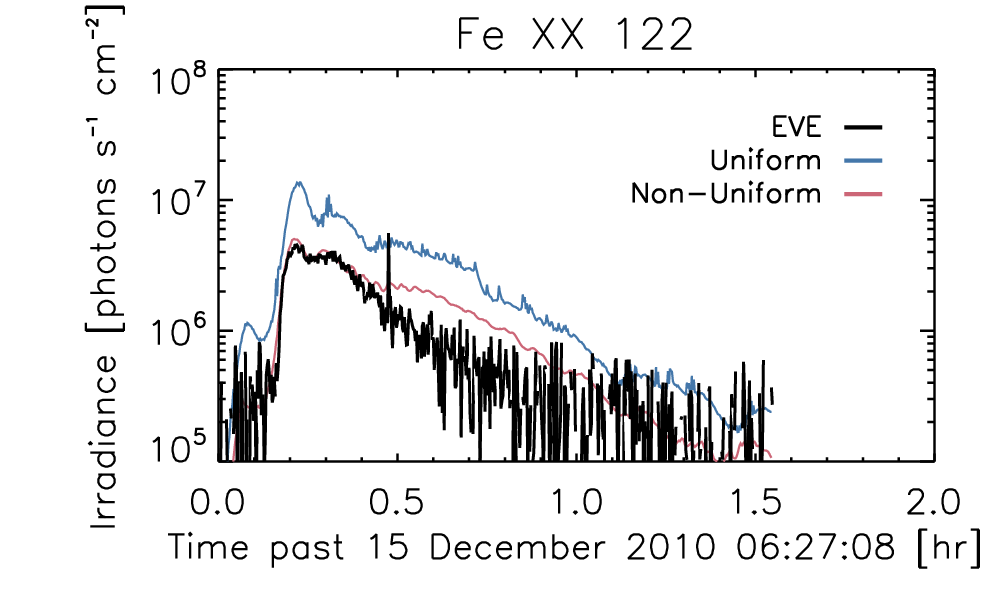}
    \includegraphics[width=0.32\textwidth]{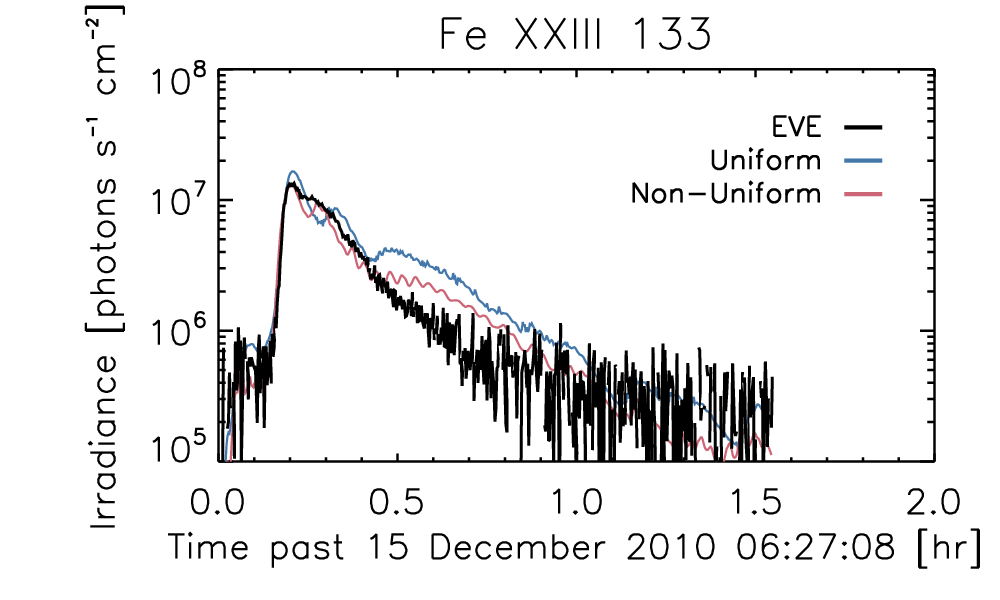}
    \includegraphics[width=0.32\textwidth]{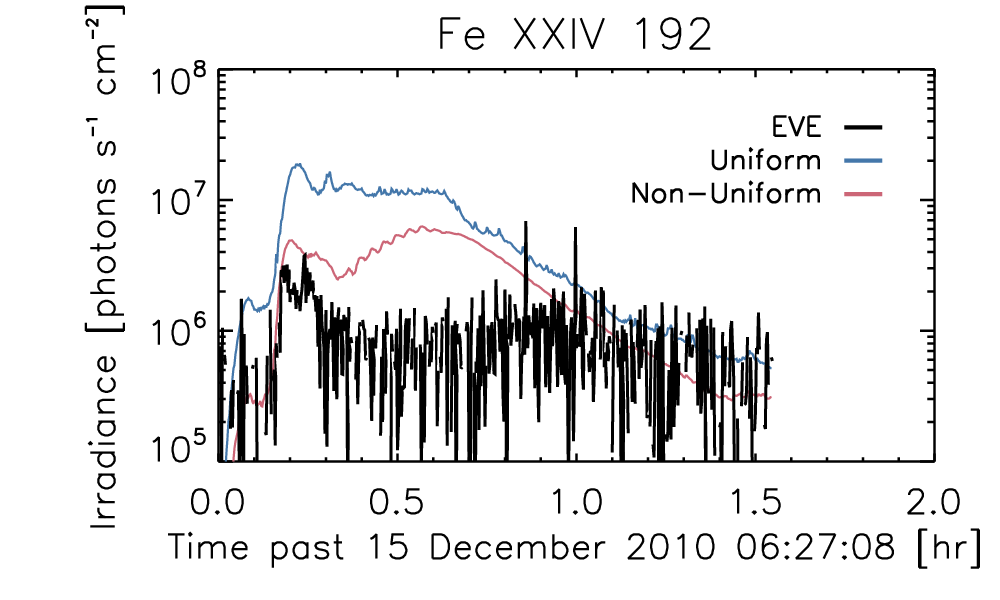}
    \caption{Modeled and observed SDO/EVE light curves for the 15 December 2010 C7.8 flare.  Similar to Figure \ref{fig:X7}.  \label{fig:C78}}
\end{figure*}

Finally, we show a small C7.8 flare that occurred on 15 December 2010 in Figure \ref{fig:C78}, with a peak GOES temperature of about 13 MK.  In this case, there is good agreement in most of the lines, though some of the lines like \ion{Ne}{8} are too noisy in the EVE data to draw strong conclusions.  However, it is apparent that the uniform area case overestimates the irradiance in the TR lines across the board.  

\subsection{Error Analysis}
\label{subsec:errors}

\begin{figure*}
\centering
\includegraphics[width=0.98\textwidth]{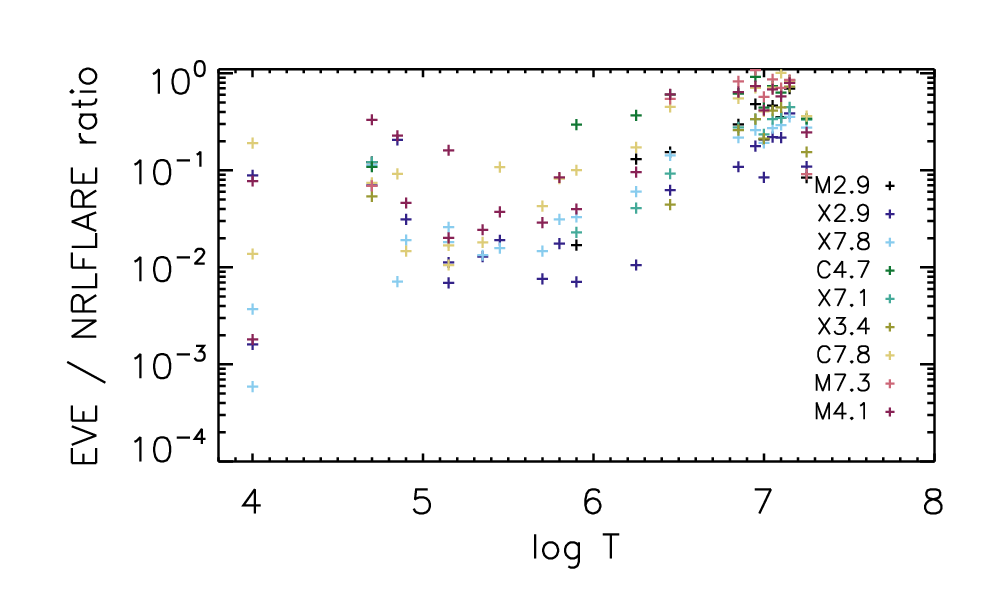}
\caption{The ratio of the peak intensity in the uniform area NRLFLARE model to the observed EVE intensity in 21 spectral lines listed in Table \ref{table:extrap} for all 9 events in Table \ref{table:events}, plotted as a function of line formation temperature.  The intensity is over-estimated in the uniform area case in all lines in all events, though the error is less significant in the hotter lines.  \label{fig:obs_model_ratio}}
\end{figure*}

We can measure the magnitude of the error in our model by plotting the ratio of the peak observed and modeled (uniform) irradiance in the various lines.  In Figure \ref{fig:obs_model_ratio}, we show the ratio of the peak intensity in the uniform area NRLFLARE model to the observed EVE intensity as a function of line formation temperature.  We show all 21 lines listed in Table \ref{table:extrap} for all 9 flares listed in Table \ref{table:events}.  In all events, all lines are overestimated in intensity.  The modeled values are fairly close to observations for the hottest lines, and the error grows with decreasing temperature.  The modeled TR lines are often 1 or 2 orders of magnitude brighter than the observations.  The chromospheric lines are similarly overestimated ($\log T \lesssim 5$), but these lines are undoubtedly optically thick, and therefore we expect poor fits using CHIANTI alone to synthesize the emission.  Additionally, the errors in some flares are worse than in others (e.g. the X7.8 flare is systematically worse than the C7.8 flare).  

\begin{figure*}
    \centering
    \includegraphics[width=0.32\textwidth]{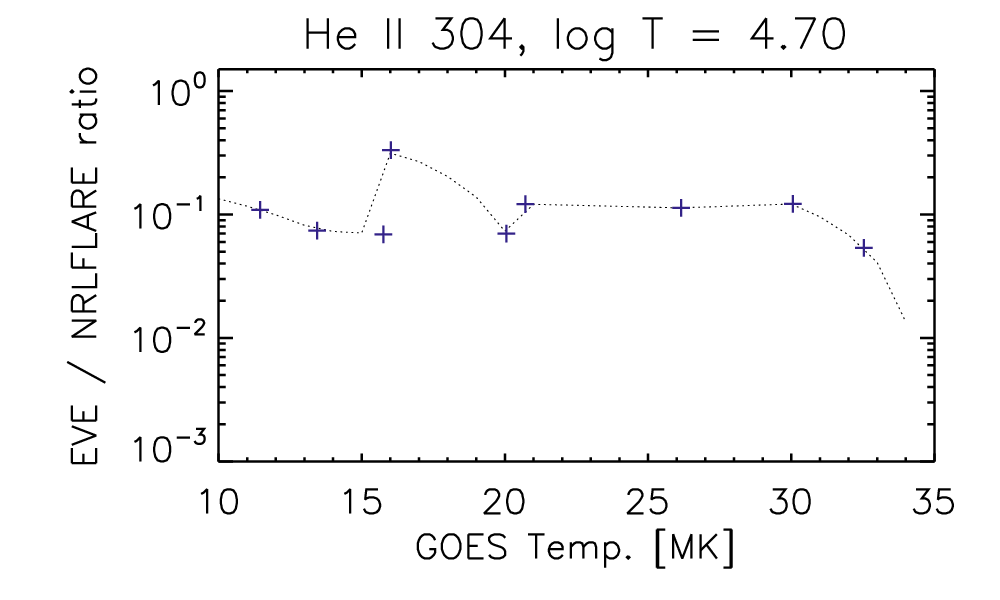}
    \includegraphics[width=0.32\textwidth]{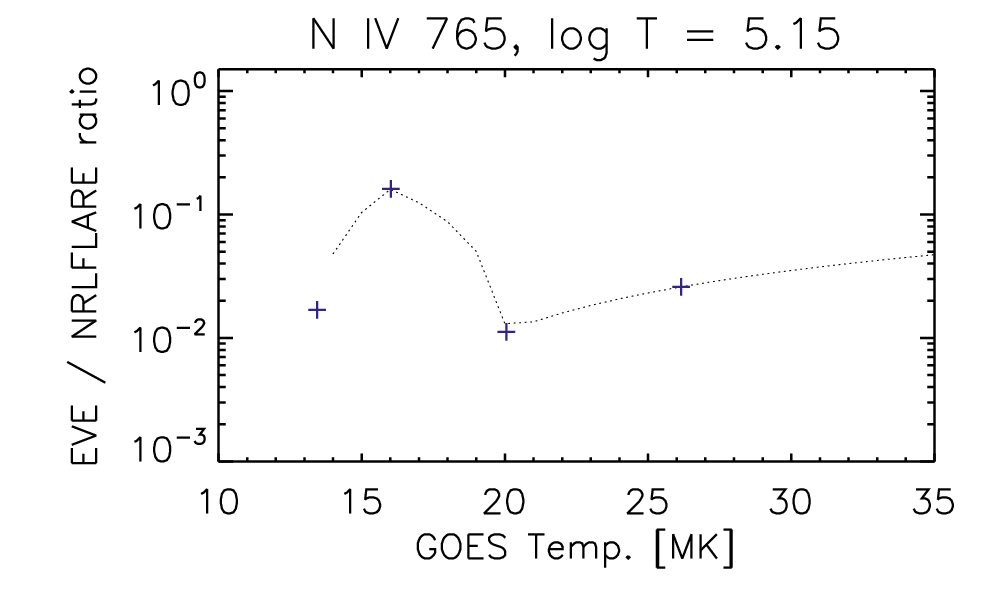}
    \includegraphics[width=0.32\textwidth]{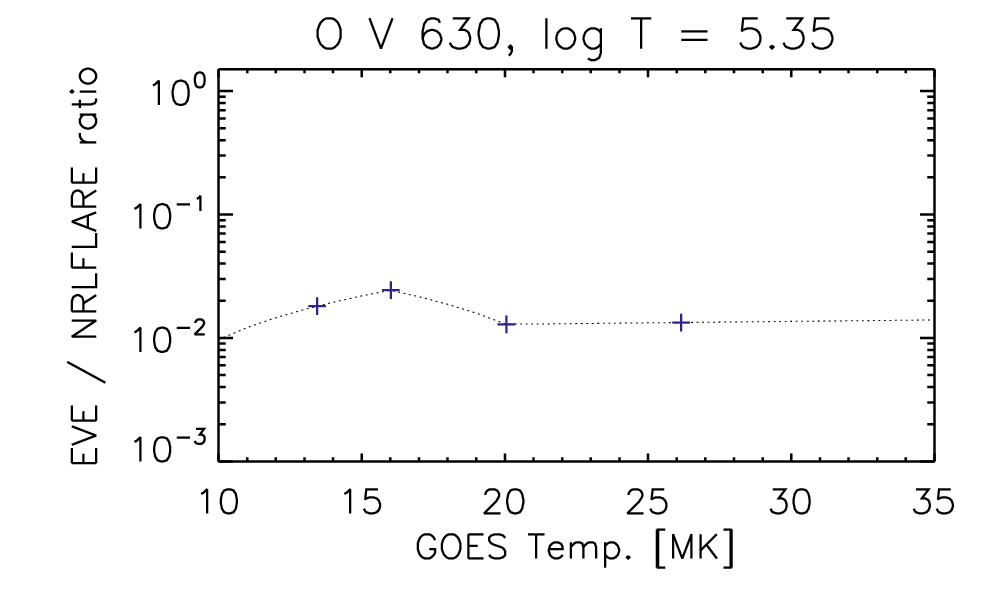}
    \includegraphics[width=0.32\textwidth]{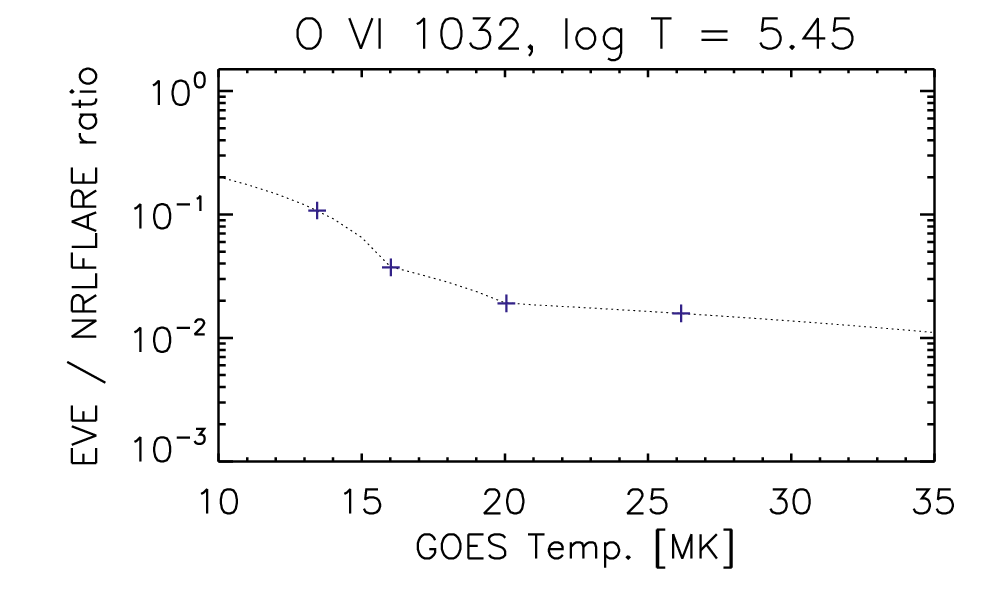}
    \includegraphics[width=0.32\textwidth]{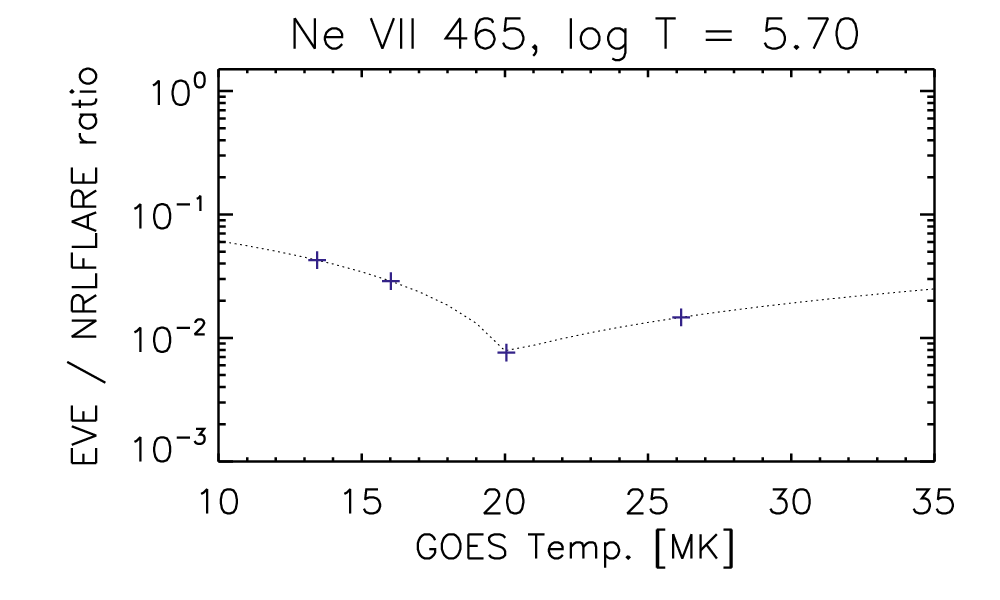}
    \includegraphics[width=0.32\textwidth]{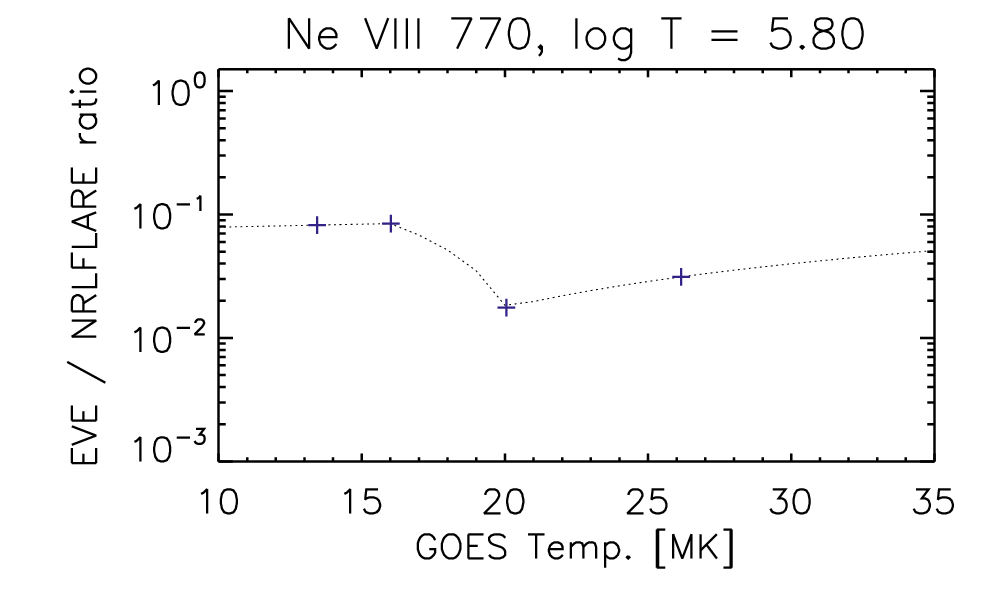}
    \includegraphics[width=0.32\textwidth]{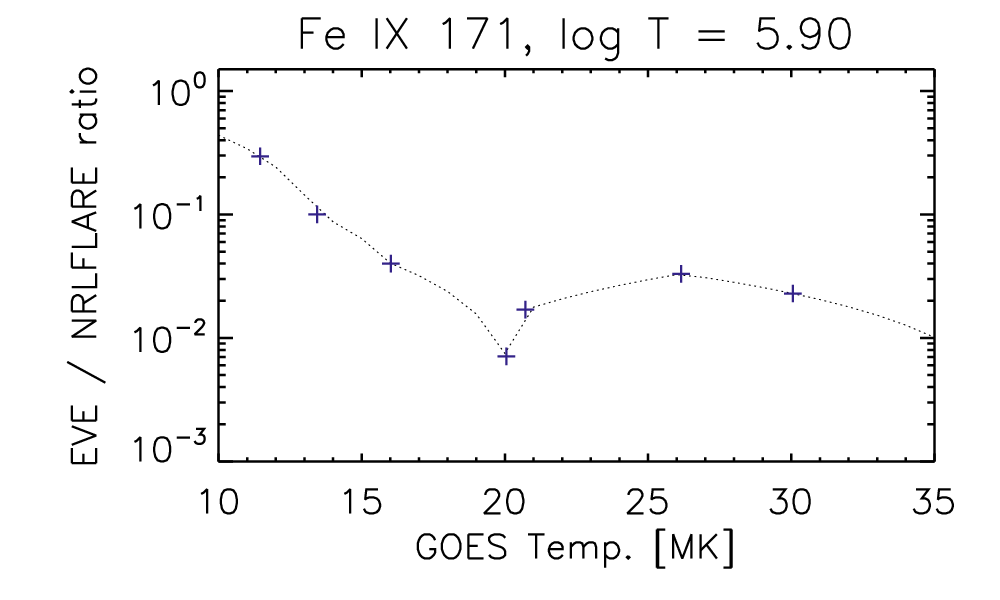}
    \includegraphics[width=0.32\textwidth]{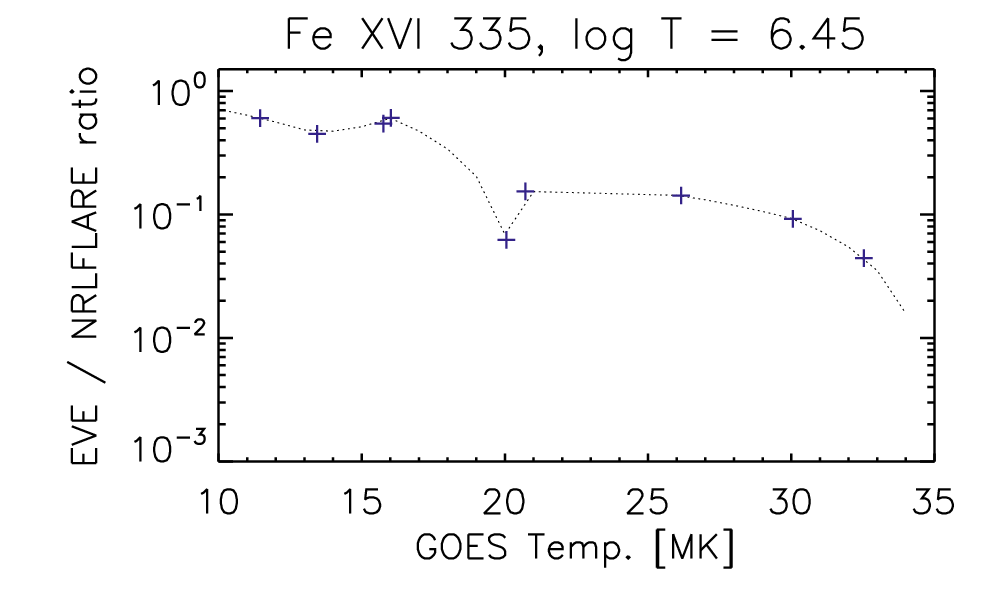}
    \includegraphics[width=0.32\textwidth]{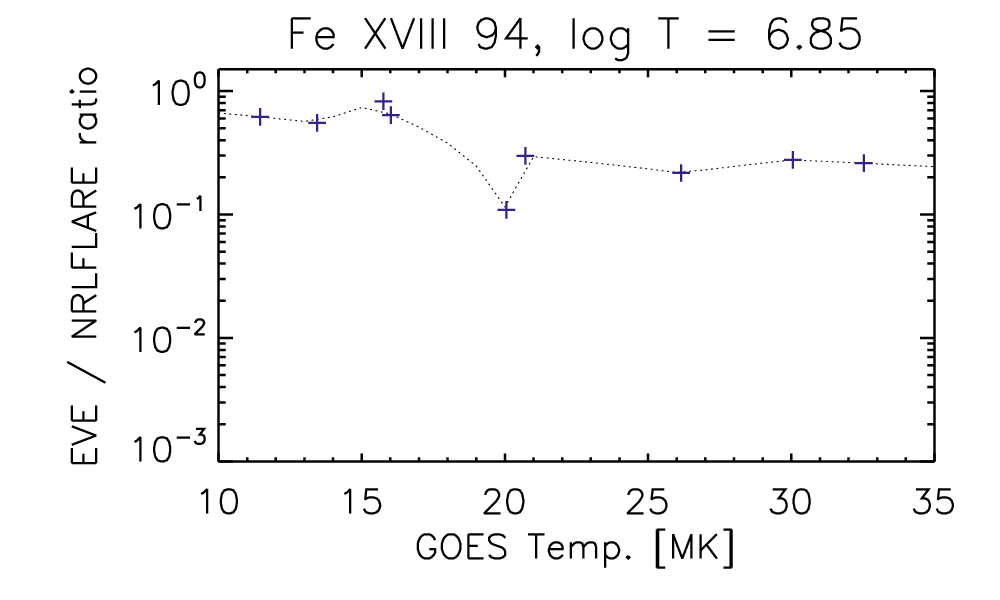}
    \includegraphics[width=0.32\textwidth]{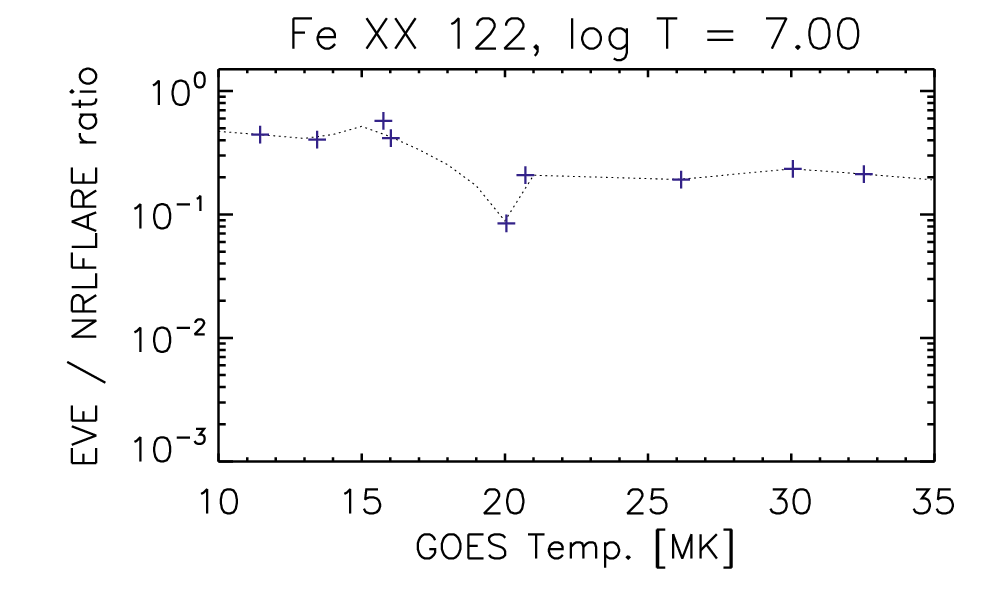}
    \includegraphics[width=0.32\textwidth]{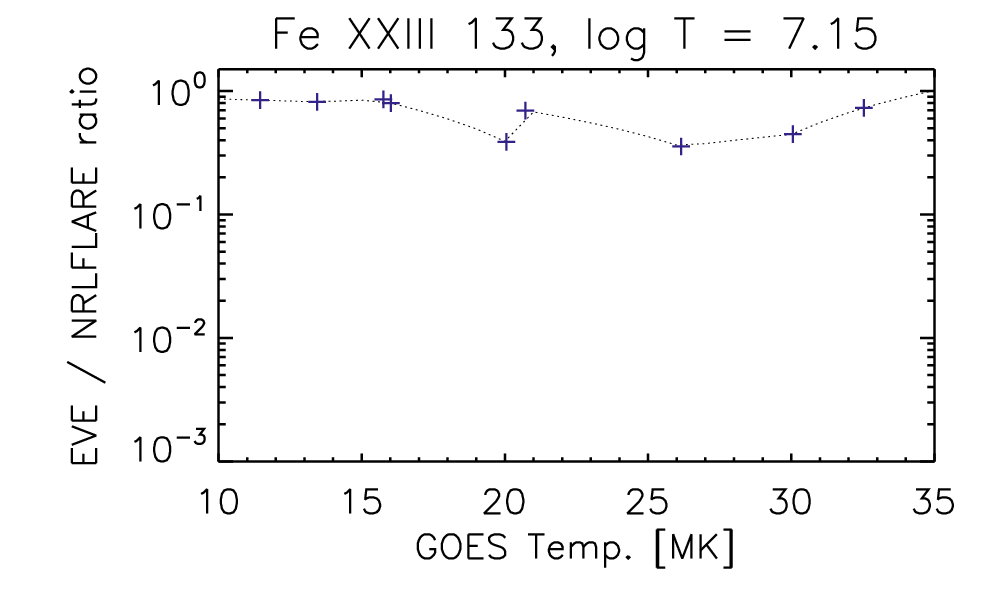}
    \includegraphics[width=0.32\textwidth]{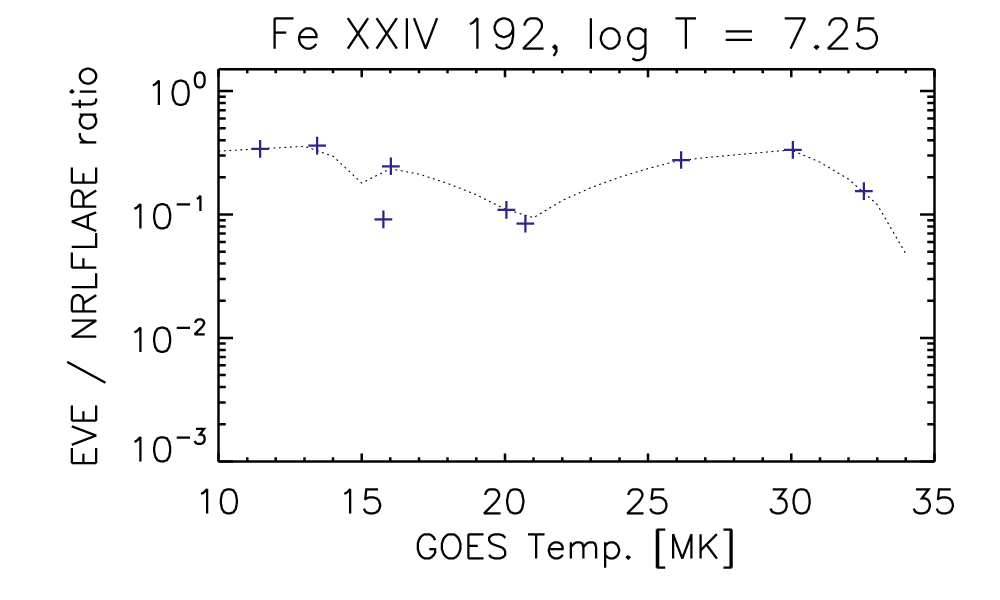}
    \caption{The ratio of observed-to-modeled irradiance for 12 lines observed with EVE, as a function of peak GOES temperature for 9 flares (tick marks).  A simple interpolation has been overlaid on each plot. \label{fig:errors}}
\end{figure*}

It is important to note that the discrepancy between model and observed line intensities depends on individual events.  Hotter flares are often worse than cooler ones in this sample (\textit{n.b.} the GOES class and GOES temperature are related, but not monotonic -- see Table \ref{table:events}).  In Figure \ref{fig:errors}, we show these ratios for all 9 flares in Table \ref{table:events}, where we have now plotted against the peak GOES temperature (measured from the ratio of XRS-A to XRS-B).  In \ion{He}{2} 304 \AA, we find that the ratio is approximately constant in all events, while in lines like \ion{Fe}{16} 335 \AA, the hottest flares deviate from observations more strongly than cooler events.   

We use these error ratios to empirically scale the loop cross-sectional area as a function of temperature to recalculate the model irradiance (e.g. the red curves in Figures \ref{fig:X7}, \ref{fig:X29}, \ref{fig:M41}, and \ref{fig:C78}).  We have done this by altering the $A(s)$ term in Equation \ref{eqn:em} when we synthesize the spectra, though future work should self-consistently treat the area expansion in the \texttt{HYDRAD} simulations.  The agreement in the non-uniform case is significantly improved over the uniform case in all events.  While the fits are still imperfect, the discrepancies have been significantly reduced, and both the impulsive phase and gradual phase show better agreement in most lines.  Extrapolations of magnetic fields show that the magnetic field is not constant along the length of the loop, which implies that the area must expand.  In flares, this is often difficult to detect directly because of saturation issues and the rapid evolution of the field.  However, the errors here give an estimate of the rate and magnitude of the expansion that we need in order to reproduce the intensities measured with EVE.  We therefore suggest that a non-uniform area expansion is a crucial element to modeling solar flares.

\section{Discussion}
\label{sec:discussion}

We have developed a physical model of solar flare irradiance.  The model is capable of reproducing emission in hot flaring and coronal lines when compared to the SDO/EVE data for the examined events, but only if there is a non-uniform cross-sectional area.  The irradiance is too bright in all lines when the loops are assumed to have uniform cross-section.  The TR emission in particular, from approximately $4.85 \leq \log{T} \leq 5.7$, is significantly over-estimated in the model, often by more than an order of magnitude.  It is possible to empirically correct this, but the physics behind this requires significantly more work in the future.  There are a few possible explanations.  

First, we find that the cross-sectional area of flaring loops cannot be constant along the length of the loops, but needs to have significant expansion from the chromosphere through the corona in order to simultaneously reproduce the intensities of coronal and transition region lines.  We have shown that one such fitting can indeed improve agreement with the light curves, but this was not done self-consistently.  Imaging observations of area expansion in (non-flaring) coronal loops have in the past shown little expansion over the length of the loop \citep{klimchuk1992,klimchuk2000,klimchuk2020}, though magnetic field extrapolations require that the magnetic field strength vary along the length, necessitating an area expansion due to conservation of magnetic flux.  \citet{dudik2014}, using extrapolations from a quiescent active region, found a wide distribution of area expansion factors ranging from uniform to over 80.  There is a fundamental disagreement between these two results, but the intensities of the transition region lines suggest that there is indeed a large and rapid expansion occurring throughout the lower atmosphere and into the corona.  Observations of the moss (loop footpoints) in active regions have similarly shown a discrepancy between modeled and observed line intensities at these temperatures \citep{warren2010}.  In the appendix, we additionally hypothesize that the cross-sectional area might depend on the local plasma $\beta$, which can increase greatly during a flare.  

Second, it is also likely that some or all of these TR lines are not optically thin.  Most modeling efforts assume that the TR lines are optically thin in general, as done here, even though the density rises sharply in this region.  In principle, this ought to be determined on a case-by-case basis for each line and even each flare.  Recent work has shown that this assumption can be wrong in \ion{C}{2} \citep{rathore2015} and \ion{Si}{4} \citep{kerr2019}, particularly in strongly heated events.  Additionally, flare observations with SDO/EVE have found center-to-limb variations in intensities of hot lines typically considered optically thin \citep{thiemann2018}, including 5 lines in this paper (\ion{Fe}{18} 94, \ion{Fe}{19} 109, \ion{Fe}{21} 129, \ion{Fe}{23} 133, and \ion{Fe}{24} 192).  The optical depth of many of these lines needs to be examined with detailed radiative transfer models to understand these possibilities better, though current transfer models cannot treat such high temperature ions.  The spectra presented in Figure \ref{fig:spectra}, for example, also show that there is a discrepancy in the helium continuum between the model and observations, which requires radiative transfer to treat correctly.  Future versions of this model will therefore build in radiative transfer, for example, using the RH1.5D \citep{pereira2015} or Lightweaver \citep{osborne2021} codes.

Third, there is the possibility that the assumption that the plasma is Maxwellian is incorrect, and that the ionization balances are therefore significantly different from those found with CHIANTI.  Under the assumption of a $\kappa$-distribution, for example, the peak formation temperatures can be shifted, and the range of formation can be significantly broadened \citep{dzifcakova2015,dudik2019}.  Observations have shown that spectral lines can often be better fit with a $\kappa$-distribution in flares \citep{polito2018}, and evidence has been possibly found for a strongly non-Maxwellian plasma in the 7 March 2012 flare examined in this work \citep{dzifcakova2018}.  This effect has been shown importantly to strongly affect the intensities of TR lines \citep{dzifcakova2018b}.  Future work should examine whether the discrepancies between the modeled and observed irradiances in this paper could be due to non-Maxwellian plasma.

Of course, it is possible and even probable that all of these effects contribute to the discrepancy between observations and modeling.  It will require significant effort to determine to what extent each affects the intensity of given lines, and we plan to examine each in turn to further improve our model of flaring irradiance.  Importantly, the model-data comparisons must focus on reproducing many lines simultaneously, as improving the model for a single line may worsen agreement in others (see, \textit{e.g.} Figure 13 of \citealt{reep2019}).  We will also examine the self-consistent modeling of an expanding cross-section on both the irradiance and cooling timescale of the plasma in coronal loops.  Since EVE is spatially unresolved, it is incapable of diagnosing dynamics on individual loops, so radiance observations from Hinode/EIS and IRIS will aid in understanding this.  

\leavevmode \newline

\acknowledgments  
The authors were supported by NASA's Living With a Star program.  The authors thank Jaroslav Dud\'ik, Yi-Ming Wang, and Graham Kerr for helpful discussions.  The authors also thank the anonymous referee for their comments which strengthened this paper.

\bibliography{apj}

\begin{thebibliography}{}
\expandafter\ifx\csname natexlab\endcsname\relax\def\natexlab#1{#1}\fi
\providecommand{\url}[1]{\href{#1}{#1}}

\bibitem[{{Asai} {et~al.}(2004){Asai}, {Yokoyama}, {Shimojo}, {Masuda},
  {Kurokawa}, \& {Shibata}}]{asai2004}
{Asai}, A., {Yokoyama}, T., {Shimojo}, M., {et~al.} 2004, \apj, 611, 557

\bibitem[{{Barnes} {et~al.}(2016){Barnes}, {Cargill}, \&
  {Bradshaw}}]{barnes2016a}
{Barnes}, W.~T., {Cargill}, P.~J., \& {Bradshaw}, S.~J. 2016, \apj, 829, 31

\bibitem[{{Bradshaw} \& {Cargill}(2013)}]{bradshaw2013}
{Bradshaw}, S.~J., \& {Cargill}, P.~J. 2013, \apj, 770, 12

\bibitem[{{Bradshaw} \& {Mason}(2003)}]{bradshaw2003}
{Bradshaw}, S.~J., \& {Mason}, H.~E. 2003, \aap, 401, 699

\bibitem[{{Bradshaw} \& {Testa}(2019)}]{bradshaw2019}
{Bradshaw}, S.~J., \& {Testa}, P. 2019, \apj, 872, 123

\bibitem[{{Brooks} {et~al.}(2021){Brooks}, {Warren}, \& {Landi}}]{brooks2021}
{Brooks}, D.~H., {Warren}, H.~P., \& {Landi}, E. 2021, \apjl, 915, L24

\bibitem[{{Carlsson} \& {Leenaarts}(2012)}]{carlsson2012}
{Carlsson}, M., \& {Leenaarts}, J. 2012, \aap, 539, A39

\bibitem[{{Chamberlin} {et~al.}(2007){Chamberlin}, {Woods}, \&
  {Eparvier}}]{chamberlin2007}
{Chamberlin}, P.~C., {Woods}, T.~N., \& {Eparvier}, F.~G. 2007, Space Weather,
  5, S07005

\bibitem[{{Chamberlin} {et~al.}(2008){Chamberlin}, {Woods}, \&
  {Eparvier}}]{chamberlin2008}
---. 2008, Space Weather, 6, S05001

\bibitem[{{Chamberlin} {et~al.}(2020){Chamberlin}, {Eparvier}, {Knoer},
  {Leise}, {Pankratz}, {Snow}, {Templeman}, {Thiemann}, {Woodraska}, \&
  {Woods}}]{chamberlin2020}
{Chamberlin}, P.~C., {Eparvier}, F.~G., {Knoer}, V., {et~al.} 2020, Space
  Weather, 18, e02588

\bibitem[{{Del Zanna} {et~al.}(2020){Del Zanna}, {Dere}, {Young}, \&
  {Landi}}]{delzanna2020}
{Del Zanna}, G., {Dere}, K.~P., {Young}, P.~R., \& {Landi}, E. 2020, arXiv
  e-prints, arXiv:2011.05211

\bibitem[{{Del Zanna} {et~al.}(2011){Del Zanna}, {Mitra-Kraev}, {Bradshaw},
  {Mason}, \& {Asai}}]{delzanna2011}
{Del Zanna}, G., {Mitra-Kraev}, U., {Bradshaw}, S.~J., {Mason}, H.~E., \&
  {Asai}, A. 2011, \aap, 526, A1

\bibitem[{{Del Zanna} \& {Woods}(2013)}]{delzanna2013}
{Del Zanna}, G., \& {Woods}, T.~N. 2013, \aap, 555, A59

\bibitem[{{Dere} {et~al.}(2019){Dere}, {Del Zanna}, {Young}, {Landi}, \&
  {Sutherland}}]{dere2019}
{Dere}, K.~P., {Del Zanna}, G., {Young}, P.~R., {Landi}, E., \& {Sutherland},
  R.~S. 2019, \apjs, 241, 22

\bibitem[{{Dere} {et~al.}(1997){Dere}, {Landi}, {Mason}, {Monsignori Fossi}, \&
  {Young}}]{dere1997}
{Dere}, K.~P., {Landi}, E., {Mason}, H.~E., {Monsignori Fossi}, B.~C., \&
  {Young}, P.~R. 1997, \aaps, 125, 149

\bibitem[{{Dud{\'\i}k} {et~al.}(2014){Dud{\'\i}k}, {Dzif{\v{c}}{\'a}kov{\'a}},
  \& {Cirtain}}]{dudik2014}
{Dud{\'\i}k}, J., {Dzif{\v{c}}{\'a}kov{\'a}}, E., \& {Cirtain}, J.~W. 2014,
  \apj, 796, 20

\bibitem[{{Dud{\'\i}k} {et~al.}(2019){Dud{\'\i}k}, {Dzif{\v{c}}{\'a}kov{\'a}},
  {Del Zanna}, {Mason}, {Golub}, {Winebarger}, \& {Savage}}]{dudik2019}
{Dud{\'\i}k}, J., {Dzif{\v{c}}{\'a}kov{\'a}}, E., {Del Zanna}, G., {et~al.}
  2019, \aap, 626, A88

\bibitem[{{Dzif{\v{c}}{\'a}kov{\'a}} \& {Dud{\'\i}k}(2018)}]{dzifcakova2018b}
{Dzif{\v{c}}{\'a}kov{\'a}}, E., \& {Dud{\'\i}k}, J. 2018, \aap, 610, A67

\bibitem[{{Dzif{\v{c}}{\'a}kov{\'a}} {et~al.}(2015){Dzif{\v{c}}{\'a}kov{\'a}},
  {Dud{\'\i}k}, {Kotr{\v{c}}}, {F{\'a}rn{\'\i}k}, \&
  {Zemanov{\'a}}}]{dzifcakova2015}
{Dzif{\v{c}}{\'a}kov{\'a}}, E., {Dud{\'\i}k}, J., {Kotr{\v{c}}}, P.,
  {F{\'a}rn{\'\i}k}, F., \& {Zemanov{\'a}}, A. 2015, \apjs, 217, 14

\bibitem[{{Dzif{\v{c}}{\'a}kov{\'a}} {et~al.}(2018){Dzif{\v{c}}{\'a}kov{\'a}},
  {Zemanov{\'a}}, {Dud{\'\i}k}, \& {Mackovjak}}]{dzifcakova2018}
{Dzif{\v{c}}{\'a}kov{\'a}}, E., {Zemanov{\'a}}, A., {Dud{\'\i}k}, J., \&
  {Mackovjak}, {\v{S}}. 2018, \apj, 853, 158

\bibitem[{{Emslie} {et~al.}(1992){Emslie}, {Li}, \& {Mariska}}]{emslie1992}
{Emslie}, A.~G., {Li}, P., \& {Mariska}, J.~T. 1992, \apj, 399, 714

\bibitem[{{Feldman} {et~al.}(1995){Feldman}, {Laming}, \&
  {Doschek}}]{feldman1995}
{Feldman}, U., {Laming}, J.~M., \& {Doschek}, G.~A. 1995, \apjl, 451, L79

\bibitem[{{Garcia}(1994)}]{garcia1994}
{Garcia}, H.~A. 1994, \solphys, 154, 275

\bibitem[{{Guidoni} {et~al.}(2016){Guidoni}, {DeVore}, {Karpen}, \&
  {Lynch}}]{guidoni2016}
{Guidoni}, S.~E., {DeVore}, C.~R., {Karpen}, J.~T., \& {Lynch}, B.~J. 2016,
  \apj, 820, 60

\bibitem[{{Haines}(2011)}]{haines2011}
{Haines}, M.~G. 2011, Plasma Physics and Controlled Fusion, 53, 093001

\bibitem[{{Hayes} {et~al.}(2017){Hayes}, {Gallagher}, {McCauley}, {Dennis},
  {Ireland}, \& {Inglis}}]{hayes2017}
{Hayes}, L.~A., {Gallagher}, P.~T., {McCauley}, J., {et~al.} 2017, Journal of
  Geophysical Research (Space Physics), 122, 9841

\bibitem[{{Hayes} {et~al.}(2020){Hayes}, {Inglis}, {Christe}, {Dennis}, \&
  {Gallagher}}]{hayes2020}
{Hayes}, L.~A., {Inglis}, A.~R., {Christe}, S., {Dennis}, B., \& {Gallagher},
  P.~T. 2020, \apj, 895, 50

\bibitem[{{Hinterreiter} {et~al.}(2018){Hinterreiter}, {Veronig}, {Thalmann},
  {Tschernitz}, \& {P{\"o}tzi}}]{hinterreiter2018}
{Hinterreiter}, J., {Veronig}, A.~M., {Thalmann}, J.~K., {Tschernitz}, J., \&
  {P{\"o}tzi}, W. 2018, \solphys, 293, 38

\bibitem[{{Kerr} {et~al.}(2019){Kerr}, {Carlsson}, {Allred}, {Young}, \&
  {Daw}}]{kerr2019}
{Kerr}, G.~S., {Carlsson}, M., {Allred}, J.~C., {Young}, P.~R., \& {Daw}, A.~N.
  2019, \apj, 871, 23

\bibitem[{{Kerr} {et~al.}(2021){Kerr}, {Xu}, {Allred}, {Polito}, {Sadykov},
  {Huang}, \& {Wang}}]{kerr2021}
{Kerr}, G.~S., {Xu}, Y., {Allred}, J.~C., {et~al.} 2021, \apj, 912, 153

\bibitem[{{Klimchuk}(2000)}]{klimchuk2000}
{Klimchuk}, J.~A. 2000, \solphys, 193, 53

\bibitem[{{Klimchuk} \& {DeForest}(2020)}]{klimchuk2020}
{Klimchuk}, J.~A., \& {DeForest}, C.~E. 2020, \apj, 900, 167

\bibitem[{{Klimchuk} {et~al.}(1992){Klimchuk}, {Lemen}, {Feldman}, {Tsuneta},
  \& {Uchida}}]{klimchuk1992}
{Klimchuk}, J.~A., {Lemen}, J.~R., {Feldman}, U., {Tsuneta}, S., \& {Uchida},
  Y. 1992, \pasj, 44, L181

\bibitem[{{Kowalski} {et~al.}(2017){Kowalski}, {Allred}, {Daw}, {Cauzzi}, \&
  {Carlsson}}]{kowalski2017}
{Kowalski}, A.~F., {Allred}, J.~C., {Daw}, A., {Cauzzi}, G., \& {Carlsson}, M.
  2017, \apj, 836, 12

\bibitem[{{Kruskal} \& {Schwarzschild}(1954)}]{kruskal1954}
{Kruskal}, M., \& {Schwarzschild}, M. 1954, Proceedings of the Royal Society of
  London Series A, 223, 348

\bibitem[{{Lin} {et~al.}(2002){Lin}, {Dennis}, {Hurford}, {Smith}, {Zehnder},
  {Harvey}, {Curtis}, {Pankow}, {Turin}, {Bester}, {Csillaghy}, {Lewis},
  {Madden}, {van Beek}, {Appleby}, {Raudorf}, {McTiernan}, {Ramaty}, {Schmahl},
  {Schwartz}, {Krucker}, {Abiad}, {Quinn}, {Berg}, {Hashii}, {Sterling},
  {Jackson}, {Pratt}, {Campbell}, {Malone}, {Landis}, {Barrington-Leigh},
  {Slassi-Sennou}, {Cork}, {Clark}, {Amato}, {Orwig}, {Boyle}, {Banks},
  {Shirey}, {Tolbert}, {Zarro}, {Snow}, {Thomsen}, {Henneck}, {McHedlishvili},
  {Ming}, {Fivian}, {Jordan}, {Wanner}, {Crubb}, {Preble}, {Matranga}, {Benz},
  {Hudson}, {Canfield}, {Holman}, {Crannell}, {Kosugi}, {Emslie}, {Vilmer},
  {Brown}, {Johns-Krull}, {Aschwanden}, {Metcalf}, \& {Conway}}]{lin2002}
{Lin}, R.~P., {Dennis}, B.~R., {Hurford}, G.~J., {et~al.} 2002, \solphys, 210,
  3

\bibitem[{{Longcope} {et~al.}(2009){Longcope}, {Guidoni}, \&
  {Linton}}]{longcope2009}
{Longcope}, D.~W., {Guidoni}, S.~E., \& {Linton}, M.~G. 2009, \apjl, 690, L18

\bibitem[{{Mandage} \& {Bradshaw}(2020)}]{mandage2020}
{Mandage}, R.~S., \& {Bradshaw}, S.~J. 2020, \apj, 891, 122

\bibitem[{{Miki{\'c}} {et~al.}(2013){Miki{\'c}}, {Lionello}, {Mok}, {Linker},
  \& {Winebarger}}]{mikic2013}
{Miki{\'c}}, Z., {Lionello}, R., {Mok}, Y., {Linker}, J.~A., \& {Winebarger},
  A.~R. 2013, \apj, 773, 94

\bibitem[{{Milligan} {et~al.}(2020){Milligan}, {Hudson}, {Chamberlin},
  {Hannah}, \& {Hayes}}]{milligan2020}
{Milligan}, R.~O., {Hudson}, H.~S., {Chamberlin}, P.~C., {Hannah}, I.~G., \&
  {Hayes}, L.~A. 2020, Space Weather, 18, e02331

\bibitem[{{Moore} {et~al.}(2018){Moore}, {Caspi}, {Woods}, {Chamberlin},
  {Dennis}, {Jones}, {Mason}, {Schwartz}, \& {Tolbert}}]{moore2018}
{Moore}, C.~S., {Caspi}, A., {Woods}, T.~N., {et~al.} 2018, \solphys, 293, 21

\bibitem[{{Osborne} \& {Mili{\'c}}(2021)}]{osborne2021}
{Osborne}, C. M.~J., \& {Mili{\'c}}, I. 2021, \apj, 917, 14

\bibitem[{{Pascoe} {et~al.}(2007){Pascoe}, {Nakariakov}, \&
  {Arber}}]{pascoe2007}
{Pascoe}, D.~J., {Nakariakov}, V.~M., \& {Arber}, T.~D. 2007, \aap, 461, 1149

\bibitem[{{Pereira} \& {Uitenbroek}(2015)}]{pereira2015}
{Pereira}, T. M.~D., \& {Uitenbroek}, H. 2015, \aap, 574, A3

\bibitem[{{Pesnell} {et~al.}(2012){Pesnell}, {Thompson}, \&
  {Chamberlin}}]{pesnell2012}
{Pesnell}, W.~D., {Thompson}, B.~J., \& {Chamberlin}, P.~C. 2012, \solphys,
  275, 3

\bibitem[{{Polito} {et~al.}(2018){Polito}, {Dud{\'\i}k}, {Ka{\v{s}}parov{\'a}},
  {Dzif{\v{c}}{\'a}kov{\'a}}, {Reeves}, {Testa}, \& {Chen}}]{polito2018}
{Polito}, V., {Dud{\'\i}k}, J., {Ka{\v{s}}parov{\'a}}, J., {et~al.} 2018, \apj,
  864, 63

\bibitem[{{Qian} {et~al.}(2010){Qian}, {Burns}, {Chamberlin}, \&
  {Solomon}}]{qian2010}
{Qian}, L., {Burns}, A.~G., {Chamberlin}, P.~C., \& {Solomon}, S.~C. 2010,
  Journal of Geophysical Research (Space Physics), 115, A09311

\bibitem[{{Qian} {et~al.}(2019){Qian}, {Wang}, {Burns}, {Chamberlin}, {Coster},
  {Zhang}, \& {Solomon}}]{qian2019}
{Qian}, L., {Wang}, W., {Burns}, A.~G., {et~al.} 2019, Journal of Geophysical
  Research (Space Physics), 124, 2298

\bibitem[{{Qiu} \& {Longcope}(2016)}]{qiu2016}
{Qiu}, J., \& {Longcope}, D.~W. 2016, \apj, 820, 14

\bibitem[{{Rathore} \& {Carlsson}(2015)}]{rathore2015}
{Rathore}, B., \& {Carlsson}, M. 2015, \apj, 811, 80

\bibitem[{{Reep} {et~al.}(2019){Reep}, {Bradshaw}, {Crump}, \&
  {Warren}}]{reep2019}
{Reep}, J.~W., {Bradshaw}, S.~J., {Crump}, N.~A., \& {Warren}, H.~P. 2019,
  \apj, 871, 18

\bibitem[{{Reep} \& {Knizhnik}(2019)}]{reep2019b}
{Reep}, J.~W., \& {Knizhnik}, K.~J. 2019, \apj, 874, 157

\bibitem[{{Reep} \& {Toriumi}(2017)}]{reep2017}
{Reep}, J.~W., \& {Toriumi}, S. 2017, \apj, 851, 4

\bibitem[{{Reep} {et~al.}(2020){Reep}, {Warren}, {Moore}, {Suarez}, \&
  {Hayes}}]{reep2020}
{Reep}, J.~W., {Warren}, H.~P., {Moore}, C.~S., {Suarez}, C., \& {Hayes}, L.~A.
  2020, \apj, 895, 30

\bibitem[{Sen(1968)}]{sen1968}
Sen, P.~K. 1968, Journal of the American Statistical Association, 63, 1379.
\newblock \url{http://www.jstor.org/stable/2285891}

\bibitem[{{Tayler}(1957)}]{tayler1957}
{Tayler}, R.~J. 1957, Proceedings of the Physical Society B, 70, 31

\bibitem[{{Tei} {et~al.}(2018){Tei}, {Sakaue}, {Okamoto}, {Kawate}, {Heinzel},
  {UeNo}, {Asai}, {Ichimoto}, \& {Shibata}}]{tei2018}
{Tei}, A., {Sakaue}, T., {Okamoto}, T.~J., {et~al.} 2018, \pasj, 70, 100

\bibitem[{Theil(1992)}]{theil1992}
Theil, H. 1992, A Rank-Invariant Method of Linear and Polynomial Regression
  Analysis, ed. B.~Raj \& J.~Koerts (Dordrecht: Springer Netherlands),
  345--381.
\newblock \url{https://doi.org/10.1007/978-94-011-2546-8_20}

\bibitem[{{Thiemann} {et~al.}(2018){Thiemann}, {Chamberlin}, {Eparvier}, \&
  {Epp}}]{thiemann2018}
{Thiemann}, E.~M.~B., {Chamberlin}, P.~C., {Eparvier}, F.~G., \& {Epp}, L.
  2018, \solphys, 293, 19

\bibitem[{{Toriumi} {et~al.}(2017){Toriumi}, {Schrijver}, {Harra}, {Hudson}, \&
  {Nagashima}}]{toriumi2017}
{Toriumi}, S., {Schrijver}, C.~J., {Harra}, L.~K., {Hudson}, H., \&
  {Nagashima}, K. 2017, \apj, 834, 56

\bibitem[{{Warren}(2006)}]{warren2006}
{Warren}, H.~P. 2006, Advances in Space Research, 37, 359

\bibitem[{{Warren} \& {Antiochos}(2004)}]{warren2004}
{Warren}, H.~P., \& {Antiochos}, S.~K. 2004, \apjl, 611, L49

\bibitem[{{Warren} {et~al.}(1998){Warren}, {Mariska}, \& {Lean}}]{warren1998}
{Warren}, H.~P., {Mariska}, J.~T., \& {Lean}, J. 1998, \jgr, 103, 12077

\bibitem[{{Warren} {et~al.}(2010){Warren}, {Winebarger}, \&
  {Brooks}}]{warren2010}
{Warren}, H.~P., {Winebarger}, A.~R., \& {Brooks}, D.~H. 2010, \apj, 711, 228

\bibitem[{{White} {et~al.}(2005){White}, {Thomas}, \& {Schwartz}}]{white2005}
{White}, S.~M., {Thomas}, R.~J., \& {Schwartz}, R.~A. 2005, \solphys, 227, 231

\bibitem[{{Woods} {et~al.}(2012){Woods}, {Eparvier}, {Hock}, {Jones},
  {Woodraska}, {Judge}, {Didkovsky}, {Lean}, {Mariska}, {Warren}, {McMullin},
  {Chamberlin}, {Berthiaume}, {Bailey}, {Fuller-Rowell}, {Sojka}, {Tobiska}, \&
  {Viereck}}]{woods2012}
{Woods}, T.~N., {Eparvier}, F.~G., {Hock}, R., {et~al.} 2012, \solphys, 275,
  115

\end{thebibliography}
\bibliographystyle{aasjournal}

\appendix

\addcontentsline{toc}{section}{Appendices}
\renewcommand{\thesubsection}{\Alph{subsection}}

\subsection{Expansion of Loop Cross-Section with Increasing Plasma $\beta$}

Although SDO/EVE is not spatially resolved, the variation of the slope with temperature found in Figure \ref{fig:scaling} can possibly inform us about the geometry of the flare loops.  In this paper, we have assumed that the change in irradiance scaling with temperature is caused by a change in cross-sectional area of the flaring loops, which  affects the EM at a given temperature directly due to the change in volume and indirectly through the change in densities, thus the line intensities can be drastically affected by a varying area.  A simple explanation of why the area should vary is the conservation of magnetic flux, that is, $A(s) B(s) = \text{const.}$, so the area scales inversely with the field strength, $A(s) \propto \frac{1}{B(s)}$.

While this could be explained by a decrease in magnetic field strength from chromosphere to corona, we do not necessarily expect a similar coronal field strength in flares across orders of magnitude in GOES class.  We speculate that there could also be variation with temperature due to a change in the magnetic tension as the plasma $\beta$ increases with temperature.  In this appendix, we derive the magnetic tension from the ideal MHD equations and show that it scales with the gas pressure $P$, and therefore with $\beta$.  The implication is that the cross-sectional area expands as the plasma is heated, although this remains unverified.  

The standard magnetohydrostatic pressure balance equation is given by (neglecting gravity):
\begin{equation}
    \nabla P = \vec{j} \times \vec{B}
    \label{eqn:hydrostatic}
\end{equation}
\noindent where $P = 2 n k_{B} T$ is the gas pressure, $\vec{j}$ the current density, and $\vec{B}$ the magnetic field.  In the limit that $\beta = \frac{P}{B^{2}/8\pi}$ is significantly less than unity, the magnetic field dominates and the gas pressure $P$ is small, so this equation reduces to the force-free field equation
\begin{equation}
    \vec{j} \times \vec{B} = 0
\end{equation}
\noindent which can be solved to give the equations describing either a potential field ($\vec{j} = 0$) or the more general case ($\vec{j} \parallel \vec{B}$).  This holds in many conditions in the solar corona.  

However, in flares, both the coronal temperature and density rise more than an order of magnitude over standard active region values.  This suggests that the plasma $\beta$ is not necessarily small.  We can solve for when the $\beta$ exceeds unity
\begin{align}
    \beta = 1 &= \frac{2 n k_{B} T}{B^{2} / 8 \pi}  \nonumber \\
     B &\geq \sqrt{16 \pi n k_{B} T} 
\end{align}
\noindent Using typical flare parameters, say a temperature of 20\,MK and number density of $\approx 10^{11}$\,cm$^{-3}$, the assumption of low-$\beta$ is no longer valid for field strengths less than approximately 120 G.  While this is a plausible coronal value, most extrapolations suggest that the average field strength is well below this.  However, even a moderate $\beta$ ($\approx 0.1$) would necessitate that the $\vec{j} \times \vec{B}$ force not be negligible.

\begin{figure*}
    \centering
    \includegraphics[width=\linewidth]{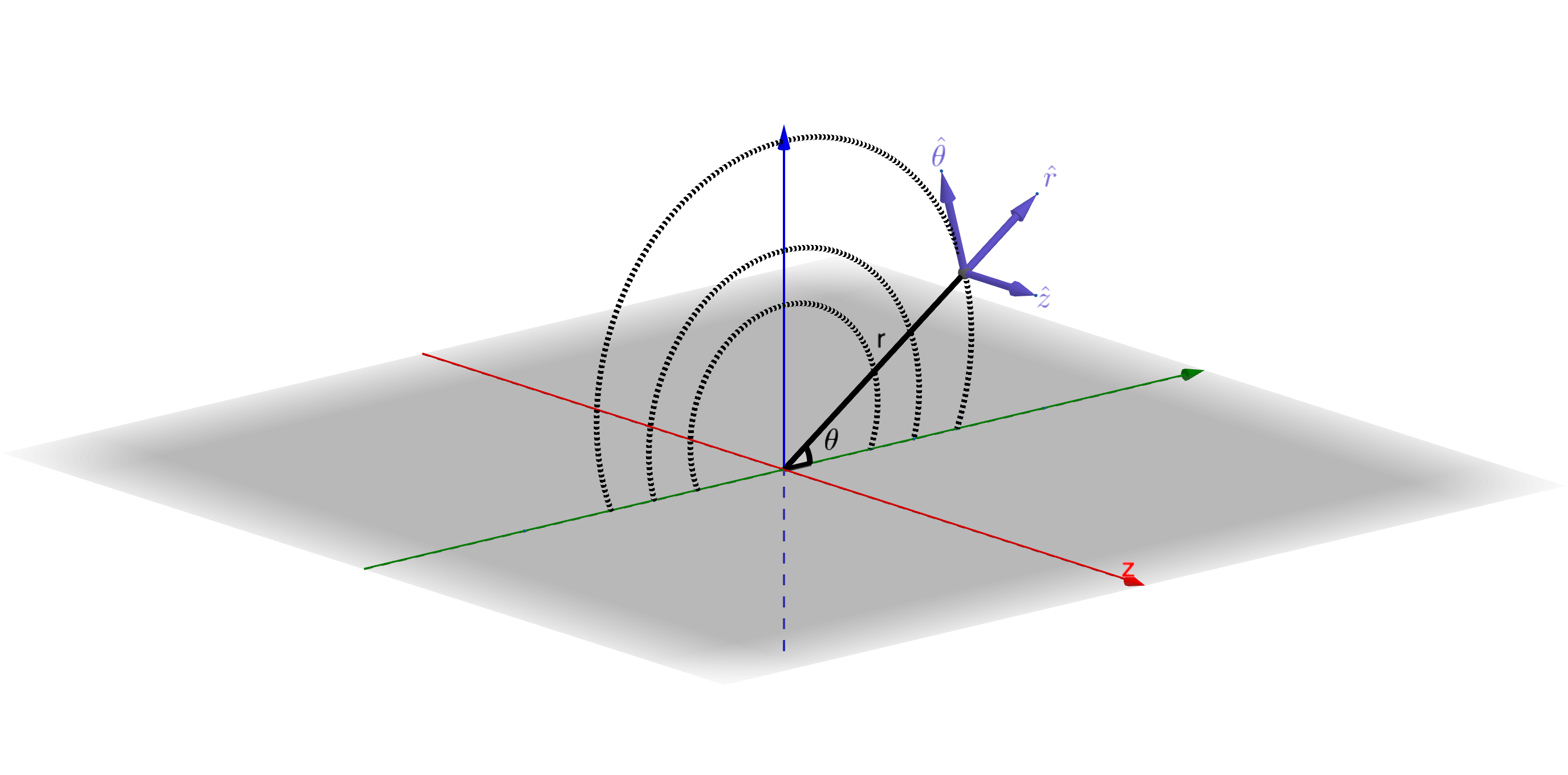}
    \caption{Geometry of the system in cylindrical coordinates.  For an arcade of loops along the $z$-axis, an individual field line can be described generally as $\vec{B} = B_{r}(r, \theta)\ \hat{r} + B_{\theta}(r, \theta)\ \hat{\theta}$.  In the special case that the field lines are semi-circular, $\vec{B} = B_{\theta}(r, \theta)\ \hat{\theta}$.  }
    \label{fig:geometry}
\end{figure*}

Supposing that the plasma $\beta$ does become relatively large, then, the $\vec{j} \times \vec{B}$ force is no longer negligible.  In cylindrical coordinates (see geometry in Figure \ref{fig:geometry}), assuming the loop inclination is small so that the $\hat{z}$-component is 0, we can write $\vec{B} = B_{r}(r, \theta)\ \hat{r} + B_{\theta}(r, \theta)\ \hat{\theta}$ at radius $r$, with $0 \le \theta \le \pi$.   

From Ampere's law, we can also write
\begin{align}
    \vec{j} &= \frac{1}{4 \pi} \Big(\nabla \times \vec{B}\Big) \nonumber \\
            &= \frac{1}{4 \pi} \Bigg[\frac{B_{\theta}}{r} + \frac{\partial B_{\theta}}{\partial r} - \frac{1}{r} \frac{\partial B_{r}}{\partial \theta} \Bigg] \hat{z}
\end{align}
\noindent Substituting back into Equation \ref{eqn:hydrostatic}, we then have:
\begin{align}
    \nabla P &= \frac{1}{4 \pi} \Big((\nabla \times \vec{B}) \times \vec{B}\Big) \nonumber \\
             &= \frac{1}{4 \pi} \Bigg[ \Big( \frac{B_{\theta}}{r} \frac{\partial B_{r}}{\partial \theta} - \frac{B_{\theta}^{2}}{r} - B_{\theta}  \frac{\partial B_{\theta}}{\partial r} \Big) \hat{r} \nonumber \\
             &+ \Big(\frac{B_{r} B_{\theta}}{r} + B_{r} \frac{\partial B_{\theta}}{\partial r} - \frac{B_{r}}{r} \frac{\partial B_{r}}{\partial \theta} \Big) \hat{\theta} \Bigg]
\end{align}
\noindent  The curl of $\vec{B}$ is in the $\hat{z}$-direction, and the cross product $(\nabla \times \vec{B}) \times \vec{B}$ is then perpendicular to the field line, acting to compress the plasma confined to that field line.  We can further split this into its $\hat{r}$ and $\hat{\theta}$ components.  First, the $\hat{r}$-component:
\begin{align}
    \frac{\partial P}{\partial r} = \frac{1}{4 \pi}\Bigg( \frac{B_{\theta}}{r} \frac{\partial B_{r}}{\partial \theta} - \frac{B_{\theta}^{2}}{r} - B_{\theta}  \frac{\partial B_{\theta}}{\partial r} \Bigg)
\end{align}
\noindent which can be rewritten as:
\begin{align}
    \frac{\partial}{\partial r} \Bigg(P + \frac{B_{\theta}^{2}}{8 \pi} \Bigg) = \frac{B_{\theta}}{4 \pi r} \Bigg(\frac{\partial B_{r}}{\partial \theta} - B_{\theta} \Bigg)
\end{align}
\noindent This equation describes a magnetic pinch, where the terms on the right-hand side are the magnetic tension, acting to confine the plasma.  In the special case where the field lines are circular, \textit{i.e.}, $B_{r} = 0$, then this equation reduces to the standard equation for a Z-pinch geometry \citep{haines2011}.  

Similarly, the $\hat{\theta}$-component can be written:
\begin{align}
    \frac{\partial}{\partial \theta}\Bigg(P + \frac{B_{r}^{2}}{8 \pi} \Bigg) = \frac{B_{r} }{4 \pi} \Bigg( r \frac{\partial B_{\theta}}{\partial r} + B_{\theta} \Bigg)
\end{align}
\noindent Together, these two equations demonstrate that there is a magnetic tension perpendicular to the direction of the field line acting to confine the plasma.  As the gas pressure increases, then, the confinement weakens and the cross-sectional area expands (akin to a sausage mode instability, \citealt{kruskal1954,tayler1957,pascoe2007}).  

Stated more directly, as the plasma $\beta$ grows, we expect there to be an effective expansion of a loop's cross-sectional area.  We note, however, that this analysis is based on the idealized MHD equations, and includes many simplifications, and therefore this would be need to validated with a full MHD simulation or with observational measurements of $\beta$ \citep{brooks2021}.

\end{document}